\documentclass[hyper]{JHEP3}
\usepackage{amsmath,amsfonts}
\usepackage{epsfig}
\usepackage{cite}
\def\IC{\mathbb{C}}
\def\IP{\mathbb{P}}
\def\IT{\mathbb{T}}

\title{$N=2$ dualities}

\author{Davide Gaiotto\\
School of Natural Sciences, Institute for Advanced Study, \\
Princeton, NJ 08540, USA
\\
{\tt dgaiotto@ias.edu} }

\abstract{We study the generalization of S-duality and Argyres-Seiberg duality
for a large class of $N=2$ superconformal gauge theories. We
identify a family of strongly interacting SCFTs and use them as
building blocks for generalized superconformal quiver gauge
theories. This setup provides a detailed description of the ``very
strongly coupled'' regions in the moduli space of more familiar
gauge theories. As a byproduct, we provide a purely four dimensional
construction of $N=2$ theories defined by wrapping M5 branes over a
Riemann surface.}

\begin{document}

\bibliographystyle{utphys}

\section{Outline}

\begin{figure}
  \begin{center}
    \includegraphics[width=2.5in]{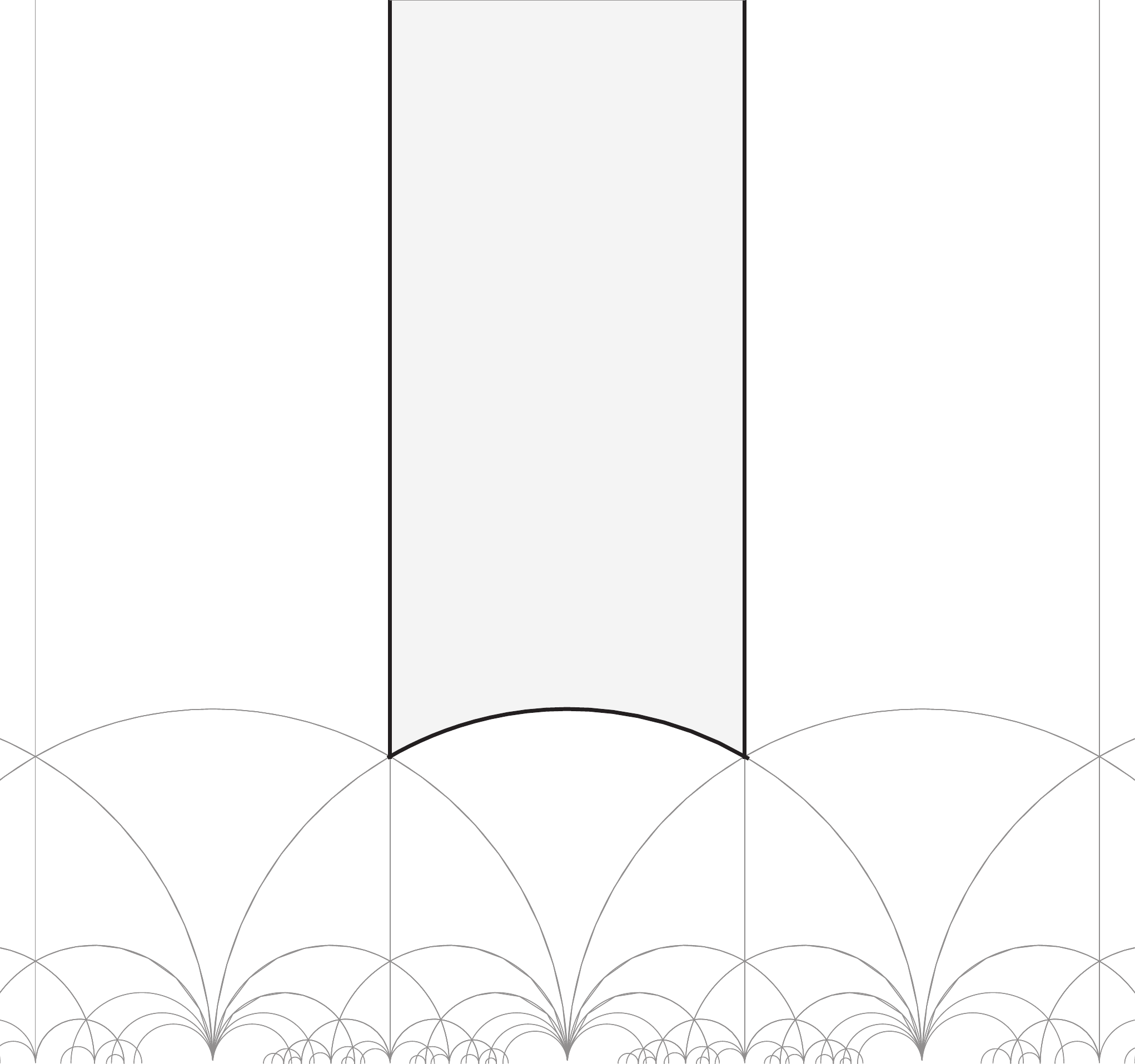} \hfill
     \includegraphics[width=2.5in]{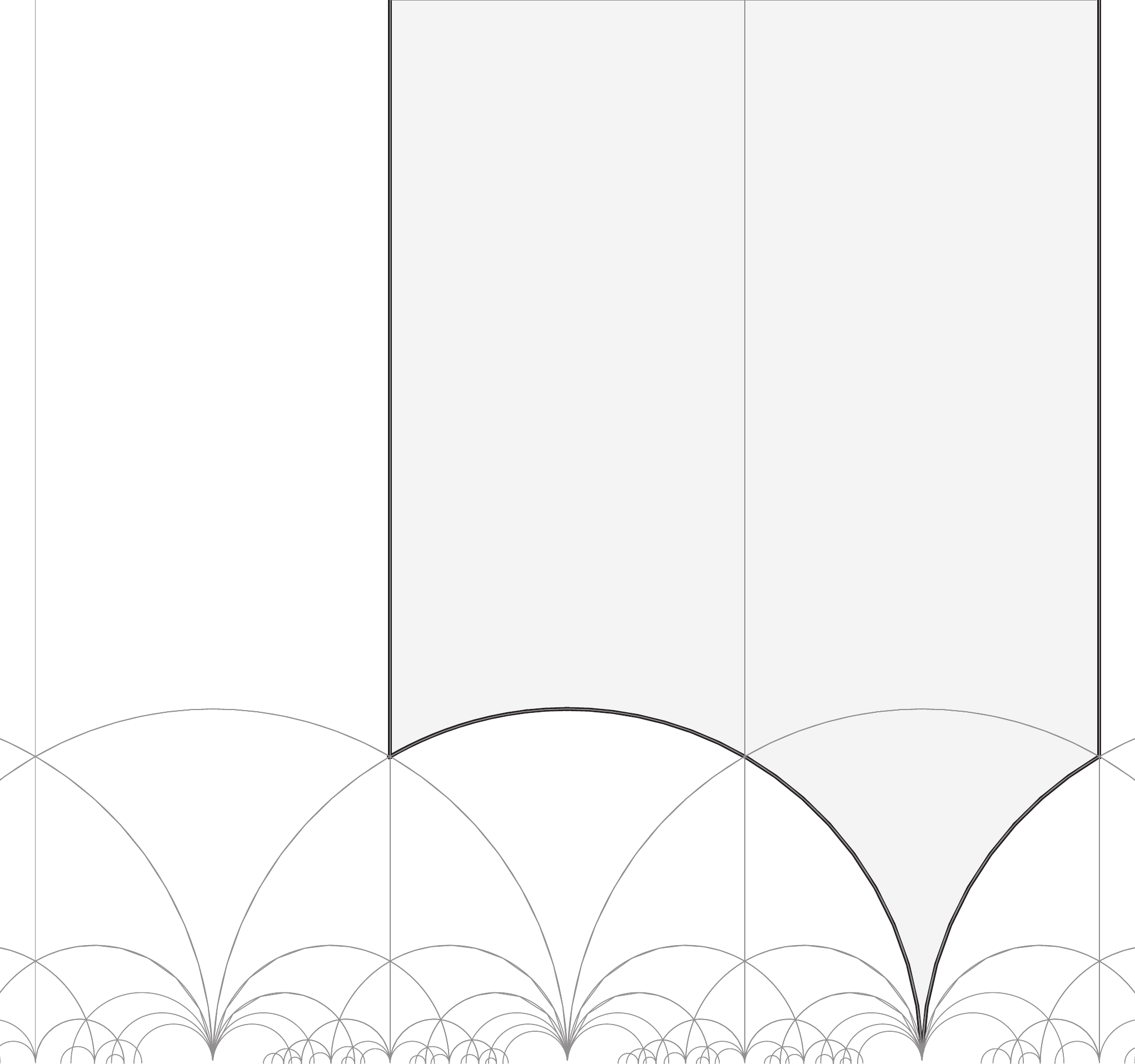}
  \end{center}
  \caption{Left: the space of gauge couplings $\tau$ modulo S-duality for $SU(2)$ $N_f=4$. The weakly coupled region is at $\tau \to i \infty$.
  Right: the space of gauge couplings $\tau$ modulo S-duality for $SU(N)$ $N_f=2N$. The weakly coupled region is at $\tau \to i \infty$. The very strongly coupled region at $\tau \to 1$}
  \label{fig:mod1}
\end{figure}

${\cal N}=2$ superconformal gauge theories have an interesting
parameter space of exactly marginal gauge couplings. $N_f=4$ $SU(2)$
gauge theory is an example: the gauge coupling
(together with the theta angle) parameterizes the upper half plane
$H$. The theory enjoys S-duality, which acts through the familiar
$SL(2,Z)$ action on the upper half plane \cite{Seiberg:1994aj}. As a
result, the actual parameter space of the SCFT is the quotient
$H/SL(2,Z)$, and has a single ``cusp'' where an $SU(2)$ gauge group
becomes arbitrarily weakly coupled. See fig. \ref{fig:mod1}

The $SU(N)$ gauge theory with $N_f = 2 N$ is also superconformal,
and has an exactly marginal gauge coupling also valued in the upper
half plane. S-duality only acts as a proper
subgroup $\Gamma_0(2)$ of $SL(2,Z)$ \cite{Argyres:1995wt}. The quotient
$H/\Gamma_0(2)$ includes both a traditional cusp where the $SU(N)$ gauge
theory becomes arbitrarily weakly coupled, and a ``very strongly
coupled'' cusp where the dynamics is in general unknown. See fig.
\ref{fig:mod1}. The authors of \cite{Argyres:2007cn} gave a surprising interpretation
of the very strongly coupled region for the $N_f=6$ $SU(3)$ gauge theory.
A dual weakly coupled $SU(2)$ gauge group emerges in the very
strongly coupled region of the $SU(3)$ gauge theory. This $SU(2)$ gauge group
is coupled to a famous rank one interacting ${\cal N}=2$ SCFT with $E_6$ symmetry.

Inspired by this result, we set up to explore the various strongly
coupled cusps in the parameter space of very general ${\cal N}=2$
superconformal gauge theories with $\prod SU(n_i)$ gauge groups. We
find that in all cases a dual weakly coupled gauge groups emerge,
coupled to a variety of ``elementary'' building blocks, interacting
${\cal N}=2$ SCFTs with no exactly marginal deformations. We learn
to combine these building blocks into a wider class of ${\cal N}=2$
``generalized quivers'' by gauging different combinations of the
flavor symmetry groups. We conjecture a Seiberg-Witten curve for
such generalized quivers and connect them into a network of S-dual
theories. Furthermore, we argue that these generalized quivers
provide a four dimensional definition of the ``theory of $N$ M5
branes wrapped on a Riemann surface''. Such theory is of some
interest, as it has a known M-theory holographic dual description
\cite{Lin:2004nb}.  This fact provided a second important motivation
for this work. Details of the holograhic correspondence will be
explored in the upcoming  \cite{GM}.

We use two main tools for our computation. The first is the general
construction of \cite{Witten:1997sc}, which provides us with the
Seiberg Witten curve for any linear quiver of unitary gauge groups.
This brane construction actually provides much more than the Seiberg
Witten curve for the theories. It truly provides a recipe for
constructing the four dimensional gauge theory as a compactification
of the six dimensional $(2,0)$ SCFT of $A_{N-1}$ type. This
information is encoded into a specific realization of the
Seiberg-Witten curve as a $N$ sheeted branched covering of a
punctured complex sphere or torus, with a canonical choice of SW
differential. Traditionally the SW curve only provides information
about the massless abelian gauge theory in the IR, and may even be
insufficient to distinguish distinct four dimensional theories. The
full fibration instead identifies the four dimensional theory
uniquely by its six-dimensional origin, encodes the BPS spectrum of
the theory, and various other properties. This useful point of view
has been developed in detail in \cite{GMN}.

While a symmetry of the SW curve is a circumstantial evidence of a
duality, a symmetry which extends to the $N$ sheeted covering structure is
essentially a proof (albeit one which depends on the existence and
properties of the $(2,0)$ theory and of certain codimension two
defect operators). A great majority of known ${\cal N}=2$ theories
(and many previously unknown) admit a description in terms of the
$(2,0)$ ADE theories, but probably not all. It is not hard to
produce certain quiver gauge theories which fall outside of the
known brane constructions (See section \ref{sec:last}). It would be
interesting to study them further. In this paper we only consider
the A-type $(2,0)$ theory. It would be interesting to extend our work to
the other types. See section \ref{sec:last} for more details and references.
We will follow an ``experimental'' approach, constructing examples
of increasing complexity and then decomposing them in simple
building blocks at the cusps of the gauge coupling moduli space.

In section \ref{sec:su2} we will consider superconformal quivers
with $SU(2)$ gauge groups. The S-duality of $SU(2)$ with $N_f=4$
extends in a surprising way to quivers of $SU(2)$ groups. Large
families of different-looking quivers with the same gauge and flavor
groups turn out to be dual descriptions of a single theory ${\cal
T}_{n,g}[A_1]$, associated to a Riemann surface of genus $g$ with
$n$ punctures $C_{n,g}$. We will identify the theory as
the twisted compactification of the $A_1$ $(2,0)$ six-dimensional CFT on
$C_{n,g}$, in the presence of $n$ defect operators.
The Seiberg-Witten curve for the theory is a ramified double cover
of $C_{n,g}$.

In section \ref{sec:su3} we will consider superconformal quivers
with $SU(3)$ (and possibly $SU(2)$) gauge groups. The
Argyres-Seiberg duality of $SU(3)$ with $N_f=6$ extends to these
quivers. Dualities relate them to various generalized quivers which include $E_6$
SCFT as building blocks. Again, we identify families of such
generalized quivers as different dual descriptions of a single
theory ${\cal T}_{(f_1,f_3),g}[A_2]$, associated to a Riemann
surface $C_{(f_1,f_3),g}$ with two types of punctures. We will identify the theory
with a twisted compactification of the $A_2$ $(2,0)$ six-dimensional
CFT on $C_{(f_1,f_3),g}$, in the presence of appropriate
defect operators. The Seiberg-Witten curve for the theory is a ramified triple cover
of $C_{(f_1,f_3),g}$.

In section \ref{sec:sun} we will extend the construction to general
superconformal linear quivers of unitary groups.
Argyres-Seiberg-like dualities will produce a variety of
``elementary'' interacting SCFTs akin to the $E_6$ theory. We will
identify families of generalized quivers ${\cal T}_{(f_a),g}[A_{N-1}]$ based on such CFTs,
and the relation to $A_{N-1}$ $(2,0)$ six-dimensional CFT on a Riemann
surface $C_{(f_a),g}$, with defects labeled by Young Tableaux with $N$ boxes.
The Seiberg-Witten curve for the theory is a ramified $N$-th cover
of $C_{(f_a),g}$.
We explore all possible very strongly coupled regions of the parameter space of
linear quivers of unitary groups. A dual weakly coupled special unitary gauge group typically emerges
in such regions, but  we also find some cases where the new gauge group is symplectic.
In particular, we find that superconformal linear quivers of unitary group with one or two terminal symplectic nodes also belong to the class  ${\cal T}_{(f_a),g}[A_{N-1}]$. The $A_3$ case includes
a second classical example of Argyres-Seiberg duality, relating a $N_f=6$ $USp(4)$ gauge theory
and the $E_7$ SCFT.

In section \ref{sec:last} we will sketch some possible
generalization of our result. Each section includes an introduction
which does not require familiarity with Seiberg-Witten curve
technology.

\section{$SU(2)$ generalized quivers}\label{sec:su2}
\subsection{S-duality}
The ${\cal N}=2$ $SU(2)$ gauge theory with four fundamental flavors
($N_f=4$) is a canonical example of ${\cal N}=2$ SCFT. The gauge
coupling
\begin{equation}
\tau = \frac{\theta}{\pi} + \frac{8 \pi i}{g^2}
\end{equation}
is exactly marginal, because the number of flavors is twice the
number of colors. The theory has an $SO(8)$ flavor symmetry.  In
general $n$ hypermultiplets in a complex representation of the gauge
group support an $SU(n)$ flavor symmetry group, which is enhanced to
$SO(2n)$ for pseudoreal representations, and $Sp(2n)$ for real
representations. The fundamental representation of $SU(2)\sim Sp(1)$
is pseudoreal. The fundamental hypermultiplets transform in a vector
$8_v$ representation of the flavor symmetry group. The ${\cal N}=2$
$SU(2)$ gauge theory with $N_f=4$ was solved in
\cite{Seiberg:1994aj}. The theory enjoys an interesting S-duality
group $SL(2,Z)$, which acts by the standard fractional linear
transformations on $\tau$ and simultaneously acts on the flavor
group $SO(8)$ by triality. The three eight dimensional
representations of $SO(8)$ are permuted the same way as $SL(2,Z)$
permutes the three even spin structures on the torus.

\begin{figure}
  \begin{center}
    \includegraphics[width=1in]{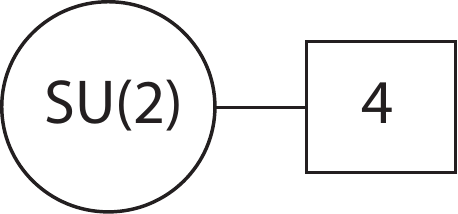} \hfill
    \includegraphics[width=1.6in]{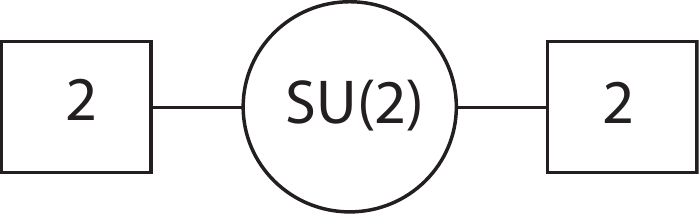}\hfill
    \includegraphics[width=1.6in]{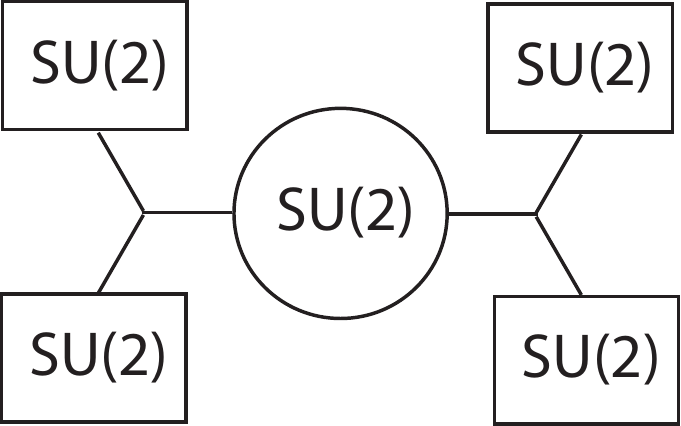}
  \end{center}
  \caption{Different useful ways to depict $SU(2)$ gauge theory with four fundamental flavors. On the left, a standard quiver diagram, where circles indicate gauge groups,
  and squares fundamental flavors; the four flavors have an $SO(8)$ flavor symmetry. In the middle, a quiver diagram where the four flavors have been split into two groups of two flavors;
  each group carries a $SO(4)=SU(2) \times SU(2)$ flavor symmetry. On the right, a generalized quiver diagram, depicting separately the two $SU(2)$
  flavor groups for each pair of fundamentals}
  \label{fig:su2nf4}
\end{figure}

\begin{figure}
  \begin{center}
    \includegraphics[width=1.5in]{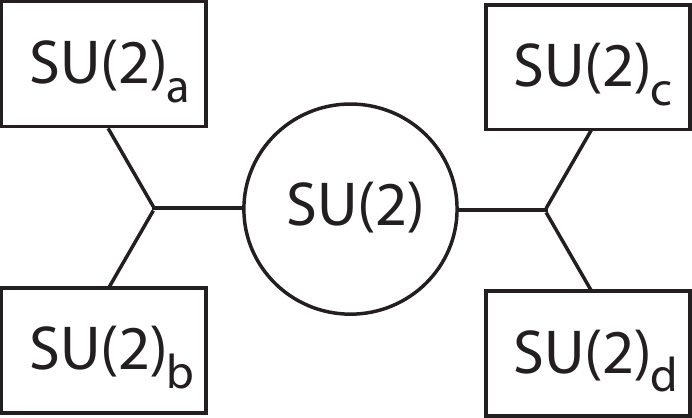}\hfill
    \includegraphics[width=1.5in]{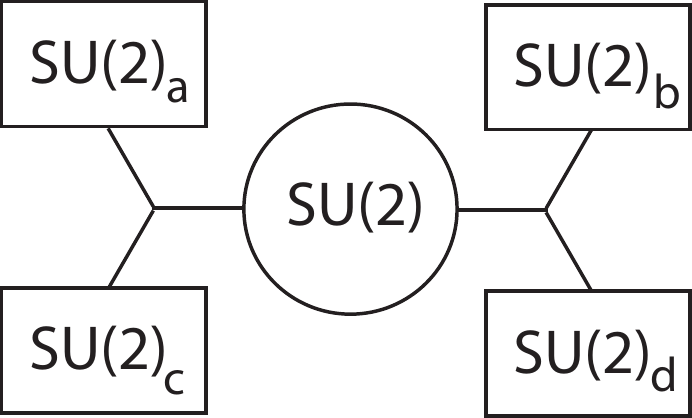}\hfill
    \includegraphics[width=1.5in]{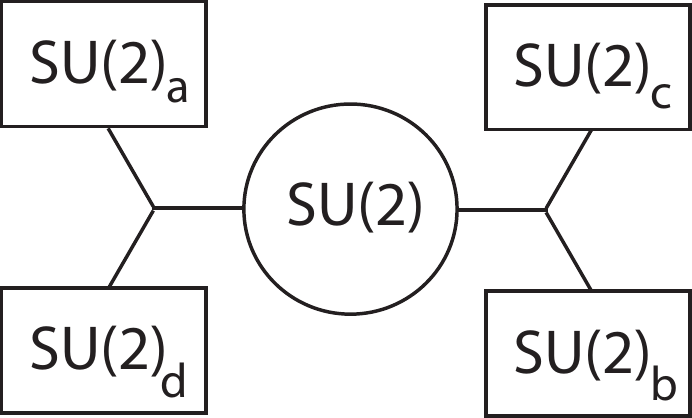}\hfill
  \end{center}
  \caption{The action of S duality on the four $SU(2)$ flavor subgroups: the three possible ways to pair them up into the $SO(4)$ symmetry of two fundamentals}
  \label{fig:su2nf4sd}
\end{figure}

It is useful to follow the action of triality on a certain subgroup
of $SO(8)$. Let us split the four fundamental hypermultiplets into
two pairs, each with a $SO(4)$ flavor symmetry. We denote one
$SO(4)$ as $SU(2)_a \times SU(2)_b$ and the other as $SU(2)_c \times
SU(2)_d$. To keep track of the four special $SU(2)$ subgroups of the
flavor group, we will denote the $SU(2)$ theory with four flavors as
in Fig. \ref{fig:su2nf4}. An interesting feature of this subgroup is that while $8_v
= (2_a \otimes 2_b) \oplus (2_c \otimes 2_d)$,  the spinorial
representations decompose as $8_s = (2_a \otimes 2_c) \oplus (2_b
\otimes 2_d)$ and $8_c = (2_a \otimes 2_d) \oplus (2_b \otimes
2_c)$. Hence triality and S-duality permute all
possible ways the four flavor symmetry subgroups $SU(2)_a \times
SU(2)_b \times SU(2)_c \times SU(2)_d$ as in Fig.
\ref{fig:su2nf4sd}.

The theory has four mass deformations, which can be presented as an
element of the Cartan subalgebra of the $SO(8)$ flavor symmetry.
Again, it is useful to restrict our attention to the $SU(2)_a \times
SU(2)_b \times SU(2)_c \times SU(2)_d$ subgroup of $SO(8)$, and
label the masses accordingly as $m_{a,b,c,d}$. In particular, if we
weakly gauge $SU(2)_a$, then $m^2_a$ will coincide with the
expectation value of the corresponding vector multiplet scalar $u_a
= Tr \Phi_a^2$. The action of triality on these mass parameters is
particularly simple, as they are permuted along with the
corresponding $SU(2)$ subgroups of $SO(8)$.

\begin{figure}
  \begin{center}
    \includegraphics[width=2.5in]{SL2plota.pdf} \hfill
     \includegraphics[width=2.5in]{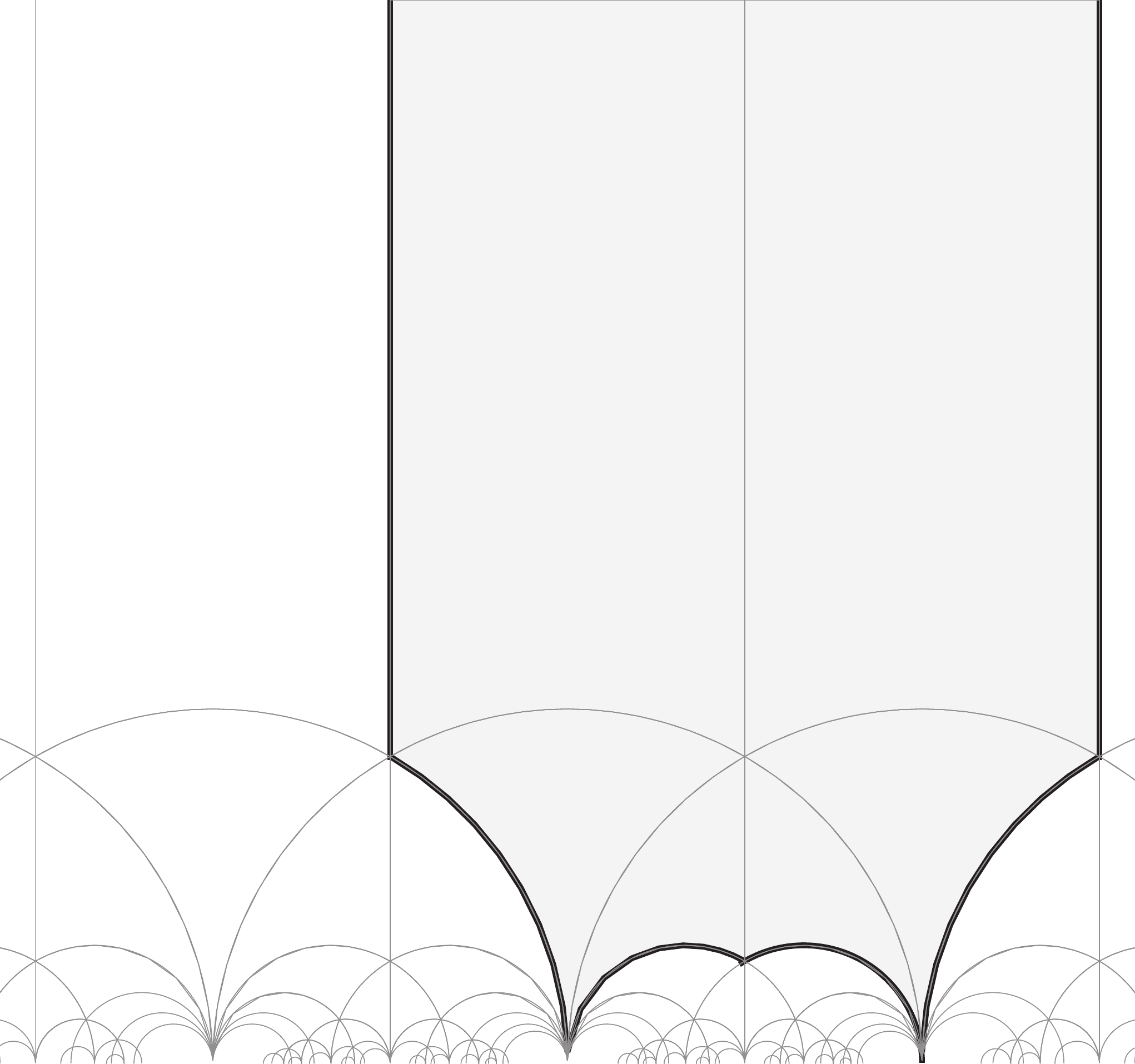}
  \end{center}
  \caption{Left: (a) the space of gauge couplings $\tau$ modulo S-duality.
  Right: (b) the space of gauge couplings $\tau$ modulo S-duality transformations which do not permute the masses. At the three cusps $\tau = i \infty$, $\tau =0$, $\tau=1$
  respectively, the weakly coupled description of the theory has flavor groups labeled as in fig. \protect \ref{fig:su2nf4sd}}
  \label{fig:mod2}
\end{figure}

\begin{figure}
  \begin{center}
    \includegraphics[width=5in]{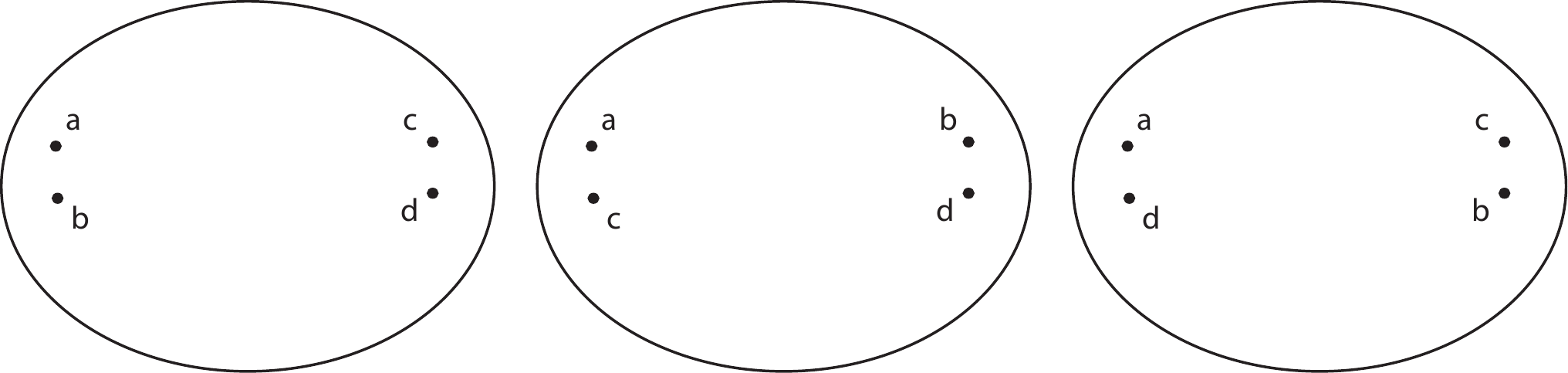}
  \end{center}
  \caption{The degeneration limits of a sphere with four marked punctures. They correspond to the three weak coupling limits of $SU(2)$ $N_f=4$ depicted in fig \protect \ref{fig:su2nf4sd}}
  \label{fig:sphere4}
\end{figure}

We want to anticipate an amusing point of view on this action, which
will be explained in better detail in the next subsection. The
parameter space of gauge couplings modulo S-duality $H/SL(2,Z)$
depicted in fig.\ref{fig:mod2}(a) coincides with the complex
structure moduli space of a sphere with four equivalent punctures.
If we only quotient by the subgroup of S-duality $\Gamma(2)$ which
does not permute the mass parameters we obtain an extended moduli
space depicted in fig.\ref{fig:mod2}(b), which coincides with the
complex structure moduli space of a sphere with four marked
punctures, labeled as $a,b,c,d$. The three cusps at $\tau =
\infty,0,1$ map to the three possible degeneration limits of the
sphere, where $a,b$, $a,c$ or $a,d$ collide (see fig.
\ref{fig:sphere4}. These are the three weak coupling limits in fig.
\ref{fig:su2nf4sd}. The full S-duality group permutes the punctures
among themselves the same way as it permutes the mass parameters
$m_{a,b,c,d}$ and the $SU(2)_{a,b,c,d}$ subgroups.

\begin{figure}
  \begin{center}
    \includegraphics[width=1.1in]{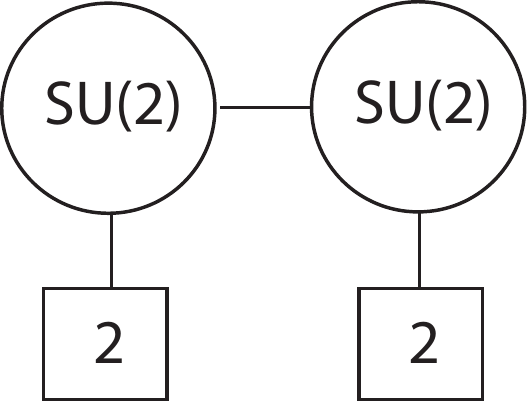} \hfill
    \includegraphics[width=2.3in]{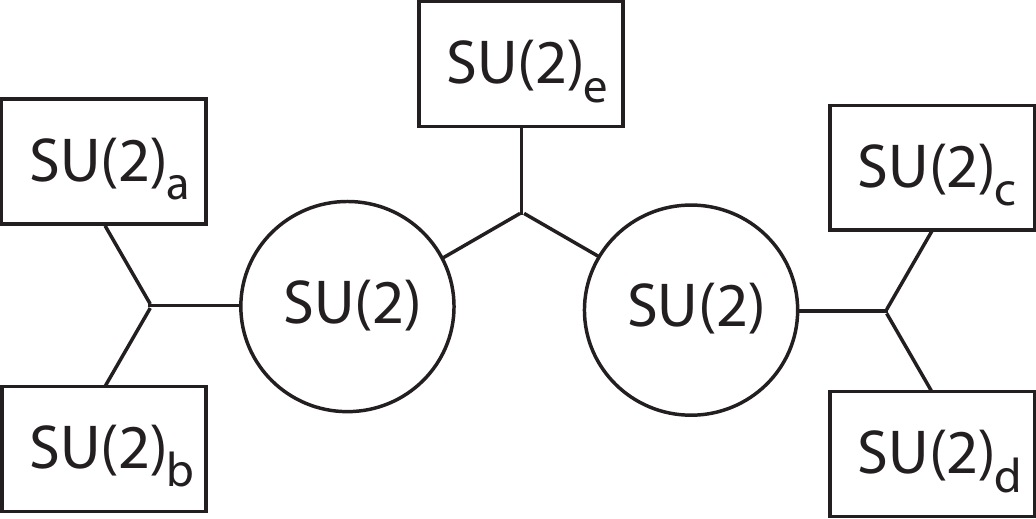}\hfill
    \includegraphics[width=1.6in]{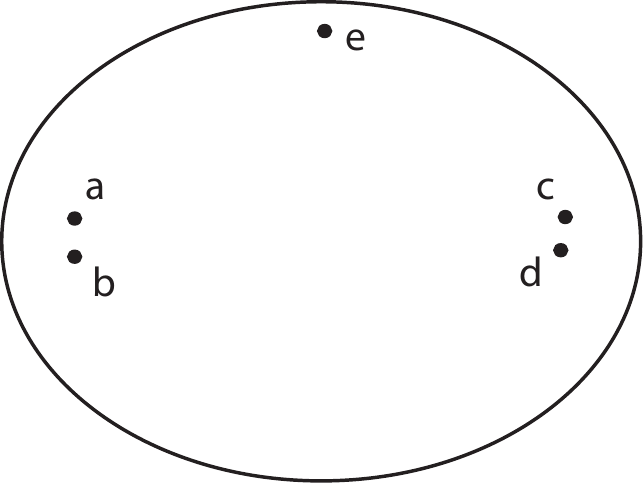}
  \end{center}
  \caption{On the left, a standard depiction of an $SU(2)$ quiver with two $SU(2)$ gauge nodes,
  and two fundamental flavors at each node. In the middle, the same theory depicted as a generalized quiver, with the $SU(2)$ factors of the flavor group made evident and labeled.
  The space of gauge couplings is parameterized by a sphere with five marked punctures. At the cusps where two pairs of punctures collide,
  there is an S-dual frame where the quiver gauge theory is weakly coupled.
  On the right, we depict the correspondence between the labeling of punctures near the cusp, and the labeling of flavor groups of the corresponding weakly coupled quiver.}
  \label{fig:su2nd2}
\end{figure}

Now we want to understand how much of this extends to more general
superconformal theories built out of $SU(2)$ gauge groups. Our first
example is a two-node quiver. The gauge group is $SU(2)_1 \times
SU(2)_2$, the matter consists of two fundamentals of $SU(2)_1$, one
bifundamental in $2_1 \otimes 2_2$ and two fundamentals of
$SU(2)_2$. Each gauge group is coupled to exactly $4$ flavors, hence
the two gauge couplings $\tau_{1,2}$ are exactly marginal and the
quiver describes a SCFT. S-duality properties of this quiver were
first analyzed in \cite{Argyres:1999fc}. The overall flavor symmetry
of this simple quiver is quite surprising: it is the product of five
factors $SU(2)_a \times SU(2)_b \times SU(2)_c \times SU(2)_d \times
SU(2)_e$. Two $SU(2)$ factors, say $a,b$, come from the $SO(4)$
flavor symmetry of the $2$ fundamentals at the first node of the
quiver. Two $SU(2)$ factors, say $c,d$ from the $SO(4)$ flavor
symmetry of the $2$ fundamentals at the second node of the quiver.
Finally, the flavor symmetry of the bifundamental hypermultiplet is
$Sp(1)\sim SU(2)_e$, as the bifundamental representation $2_1
\otimes 2_2$ is real. See Fig. \ref{fig:su2nd2}

There is a simple strategy to analyze the duality properties of this
quiver: explore regions of the moduli space where one gauge group is
extremely weakly coupled, and vary the coupling for the other gauge
group. If we completely turn off the gauge coupling at the second
node, the first gauge theory coincides with the $SU(2)$ gauge theory
with four flavors, which sit in a $(2_a \otimes 2_b) \oplus (2_e
\otimes 2_2)$ representation ($2_2$ denotes the fundamental
representation of the second gauge group). If we keep the second
gauge coupling arbitrarily weak, we expect S-duality to be still
valid for the first gauge group. If we start from the weakly coupled
quiver theory and drive the first gauge coupling to some other cusp
in the upper half plane, we should use a different S-dual
description for the first node, and ask what is the matter content
of the resulting $SU(2) \times SU(2)$ weakly coupled gauge theory.

Because of the action of triality, the first gauge group may now be
coupled to fundamental hypermultiplets in a $(2_a \otimes 2_e)
\oplus (2_b \otimes 2_2)$ or $(2_e \otimes 2_b) \oplus (2_a \otimes
2_2)$ representation. In either case we see a bifundamental
representation for the two gauge groups, hence our tentative
S-duality has brought the quiver gauge theory back to itself. In the
process, the three flavor groups $SU(2)_a \times SU(2)_b \times
SU(2)_e$ have been permuted among each other. For example, $SU(2)_a$
or $SU(2)_b$ may be the flavor symmetry of the new bifundamental
fields. Now that the gauge coupling of the first gauge group is
arbitrarily weak, albeit in a new S-dual frame, we should be able to
repeat the exercise on the second node. We could keep the first
gauge coupling fixed at this S-dual weak coupling region, and
increase the second gauge coupling, until it undergoes S-duality to
a different weakly coupled description. The four flavor symmetries
of the $SU(2)$ $N_f=4$ theory at the second node are, say, $SU(2)_a
\times SU(2)_1 \times SU(2)_c \times SU(2)_d$. In the new S-dual
frame, the three $SU(2)_a \times SU(2)_c \times SU(2)_d$ flavor
groups will be permuted.

We see that the S-dualities of single nodes of the quiver have been
inherited by the full quiver theory. Although this is not a surprise
per se, the action of S-dualities on the flavor symmetry group shows
a surprising feature: the two $SL(2,Z)$ S-duality groups do not
commute! They must be subgroups of a full S-duality group which acts
on the flavor symmetry groups as the permutation group of five
objects. Our analysis only explored the boundary of the gauge
coupling moduli space. A full analysis based on the Seiberg-Witten
curve is given in the next subsection.

Again, we find it useful to anticipate a few results. Our simple
reasoning captured all the weakly coupled cusps of the moduli space.
At all cusps the theory is described by the two-node quiver gauge
theory, and the five flavor symmetry groups appear in all possible
combinations. The gauge coupling parameter space modulo S-dualities
coincides with the complex structure moduli space of a sphere with
five punctures. If we only quotient by S-dualities which do not
permute the mass parameters $m_{a,b,c,d,e}$ and flavor symmetry
groups $SU(2)_{a,b,c,d,e}$, then we should consider instead the
moduli space of a sphere with five marked punctures, labeled by
$a,b,c,d,e$. The weakly coupled cusps of moduli space correspond to
the degeneration limits where punctures collide in different
possible ways. See fig. \ref{fig:su2nd2} for an example.

\begin{figure}
  \begin{center}
    \includegraphics[width=2in]{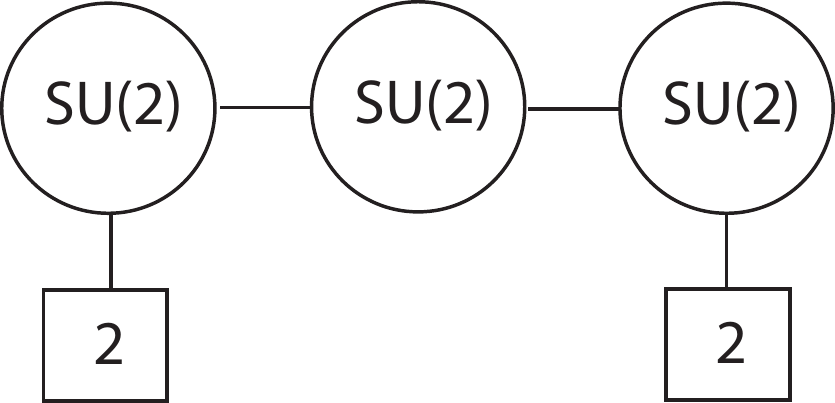} \hfill
    \includegraphics[width=3in]{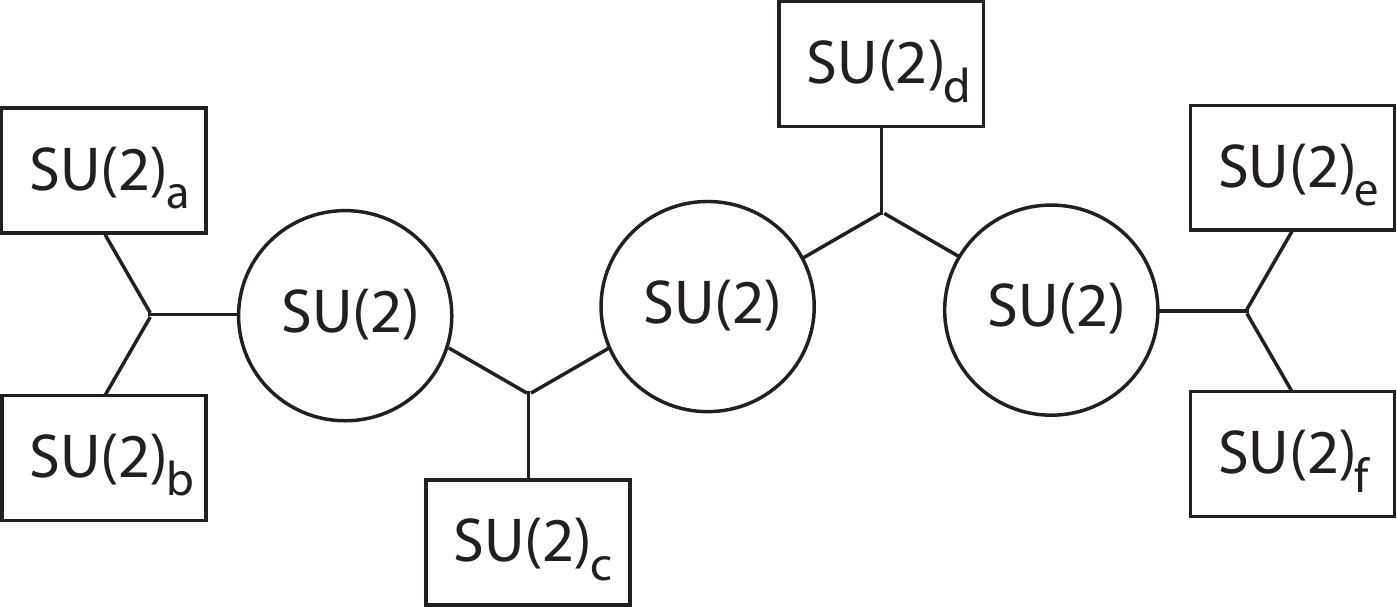}
  \end{center}
  \caption{On the left, a standard depiction of an $SU(2)$ quiver with three $SU(2)$ gauge nodes,
  and two fundamental flavors at each terminal node. On the right, the same theory depicted as a generalized quiver, with the $SU(2)$ factors of the flavor group made evident and labeled.}
  \label{fig:su2nd3}
\end{figure}

Next, we consider a superconformal linear quiver of three $SU(2)$
gauge groups, with two fundamental flavors at each end node, so that
all nodes have $N_f=4$. A novel phenomenon will become apparent. We
will denote the gauge groups as $SU(2)_1 \times SU(2)_2\times
SU(2)_3$. The reader may guess already the form of the flavor
symmetry group: six $SU(2)$ factors, $SU(2)_{a,b}$ from the first
two fundamental hypers, $SU(2)_c$ and $SU(2)_d$ from the two sets of
bifundamental hypers and $SU(2)_{e,f}$ from the fundamental hypers
at the last node. See Fig. \ref{fig:su2nd3}

\begin{figure}
  \begin{center}
    \includegraphics[width=3in]{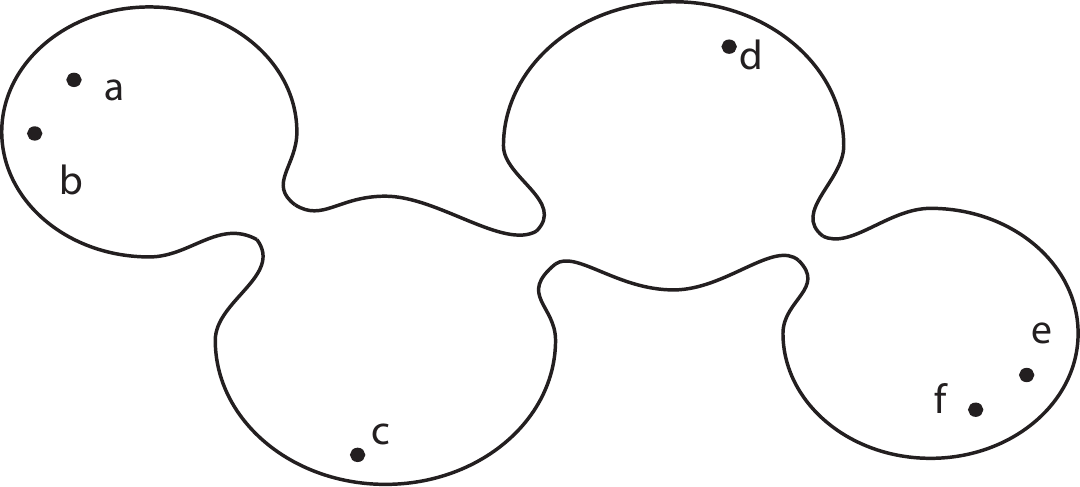}
  \end{center}
  \caption{
  The space of gauge couplings for the theory in fig. \protect \ref{fig:su2nd3} is parameterized by a sphere with six punctures. The six punctures can collide in two distinct ways. On the right, we depict
  the type of degeneration which corresponds to the weak coupling limit for the quiver.
  Notice the relation between labeling of punctures and of $SU(2)$ flavor groups.}
  \label{fig:su2nd3b}
\end{figure}
If the middle node is very weakly coupled, we expect the S-duality
at the side nodes to survive, and permute freely the roles of
$SU(2)_{a,b,c}$ or of $SU(2)_{d,e,f}$.
If we take both end nodes to be very weakly coupled, we may also
want to apply S-duality on the middle node. The matter fields at
that node sit in a $2_2 \otimes [(2_1 \otimes 2_c) \oplus (2_3
\otimes 2_d)]$ representation, where each summand is a bifundamental
hypermultiplet. If we apply S-duality, the new matter fields will
either be in $2_2 \otimes [(2_1 \otimes 2_d) \oplus (2_3 \otimes
2_c)]$ or in $2_2 \otimes [(2_1 \otimes 2_3) \oplus (2_c \otimes
2_d)]$. The first possibility gives us back the original three nodes
linear quiver. The main effect of this S-duality operation is to
permute the flavor symmetries of the two bifundamentals. Together
with S-duality at the other nodes, this is enough to generate the
full permutation group of the six gauge groups.

\begin{figure}
  \begin{center}
    \includegraphics[width=2.5in]{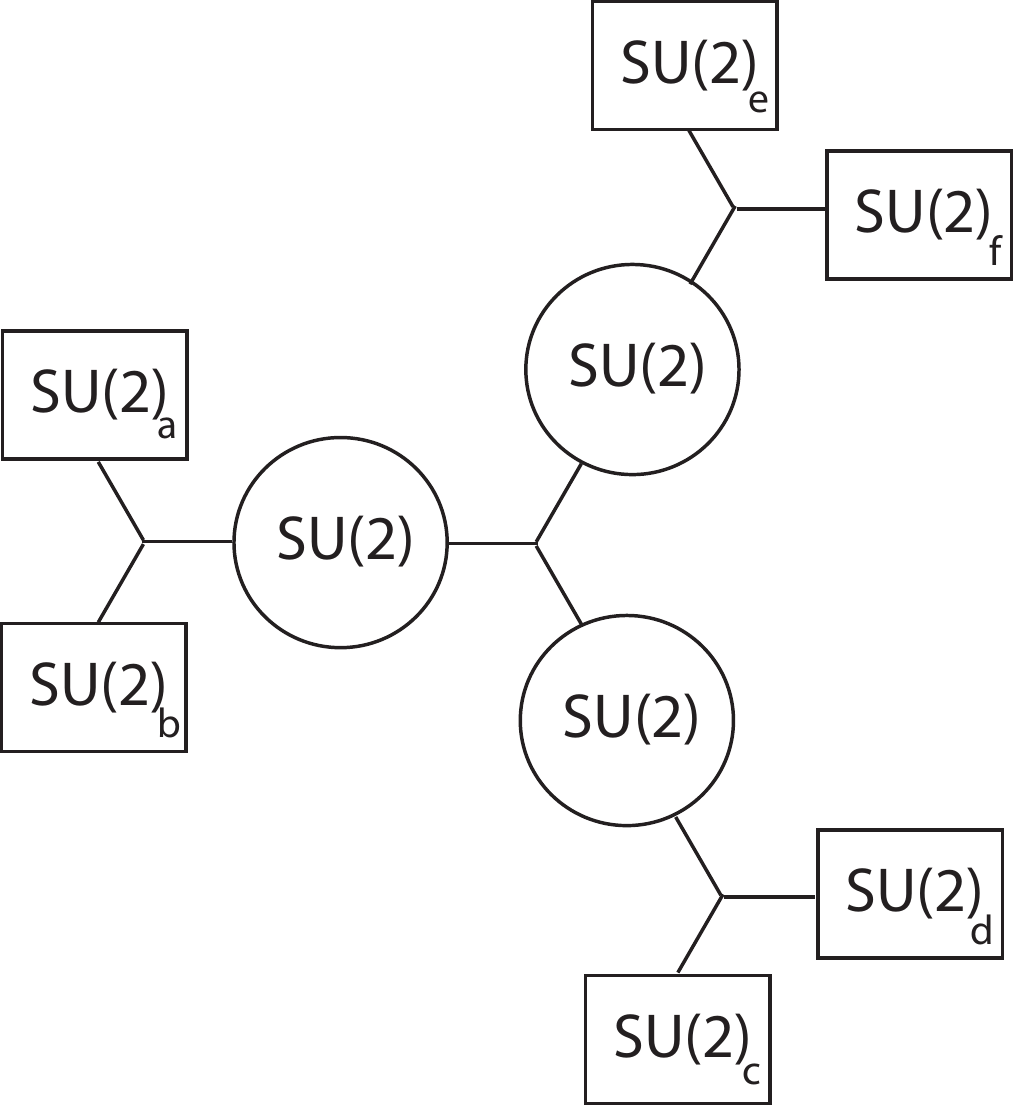}\hfill
    \includegraphics[width=2.3in]{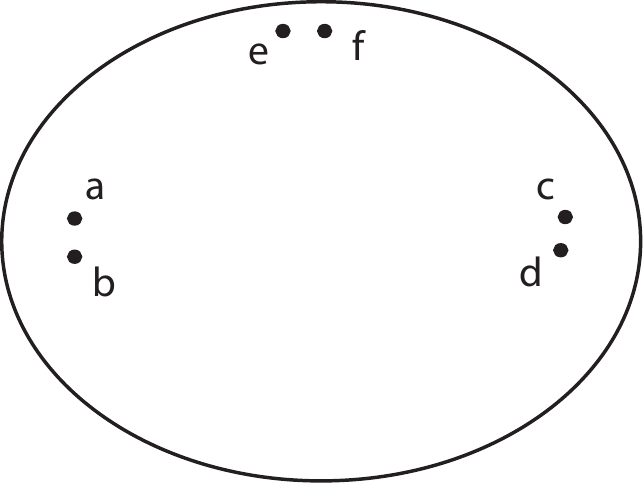}
  \end{center}
  \caption{
  From left to right. The result of S-duality at the middle node of the three-node quiver: a generalized quiver where three $SU(2)$
  gauge groups are coupled independently to the same group of four hypermultiplets. The region in gauge coupling parameter space corresponding to this quiver and this labeling of flavor groups. }
  \label{fig:su2nd3s}
\end{figure}

The second possibility is the real surprise. In that duality frame,
the theory is not a standard quiver anymore, but it is the first
example of a ``generalized $SU(2)$ quiver''. The gauge group is
still $SU(2)_1 \times SU(2)_2\times SU(2)_3$. Each of the gauge
groups now is coupled to two fundamentals, with flavor symmetries
$(2_a \otimes 2_b)$, $(2_c \otimes 2_d)$, $(2_e \otimes 2_f)$
respectively. Furthermore, they couple simultaneously to four
hypermultiplets which carry an overall $2_1 \times 2_2 \times 2_3$
representation of the gauge groups, and no flavor symmetry
\ref{fig:su2nd3s}. This object can be seen as a bifundamental
hypermultiplet of two of the gauge groups, whose $SU(2)$ flavor
symmetry has been further gauged.

This exhausts the set of weakly coupled duality frames which can be
reached by the elementary S-duality at each node. Any further
S-duality transformation would bring us back to the original linear
quiver with some rearrangement of the flavor symmetry groups. We
will show in the next subsection that this also exhausts the list of
possible cusps of the gauge coupling parameter space. The exact
gauge coupling parameter space coincides with the moduli space of a
sphere with six marked points. There are two types of degeneration
limits, depicted in figures \ref{fig:su2nd3} and \ref{fig:su2nd3s}.
They correspond to the two distinct realizations of the theory as a
weakly coupled generalized quiver.

\begin{figure}
  \begin{center}
    \includegraphics[width=2.7in]{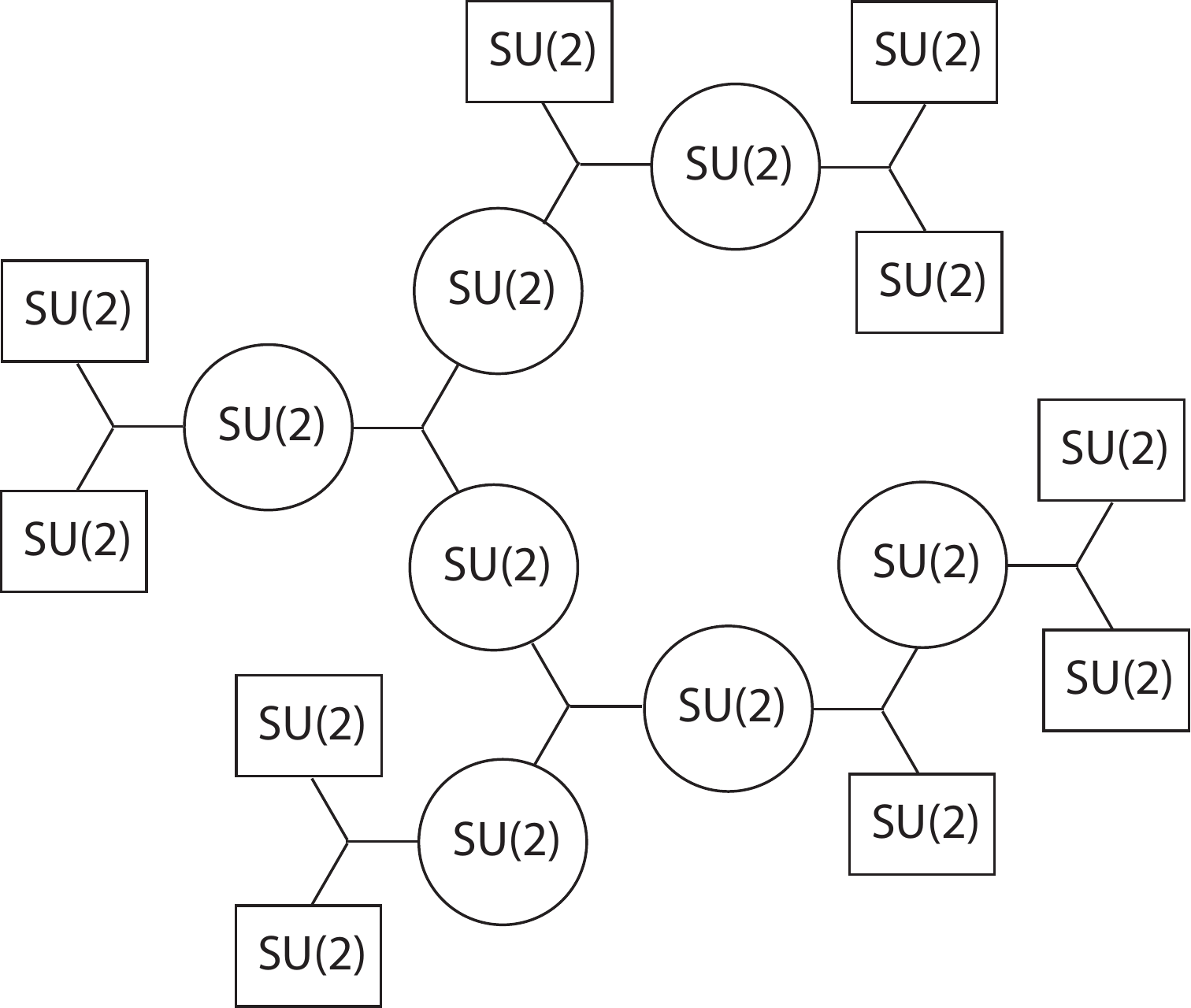}\hfill
    \includegraphics[width=2.7in]{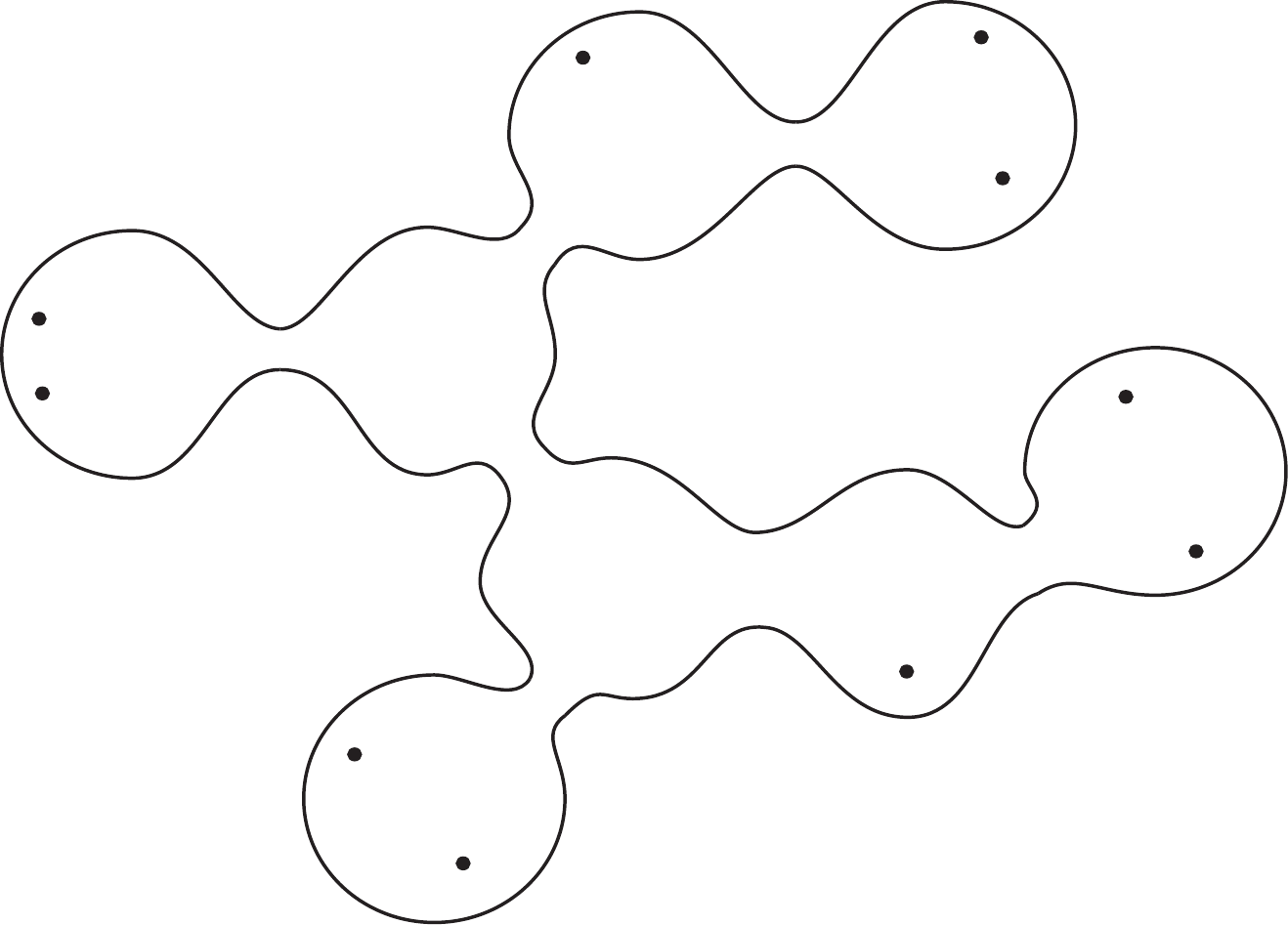}
  \end{center}
  \caption{
  A generalized quiver with no loops. The space of gauge couplings is parameterized by a sphere with $10$ punctures. The quiver is weakly coupled
  at the cusp where the punctured sphere degenerates as in figure. Punctures correspond to flavor $SU(2)$ groups.
  Long/thin necks correspond to weakly coupled $SU(2)$ gauge groups.}
  \label{fig:su2g}
\end{figure}

Finally we can consider a linear superconformal quiver theory with
$n$ $SU(2)$ nodes. There will be $n+3$ $SU(2)$ flavor groups. As
before we expect that if all nodes but one are very weakly coupled,
S-duality at that node remains a valid symmetry of the theory.
Repeated action of S-duality at different nodes can rearrange the
matter content profoundly. The result will be a variety of
generalized quivers, where some blocks of four hypermultiplets play
the role of two fundamentals for a single $SU(2)$ gauge group, some
play the role of a bifundamental for two $SU(2)$ gauge groups, and
the others are coupled simultaneously to three $SU(2)$ gauge groups.
In each duality frame there will always be $n$ $SU(2)$ gauge groups,
each with $N_f=4$. There are always $n+1$ sets of four
hypermultiplets. Finally, there must be $n+3$ flavor symmetry
groups. What is the set of possible topologies for such generalized
quivers? It will be convenient to represent each generalized quiver
as a graph, with a trivalent vertex for each hypermultiplet, an
internal edge for each gauge group, and an external edge for each
flavor group. The number of loops of such a graph is the number of
internal edges minus the number of nodes plus one, i.e. $0$. See Fig
\ref{fig:su2g}
S-duality at any edge rearranges the other four edges attached to
its endpoints. Under the action of S-duality, the graph will always
remain a (binary) tree, and it is easy to argue that every possible
topology for that tree can be reached by elementary S-duality moves.
Hence we are lead to postulate the existence of a global ${\cal
N}=2$ superconformal theory, denoted as ${\cal T}_{n+3,0}[A_1]$,
with an intricate $n$ dimensional parameter space of gauge couplings
with several weak coupling cusps, labeled by all the possible
loopless generalized quivers of $n$ $SU(2)$ gauge groups.

\begin{figure}
  \begin{center}
    \includegraphics[width=2.7in]{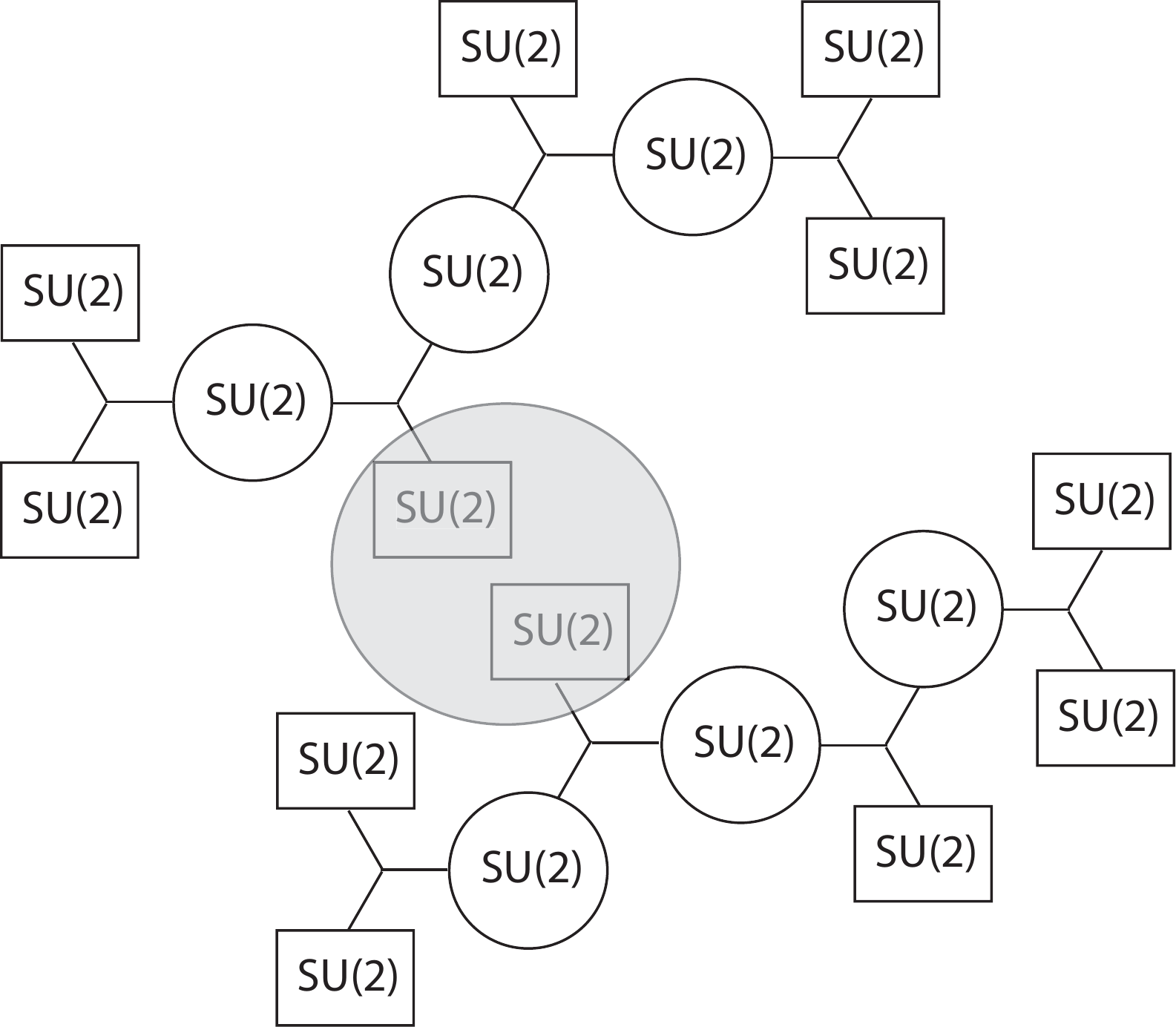}\hfill
    \includegraphics[width=2.7in]{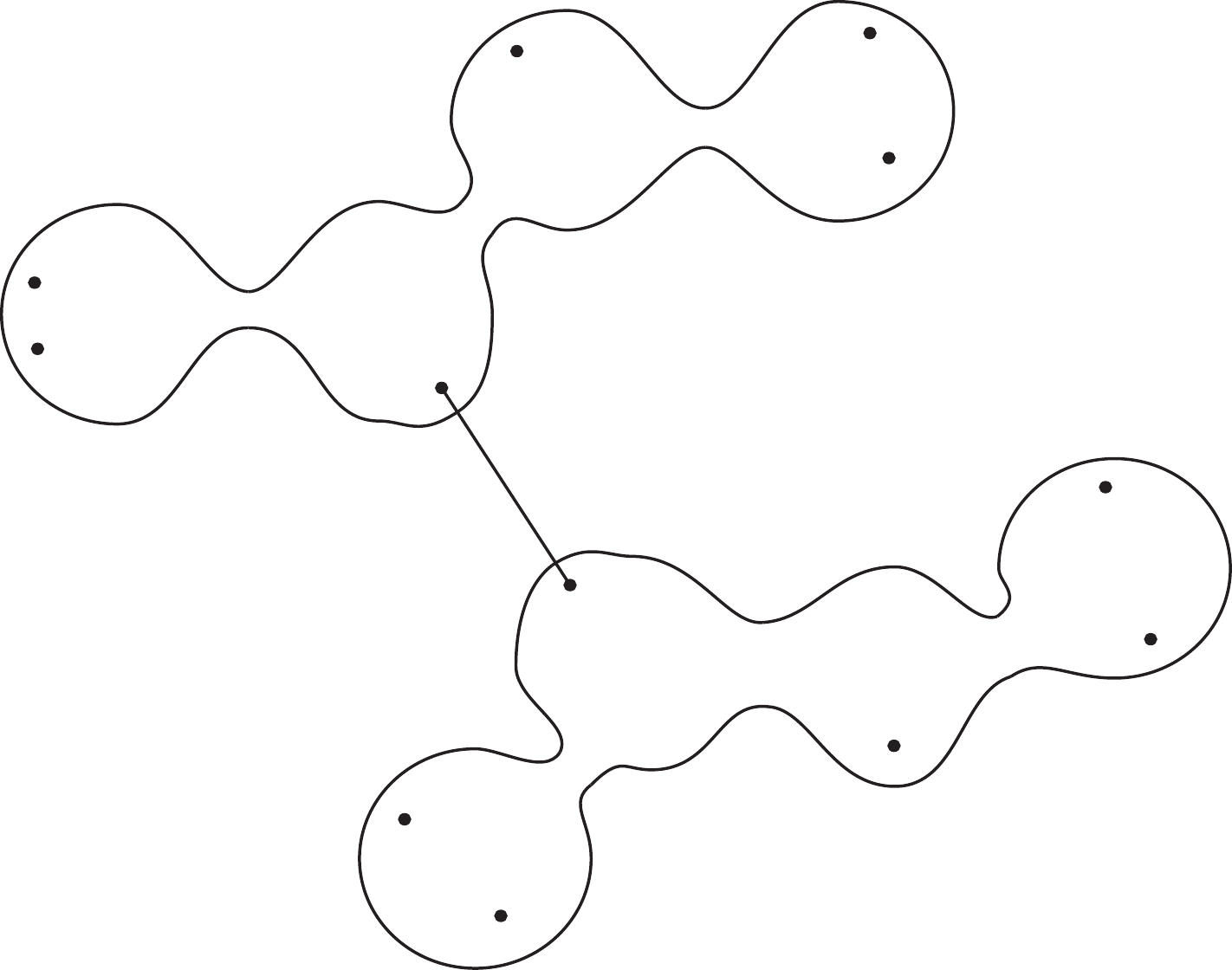}
  \end{center}
  \caption{
  As one gauge coupling is brought to zero, the $SU(2)$ gauge theory decouples and leaves behind two $SU(2)$ flavor groups.
  This corresponds to the full degeneration limit of the punctured sphere, which leaves behind two punctured spheres}
  \label{fig:su2gb}
\end{figure}

The moduli space should have boundaries where a single $SU(2)$ gauge
group becomes infinitely weak. As the generalized quiver is a tree,
decoupling a gauge group leads to two disconnected quivers. The
theory becomes the sum of two decoupled theories, which are ${\cal
T}_{m+3,0}[A_1]$ and ${\cal T}_{n-m+2,0}[A_1]$ for some $m$ (see
fig. \ref{fig:su2gb}). In the next subsection we will confirm this
picture, by computing the Seiberg-Witten curve and differential for
the linear quiver gauge theories. The parameter space of gauge couplings
turn out to coincide with the moduli space of $n+3$ points on
the sphere ${\cal M}_{n+3,0}$. The various weakly coupled S-dual
frames of the theory coincide with the different ways a sphere with
$n+3$ points can degenerate completely to a set of $n+1$
three-punctured spheres attached together at $n$ nodes to form a
generic binary tree. The S-duality group is the fundamental group
$\pi_1$ of ${\cal M}_{n+3,0}$. The action of S-duality on the $n+3$
flavor symmetry groups is a permutation action, and coincides with
the permutation action on the $n+3$ punctures on the sphere.

\begin{figure}
  \begin{center}
    \includegraphics[width=2.7in]{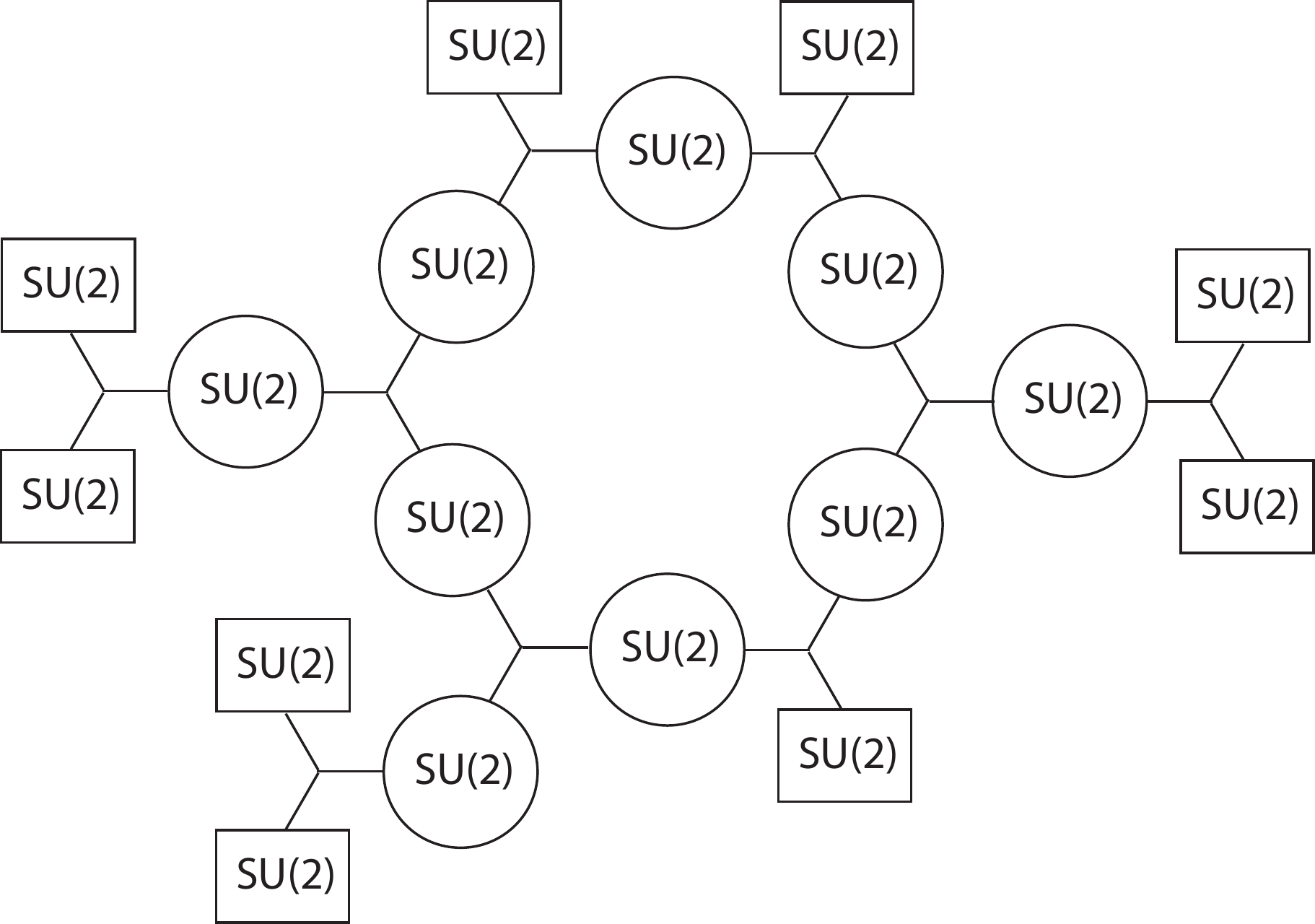}\hfill
    \includegraphics[width=2.7in]{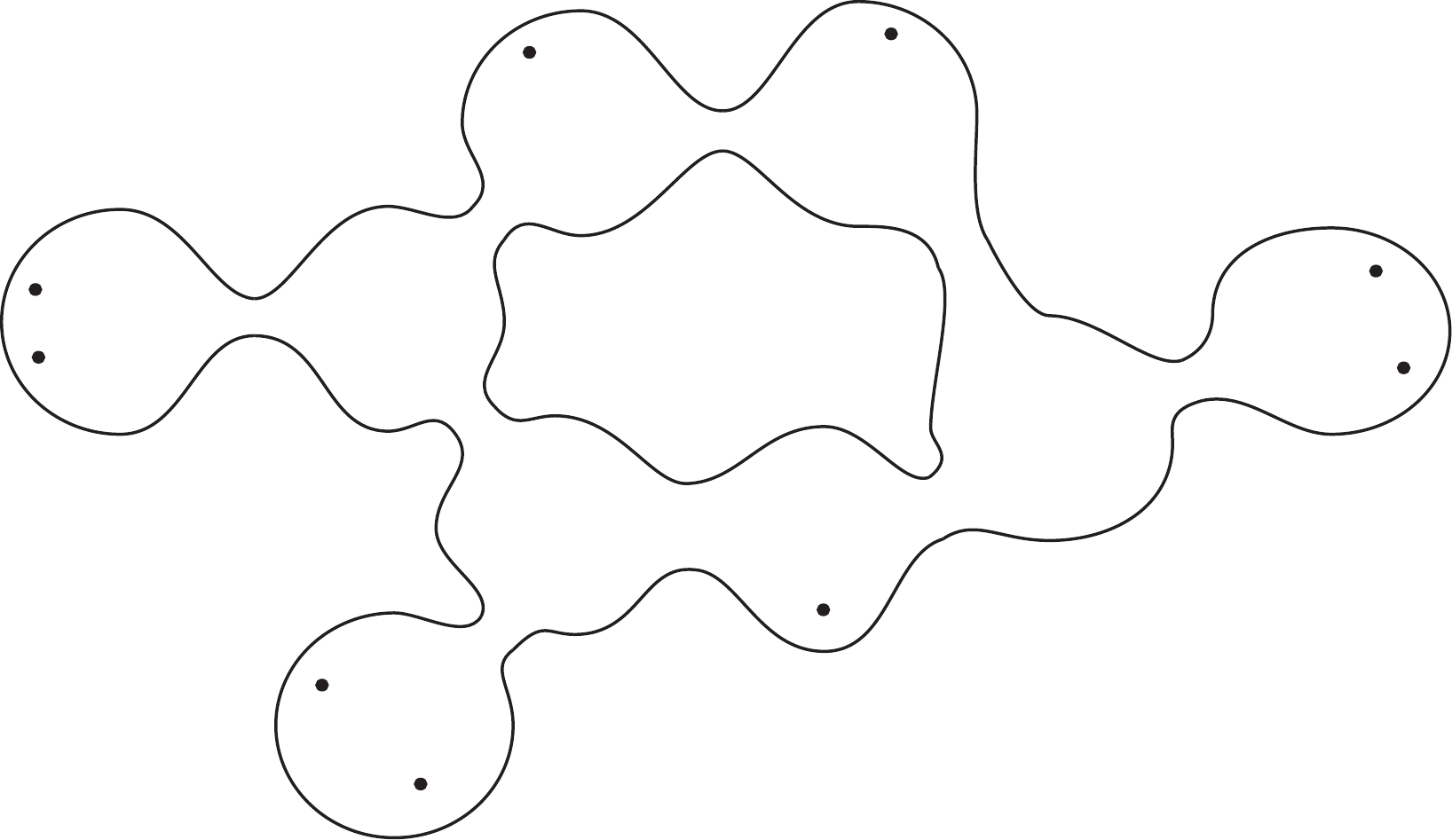}
  \end{center}
  \caption{
  A generalized quiver with one loop. The space of gauge couplings is parameterized by a torus with $9$ punctures. The quiver is weakly coupled
  at the cusp where the punctured sphere degenerates as in figure.}
  \label{fig:su2gl}
\end{figure}

\begin{figure}
  \begin{center}
   \includegraphics[width=2.2in]{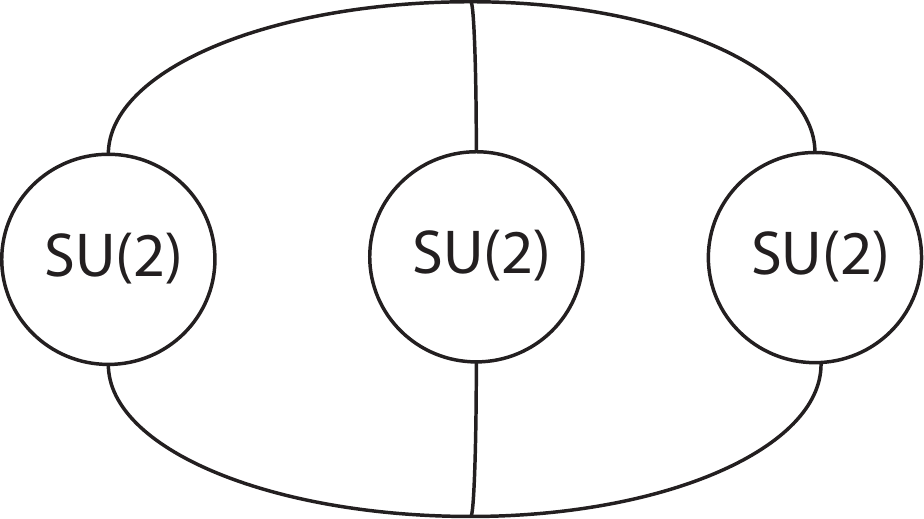} \hfill
    \includegraphics[width=2.7in]{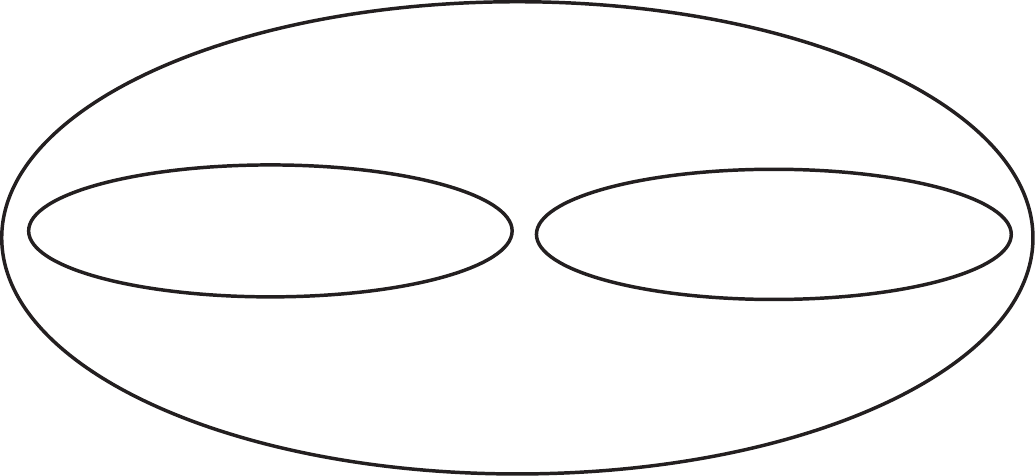}
  \end{center}
  \caption{
  A generalized quiver with two loops and no flavor groups, in the first of  two distinct S-duality frames. The gauge couplings are parameterized by a genus two Riemann surface.
  We depict the degeneration limit of the genus two Riemann surface which map to the weak coupling limit for the generalized quiver.}
  \label{fig:su2gl2}
\end{figure}

\begin{figure}
  \begin{center}
    \includegraphics[width=2.6in]{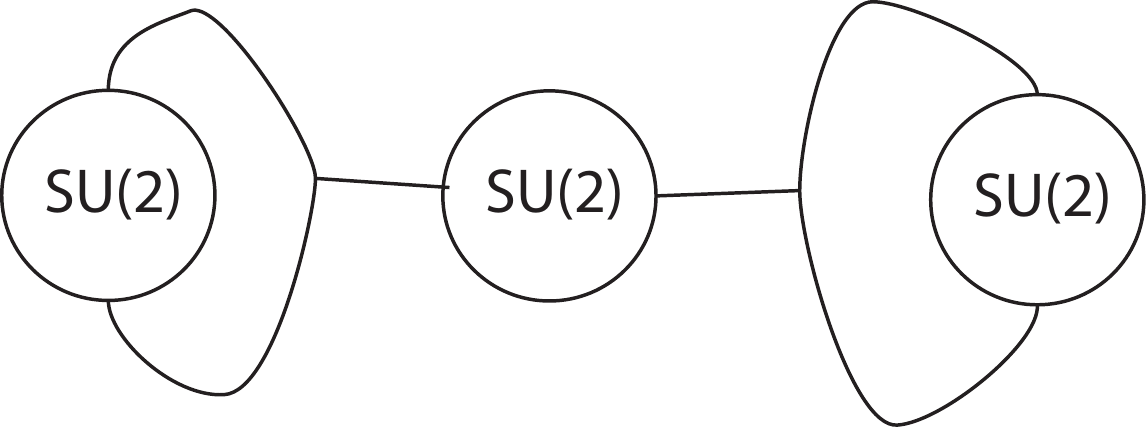}  \hfill
    \includegraphics[width=2.7in]{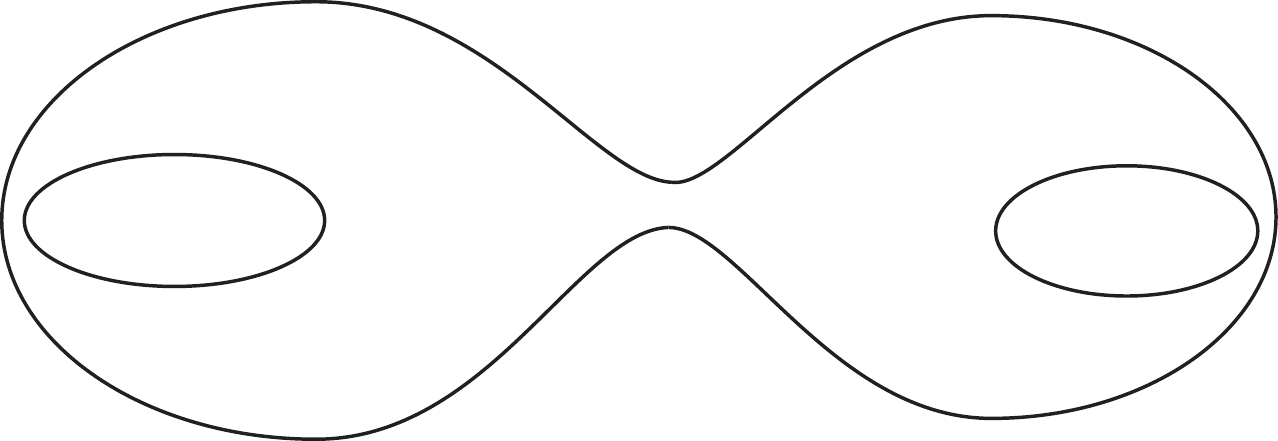}
  \end{center}
  \caption{
  A generalized quiver with two loops and no flavor groups, in the second of  two distinct S-duality frames. The gauge couplings are parameterized by a genus two Riemann surface.
  We depict the degeneration limit of the genus two Riemann surface which map to the weak coupling limit for the generalized quiver.}
  \label{fig:su2gl2b}
\end{figure}

There is no reason to limit ourselves to generalized quivers with
the topology of a tree. In general we can build a generalized quiver
with $n$ flavor groups and $g$ loops by connecting $n+2g-2$ sets of
four hypers with $n+ 3g-3$ gauge groups. Elementary S-dualities at
single nodes connect all the generalized quivers with the same $g$
and $n$ into a single theory which we will label as ${\cal
T}_{n,g}[A_1]$. See Fig \ref{fig:su2gl} and \ref{fig:su2gl2},\ref{fig:su2gl2b}

We can make gauge couplings infinitely weak to either open up a
loop, or disconnect the generalized quiver into two subquivers. We
do not have a direct brane construction of the SW curve for these
generalized quiver gauge theories. We will conjecture an answer
which satisfy several consistency checks. There is an evident similarity
between the weak coupling limits of a generalized quiver and the and
the degeneration limits of a Riemann surface of genus $g$ and $n$
punctures. Indeed the weak coupling S-dual frames we can identify
from S-duality at each node of the generalized quivers in ${\cal
T}_{n,g}[A_1]$ precisely coincide to the different ways a punctured
Riemann surface can degenerate to a set of $n+2g-2$ three-punctured
spheres joined by $n+3g-3$ thin tubes. We are lead to conjecture
that the parameter space of gauge couplings of ${\cal T}_{n,g}$ coincides with
the Teichmuller moduli space of a genus $g$ Riemann surface with $n$
punctures $\tilde {\cal M}_{n,g}$, reduced by S-dualities to the moduli space ${\cal M}_{n,g}$

\begin{figure}
  \begin{center}
    \includegraphics[width=2.5in]{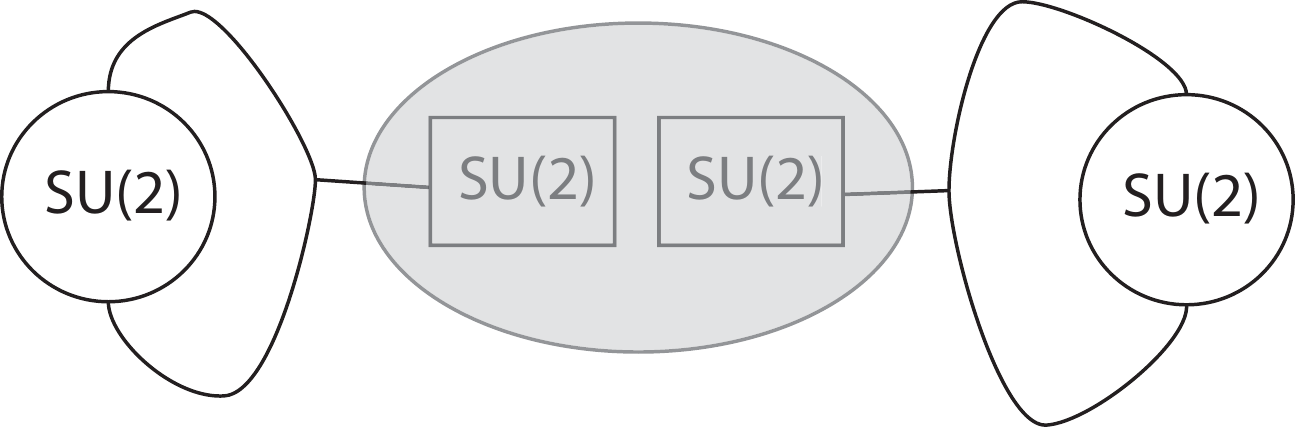} \hfill
    \includegraphics[width=2.5in]{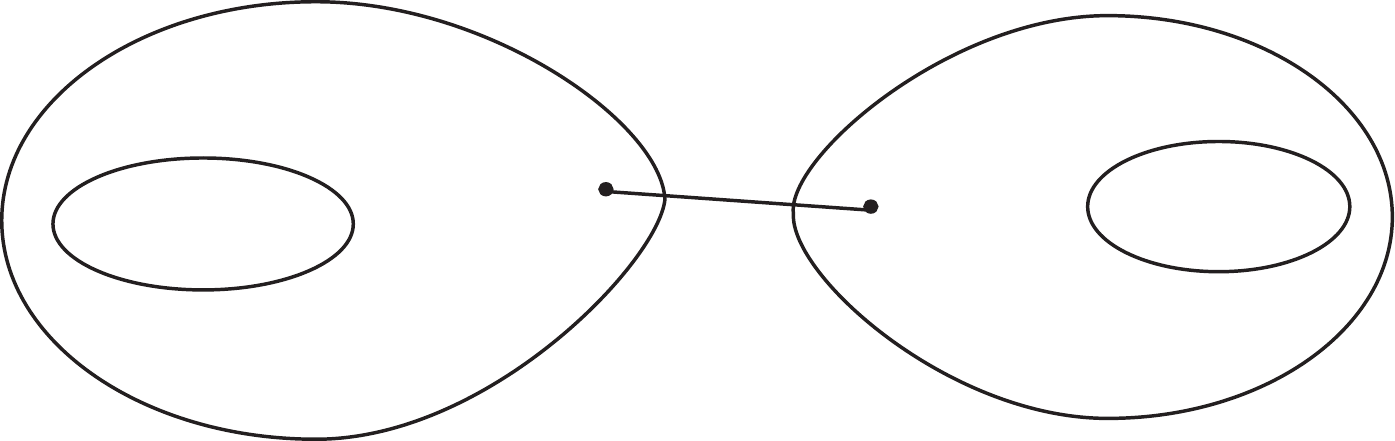}
  \end{center}
  \caption{The first of two decoupling limits of a generalized quiver with two loops and no flavor groups, and the corresponding complete degeneration of a genus two Riemann surface}
  \label{fig:su2g2deq}
\end{figure}

\begin{figure}
  \begin{center}
     \includegraphics[width=2.5in]{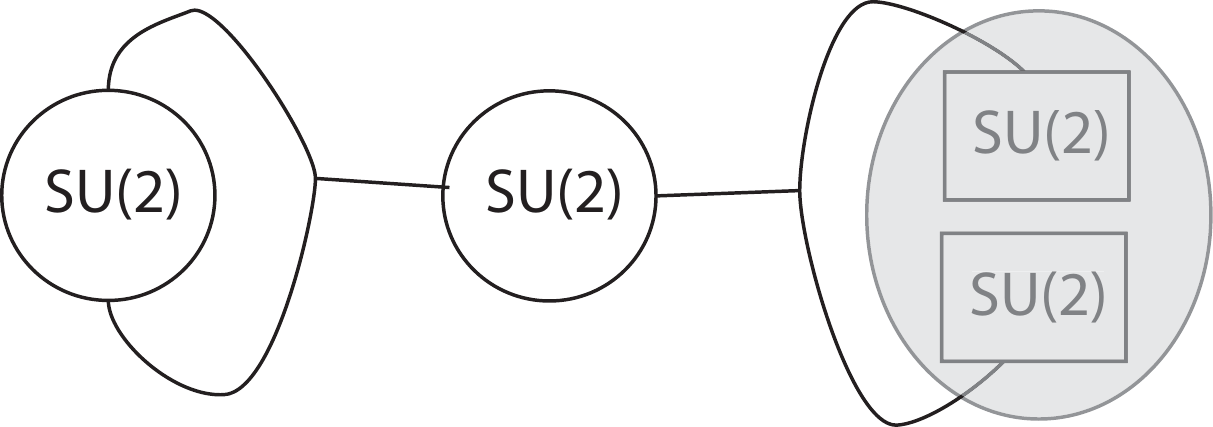} \hfill
    \includegraphics[width=2.5in]{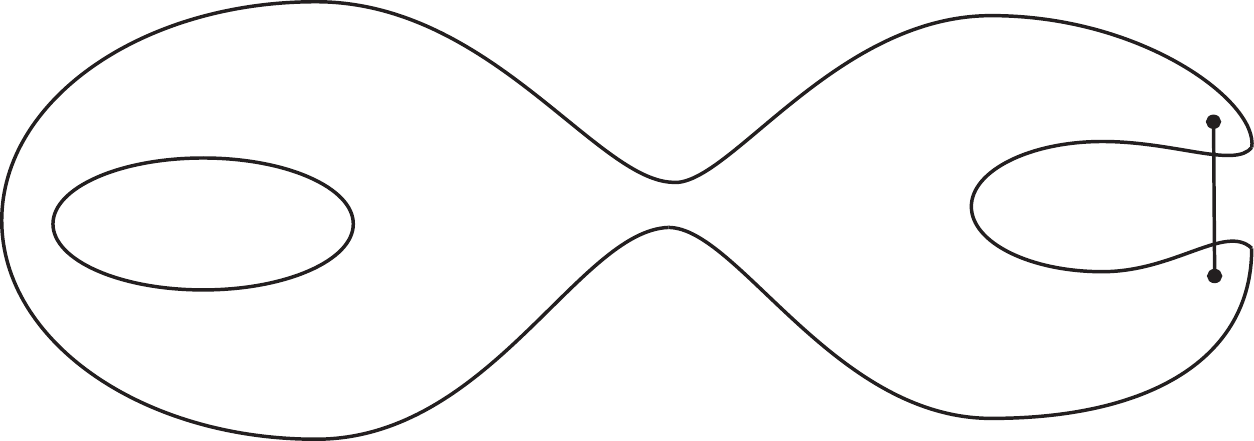}
  \end{center}
  \caption{The second of two decoupling limits of a generalized quiver with two loops and no flavor groups, and the corresponding complete degeneration of a genus two Riemann surface}
  \label{fig:su2g2deqb}
\end{figure}

In the next subsection we will conjecture a Seiberg-Witten curve and
differential for these theories ${\cal T}_{g,n}[A_1]$, which has
${\cal M}_{n,g}$ as the moduli space of exactly marginal
deformations. The SW curve is a ramified double cover of the genus
$g$ Riemann surface $C_{g,n}$. The SW differential has poles at the
preimage of the $n$ punctures on $C_{g,n}$. At the boundaries of
${\cal M}_{n,g}$, $C_{g,n}$ degenerates either to a curve of
lower genus $C_{g-1,n+2}$ or to the union of two curves $C_{g',n'+1}
\cup C_{g-g',n-n'+1}$ (see fig. \ref{fig:su2g2deq}, \ref{fig:su2g2deqb}). The SW curve
similarly degenerates to the curve for the theory ${\cal
T}_{g-1,n+2}[A_1]$ or ${\cal T}_{g',n'+1}[A_1] \times {\cal
T}_{g-g',n-n'+1}[A_1]$ exactly in the way associated with the decoupling limit of an $SU(2)$ gauge group. In the last subsection we will present a
direct definition of the theory ${\cal T}_{n,g}$ as the (twisted)
compactification of the $A_1$ $(2,0)$ theory on a Riemann surface of
genus $g$, in the presence of $n$ codimension $2$ defect operators.
This will give a direct construction of the conjectured SW curve.

\subsection{SW curves}
The Seiberg-Witten curve for linear quivers of unitary gauge groups,
possibly with fundamental matter at one end, has been computed in
section $2$ of \cite{Witten:1997sc}. We will simply take the final
result, specialized to a product of $n$ $SU(2)$ gauge groups with
two hypermultiplets at each end. The curve is given as a polynomial
equation in two variables $(v,t)$, defined in $\IC \times
(\IC^*-\{t_0, \cdots, t_n\})$. The polynomial is of second order in
$v$ and order $n+1$ in $t$. If we collect the same powers of $t$ it
can be written as
\begin{equation} \label{eq:su2line}
v^2 t^{n+1} + c_1 (v^2 - u_1) t^n + \cdots + c_{n} (v^2 - u_{n}) t + c_{n+1}v^2=0
\end{equation}

At weak coupling, the coefficient of each power of $t$ is associated
with a gauge group in the quiver. The degree in $v$ of each
coefficient is two because all gauge groups are $SU(2)$. The $u_i$
coefficients parameterize the Coulomb branch of the theory. At weak
coupling, $u_i$ should be identified with the expectation value $Tr
\Phi_i^2$ for the $i$-th $SU(2)$ gauge group in the quiver. The
$c_i$ parameterize the space of gauge couplings of the theory. More
precisely, if we collect the same powers of $v$ as (in general the
subscript of a polynomial will denote its degree)
\begin{equation} \label{eq:linsu2}
\prod_{a=0}^{n} (t-t_a) v^2 = U_{n-1}(t)t
\end{equation}
then at weak coupling
\begin{equation}\label{eq:coupling} \tau_a = \frac{1}{ i \pi} \log \frac{t_{a-1}}{t_{a}}.\end{equation}
The overall scale of $t_a$ is immaterial. We can set, for example,
$t_0=1$.

The Seiberg Witten differential is not written down explicitly in
\cite{Witten:1997sc}, but it is implicit in the analysis of the low
energy effective action and BPS states. See \cite{Mikhailov:1997jv} for a
detailed discussion. It a has a very simple expression, which is
valid for all the SW curves described in \cite{Witten:1997sc}.
\begin{equation}
\lambda = v \frac{dt}{t}
\end{equation}
Indeed the embedding space has a natural complex two form $dv
\frac{dt}{t}$, and if a curve is defined by a polynomial equation
$F(v,t)=0$, then
\begin{equation}
\frac{\partial \lambda}{\partial u_i} = - \frac{\partial F}{\partial u_i} \frac{dt}{t \frac{\partial F}{\partial v}}
\end{equation}
gives a basis of holomorphic differentials whenever $u_i$ is a
normalizable deformation. In general the normalizable deformations
are the coefficient of a monomial in $F$ whose degree in $v$ is
smaller than the degree of $F$ minus $1$, and whose degree in $t$ is
positive and smaller than the degree of $F$. We can easily see from
\ref{eq:su2line} that the $u_a$ are the only marginal deformations.

Let us specialize further to the case $n=1$, to recover the SW curve
for $SU(2)$ $N_f=4$. We warn the reader that we do not plan to bring
it to a more traditional form, but quite the opposite. The curve
equation is
\begin{equation}
(t-1)(t-t_1) v^2 = u t
 \end{equation}
The central charges are the period integrals of
\begin{equation}
\lambda = \frac{\sqrt{u}}{\sqrt{t (t-1) (t-t_1)}} dt
\end{equation}
on appropriate cycles of the SW curve. The trivial $u$ dependence is
a consequence of conformal invariance. The gauge coupling $\tau$ is
traditionally identified with the modular parameter of an auxiliary
torus of equation.
\begin{equation} \label{eq:auxiliary}
y^2 = t (t-1) (t-t_1)
\end{equation}

There still is a certain amount of redundancy in the
parameterization of the gauge coupling by $t_1$. The torus
degenerates when $t_1\to \infty$, $t_1 \to 1$, $t_1 \to 0$. The
first degeneration is the naive weak coupling limit for the $SU(2)$
gauge theory at $\tau \to i \infty$, the other two correspond to the
S-dual weakly coupled regions $\tau \to 1,0$ respectively. It is
useful to modify our coordinate system slightly, to make the action
of S-duality more manifest. If we write simply $v= t x$, so that
$\lambda = x dt$, the curve becomes
\begin{equation}
t (t-1)(t-t_1) x^2 = u
\end{equation}
S-duality acts by fractional linear transformations, generated by
$(t,x) \to (1-t,x)$ and $(t,x) \to (1/t, - t^2 x)$. The S-duality
invariant moduli space is simply the complex structure moduli space
of a sphere with four punctures, at $0, 1, t_1, \infty$.

The four punctures can be put on an even footing by a further
fractional linear transformation $t\to \frac{a z+b}{c z +d} , x\to
(c z + d)^2 x$ to a final
\begin{equation}
x^2 = \frac{u}{\Delta_4(z)}
\end{equation}
Then $t_1$ is the cross-ratio of the four roots of $\Delta_4(z)$.
The SW differential is $\lambda = x dz$. Here we consider $z$ as a
variable on a four-punctured sphere, and $x$ is naturally a
coordinate on the cotangent bundle of the sphere. We see here the
first example of a certain canonical form for the SW curve, which we
will meet over and over in the paper. For generalized $SU(2)$
quivers the canonical form is
\begin{equation}
x^2 = \phi_2(z)
\end{equation}
$(x,z)$ are local coordinates in the cotangent bundle of some
punctured Riemann surface. The SW differential is the canonical one
form $x dz$. The expression $\phi_2(z) dz^2$ is a quadratic
differential on the punctured Riemann surface with appropriate poles
at the puncture.

For $SU(2)$ $N_f=4$ the differential $\phi_2 dz^2$ on the
four-punctured sphere has simple poles at all punctures. The gauge
coupling moduli space can be identified with the moduli space of a
four punctured sphere ${\cal M}_{4,0}$, or better with the
Teichmuller moduli space $\tilde {\cal M}_{4,0}$ subject to the
action of the S-duality group the fundamental group $\pi_1({\cal
M}_{4,0})$.

In the present situation, with no mass deformations, the parameter
$\tau$ coincides with the gauge coupling of the IR $U(1)$ gauge
theory, but that will not be true in presence of mass deformations.
The parameter $\tau$ should be really considered a convenient
parametrization of the exactly marginal deformations of the theory.
It does not really coincide with the UV gauge coupling, at least in
standard renormalization schemes (see for example \cite{Dorey:1996bn}), except
asymptotically at weak coupling. The Teichmuller space $\tilde {\cal
M}_{4,0}$ is known to coincide with the upper half plane, i.e. the
moduli space of a two-torus defined by the auxiliary equation
\ref{eq:auxiliary}. Hence $\tau$ is really nothing else but a
coordinate on the Teichmuller space $\tilde {\cal M}_{4,0}$. This
distinction is important when we study general linear quivers, where
this fortuitus coincidence between the ambiguously-defined space of
``UV gauge couplings'' and the true gauge coupling parameter space
visible from the SW curve does not happen.

 If we go back to the general $n$ case
\ref{eq:linsu2}, and we apply the same coordinate transformation as
for $n=1$, the symmetry between the $n+3$ punctures at $t=0, \cdots
t_a, \infty$ becomes manifest:
\begin{equation}
x^2 = \frac{U_{n-1}(z)}{\Delta_{n+3}(z)}= \phi_2(z)
\end{equation}
The quadratic differential $\phi_2 dz^2$ has simple poles at all
punctures. The parameter space of gauge couplings is parameterized by
the cross-ratios of the roots of $\Delta_{n+3}(z)$, and coincides
with the moduli space of a $n+3$ punctured sphere ${\cal
M}_{n+3,0}$. The S-duality group of the theory is the fundamental
group $\pi_1({\cal M}_{n+3,0})$

To see the action of S-duality on the flavor symmetry groups, we
need to introduce some mass deformations. The mass deformed curve
for the general quiver is also computed in  \cite{Witten:1997sc}. We
collect immediately the same powers of $v$
\begin{equation}
\prod_{a=0}^{n} (t-t_a) v^2 = M_{n+1}(t) v + U_{n+1}(t)
\end{equation}
All the $n+2$ coefficients of $M_{n+1}$, the first and the last
coefficients of $U_{n+1}$ are mass parameters. Because of the
freedom $v \to v+ v_0$ only $n+3$ parameters are physical. The
Seiberg-Witten differential has now simple poles, at $t=0, t=t_a,
t=\infty$. Near $t=t_a$ one of the roots $v_\pm(t)$ diverges as
\begin{equation}
\lambda = v \frac{dt}{t} \sim \frac{M_{n+1}(t_a)}{t_a \prod_{b=0}^{n'} (t_a-t_b)}\frac{dt}{t-t_a}
\end{equation}
At $t=0$ there is a pole for both roots of $v$
\begin{equation}
\lambda = v \frac{dt}{t} \sim v_\pm(0) \frac{dt}{t}
\end{equation}
and similarly at $t \to \infty$. In the brane setup of
\cite{Witten:1997sc} the full flavor symmetry for $SU(2)$ gauge
groups is not visible. The manifest flavor symmetry for
bifundamentals is the naive $U(1) \in SU(2)$, and for the
fundamentals is $U(2) \in SO(4)$.

The residues at $t_a$ have been identified with the $U(1)$ mass
parameters for the $a$-th set of bifundamentals ($a=0$ and $a=n$ are
the fundamentals, and the $U(1)$ is the diagonal subgroup of
$U(2)$). All these $U(1)$ are the Cartans of appropriate $SU(2)$
true flavor symmetry groups. Furthermore the difference between the
asymptotic values $v_\pm(0)$ is identified with the mass parameter
for the Cartan of the $SU(2)$ subgroup in the $U(2)$ flavor symmetry
at the first node. Similarly the difference between the asymptotic
values $v_\pm(\infty)$ is identified the mass parameter for the
Cartan of the $SU(2)$ subgroup in the $U(2)$ flavor symmetry at the
last node. We see that each puncture in the $t$ plane, be it $t=0,
\cdots,t_a, \cdots \infty$ is associated to the mass parameter of a
specific $SU(2)$ flavor symmetry group, but in the variables $v,t$,
the identification is still unpleasantly asymmetric.

Things become a bit more transparent after a change of coordinates.
First of all, we should shift away the linear term in $v$ to get
\begin{equation}
\prod_{a=0}^{n} (t-t_a)^2 v^2 =  U_{n+1}(t)\prod_{a=0}^{n} (t-t_a)- \frac{1}{4}M_{n+1}(t)^2
\end{equation}
After the shift in $v$ the Seiberg-Witten differential takes a form
$\lambda = v \frac{dt}{t}-\frac{M}{t\Delta} dt$. We now modify the
differential, and bring it back to $\lambda = v \frac{dt}{t}$. This
modification is harmless: the residues of $\frac{M}{t\Delta} dt$ are
linear combinations of the mass parameters of the theory, and the
the first derivatives $\lambda_{u_i}$ are unchanged, and still
coincide with the holomorphic one forms on the SW curve. The central
charges are shifted by a certain linear combination of the mass
parameters. This amounts to a shift of the flavor charges of BPS
particles by multiples of their gauge charges, i.e. to a
redefinition of the flavor currents.

Notice that if the flavor symmetries are nonabelian, the charges of
the BPS particles in the spectrum under the Cartan subalgebra of the
flavour group must be the weights of the appropriate irreps of the
flavor group. In the original brane setup only the $U(1)$ Cartan was
visible for $n+1$ of the $SU(2)$ flavor groups, and the original SW
differential gave an assignment of flavor charges which was possibly
incompatible with the full non-abelian flavor symmetry. Inspection
of a few examples shows that the shift of the SW differential
automatically cures that unpleasantness. Now at each of the $n+3$
punctures in the $t$ plane $\lambda$ has poles on both branches
$v_\pm(t) \frac{dt}{t}$, with equal and opposite residue given by
the mass parameter of the corresponding $SU(2)$ flavor symmetry
group.

Finally we can do our usual transformation to coordinates $(x,z)$,
to bring the punctures on the same footing. The final result is
\begin{equation}
x^2 =  \frac{P_{2n+2}(z)}{\Delta_{n+3}(z)^2}= \phi_2(z)
\end{equation}
The mass parameters $m_a^2$ of the $SU(2)$ flavor symmetry groups
are the coefficients of the double poles in the quadratic
differential $\phi_2 dz^2$. In presence of mass deformations it is a
bit arbitrary to define an origin for the $u$ parameters. If we set
an origin at $P_{2n+2} = P^{(0)}_{2n+2}$, we can write the
deformation $P_{2n+2}(z) = P^{(0)}_{2n+2}(z)+\Delta_{n+3}(z)
U_{n-1}(z)$ which does not change the coefficients of the double
poles. The $n$ coefficients of $U_{n-1}(t)$ parameterize the Coulomb
branch.

The punctures on the $z$ plane are now associated to the factors of
the flavor group. The moduli space of a sphere with $n+3$
distinguished punctures ${\cal M}'_{n+3,0}$ is a cover  of the gauge
moduli space ${\cal M}_{n+3,0}$. The S-duality group $\pi_1({\cal
M}_{n+3,0})$ can be factored as the semi-direct product
\begin{equation}
0 \to \pi_1({\cal M}'_{n+3,0})\to \pi_1({\cal M}_{n+3,0})\to S \to 0
\end{equation}
The quotient $S$ acts as the permutation group on the punctures,
i.e. on factors of the flavor group.

It is important not to confuse the gauge coupling parameter space, i.e.
the space of exactly marginal deformations of the ${\cal N}=2$ SCFT,
${\cal M}_{n+3,0}$, with the space of possible IR gauge couplings,
which are the period matrix of the SW curve. The complex structure
of the SW curve is determined by the polynomial $P_{2n+2}(z)$,
through the auxiliary equation
\begin{equation}
y^2 =  P_{2n+2}(z)
\end{equation}
and depends on everything, including $u_a$ parameters and mass
parameters. The space ${\cal M}_{n+3,0}$ is parameterized by the
(cross-ratios of the) roots of $\Delta_{n+3}(z)$ only.

\begin{figure}
  \begin{center}
    \includegraphics[width=5.5in]{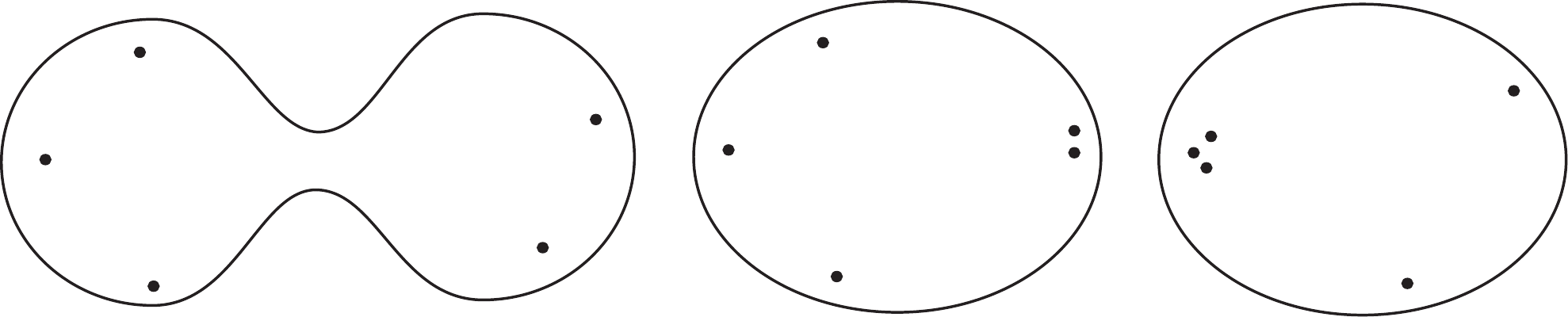}
  \end{center}
  \caption{A degeneration limit of a five punctured sphere to a four punctured sphere and a three punctured sphere. Shown as the collapse of a thin tube.
  From the point of view of the first sphere, two punctures are collapsing to one. From the point of view of the second sphere, three punctures are collapsing to one}
  \label{fig:su2split}
\end{figure}

There is an important and simple consequence of the equivalence
between different punctures in the $z$ plane. If we go to the
boundary component of moduli space where $m$ punctures collide, we
can always pick an S-duality frame where the punctures are $0, t_0,
t_1,\cdots t_{m-2}$. Then the formula \ref{eq:coupling} tells us
that the gauge coupling at the $m-1$-th node is becoming very weak,
and the quiver is breaking down to two decoupled subquivers, just as
the punctured sphere is degenerating to two sub-spheres joined by a
nodal singularity (See figure \ref{fig:su2split} for a case with
$n=2$ and $m=3$). It is easy to see how the process affects the SW
curve. We can focus on one sub-sphere, scaling all
$t_{0,\cdots,m-2}$ uniformly to $0$ by a factor of $\epsilon$. We
want to keep all masses finite in the degeneration process. The
coefficients of double poles at $t_a$ for $a\leq m-2$ receive a
factor $\epsilon^{-2(m-1)}$ from $\Delta_{n+3}(z)$. We should adjust
$P_{2n+2}(z)$ so that it would scale as $\epsilon^{2(m-1)}$ at $t=t_a$, so that the
coefficient of the double poles remain finite as $\epsilon \to 0$.

This implies that $P_{2n+2}(t) \to t^{2(m-1)} P_{2n-2m+4}(t)$ after the $\epsilon \to 0$ limit, and
cancels enough factors of $t$ in $\Delta_{n+3}(t)\to t^{2m}
\Delta_{n-m+3}(z)$ to leave a second order pole in $\phi_2$ at $t=0$. Hence the
SW curve reduces properly to the curve for the linear quiver of
$n-m+1$ nodes.

The coefficient of the new double pole at $0$ is the mass parameter
of the new flavor symmetry group left by the decoupling gauge group.
It could have been read in the original sphere by integrating
$\sqrt{\phi_2}$ around the $m$ collapsing punctures. If we want to study the
second fragment of the SW curve, we need to bring instead the other $n-m+3$ punctures
to infinity. The same path will compute the coefficient of the double pole at infinity,
hence the mass parameters for
the two new $SU(2)$ flavor groups in the two subquivers coincide.
This was expected: the mass parameters are simply the $u$ parameter
of the decoupling gauge group.

We conclude that as one
approaches boundaries of ${\cal M}_{n+3,0}$ where the sphere
degenerates and splits, the theory ${\cal T}_{n+3,0}[A_1]$ also
decouples to the product theory ${\cal T}_{m+1,0}[A_1] \times {\cal
T}_{n-m+4,0}[A_1]$. If we let as many handles degenerate as possible, to go to a
generic cusp in the parameter space ${\cal M}_{n+3,0}$
the sphere approaches a collection of three-punctured spheres connected into a binary tree.
Weakly coupled $SU(2)$ gauge groups appear in the same pattern.
The canonical SW curve corresponding to a three
punctured sphere \begin{equation} x^2 = \frac{P_2(z)}{\Delta_3(z)^2}
\end{equation} is the SW curve of a block of four hypermultiplets. All coefficients in $P_2(z)$ are mass
parameters for the three important $SU(2)$ flavor subgroups. Hence at the cusp of the parameter space, the theory ${\cal T}_{n+3,0}[A_1]$ takes the form of a weakly coupled generalized $SU(2)$ quiver.
The shape of the quiver coincides with the shape of the binary tree of three punctured spheres.

Next we would like to conjecture a SW curve for generalized $SU(2)$
conformal quiver gauge theories with $g$ loops, the ${\cal
T}_{n,g}[A_1]$. Notice that \cite{Witten:1997sc} presents the
solution for a close linear loop of unitary groups, with
bifundamental hypermultiplets between consecutive nodes. The closed
linear loop of $n$ $SU(2)$ groups is the perfect starting point to
study the genus one case,  ${\cal T}_{n,1}[A_1]$. For zero masses, the SW equation for
the closed loop of $n$ $SU(2)$ gauge groups is given simply as curve
in $\IC \times \IT^2$, which we parameterize with coordinates $v$
and $z$. The curve is defined by a degree $2$ polynomial in $v$
\begin{equation}
v^2 = f_2(z)
\end{equation}
The function $f_2(z)$ is a meromorphic function on $\IT^2$, with
simple poles at the $n$ punctures $z=z_a$. The SW differential is
again $v dz$.

The space of couplings of the theory is the moduli space of a torus
with $n$ marked points. In the weak coupling limit, the gauge
couplings are simply given as $\tau_a \sim z_{a+1}-z_a$. The sum of
gauge couplings is the modular parameter of the torus. The space of
possible $f_2(z)$ is $n$ dimensional: constant shifts are allowed,
but the sum of the residues of $f_2$ is zero, because the torus is
compact.

The solution with mass deformations given in \cite{Witten:1997sc}
involves a slight complication: morally, one would just set
\begin{equation}
v^2 = f_1(z) v + f_2(z)
\end{equation}
and identify the residues of $f_1$, which coincide with the residues
of $\lambda$, with the mass parameters of the theory. Unfortunately
the sum of the residues of $f_1$ is zero if $f_1$ is a standard
meromorphic function. To have a non-zero sum of residues, $f_1$ has
to be a section of a non-trivial affine bundle on the torus, i.e. it
must be allowed to shift by a constant around a cycle of the torus.
$v$ and $f_2$ must also transform appropriately for the shift to
make sense. Fortunately for the case of $SU(2)$ this unpleasantness
is immediately eliminated a the shift of $v \to x+1/2 f_1(t)$ which
eliminates the linear term, bringing us back to the form
\begin{equation}
x^2 = \phi_2(z)
\end{equation}
Again we keep $\lambda=x dz$, as the shift produces a useful shift
in the definition of the flavor charges. We changed the notation
from $f_2$ to $\phi_2$ to remark that $\phi_2(z)$ is morally a
quadratic differential, which is allowed double poles at $z=z_a$,
with coefficients $m_a^2$ which are naturally identified with the
mass parameters of the $a$-th $SU(2)$ flavor group in the chain.
Indeed $(x,z)$ live in $\IC\times \IT^2$, naturally identified with
$T^* \IT^2$.

Normalizable deformations can be parameterized by a shift
$\phi_2(z)=\phi^{(0)}_2(z) + U(z)$ where $U$ has simple poles only.
Notice that there is no constraint on the coefficients of double
poles of a meromorphic function on the torus. $\phi^{(0)}_2$ could
be given by a simple sum of Weierstrass functions $\sum_a m^2_a
\rho(z-z_a)$. There are truly $n$ independent mass parameters. Hence
we have identified the gauge coupling parameter space as the
Teichmuller moduli space $\tilde {\cal M}_{n,1}$ of a torus with $n$
punctures. The S-duality group is $\pi_1({\cal M}_{n,1})$. Again
$\pi_1({\cal M}_{n,1})$ acts as the permutation group on the $n$
punctures, and acts in the same fashion on the $SU(2)$ factors of
the flavor symmetry group, which are in one-to-one correspondence
with the punctures.

A $n$ punctured torus can degenerate to an $n+2$ punctured sphere by
pinching off a handle. Then the distance between a certain pair of
consecutive punctures diverges, and by $\tau_a \sim z_{a+1}-z_a$ we
see that one gauge group of the quiver has become very weakly
coupled. The quiver degenerates to a simple linear quiver, indeed
associated to a $n+2$ punctured sphere. The quadratic differential
on the torus naturally degenerates to a quadratic differential on
the punctured sphere, with double poles at the two new punctures.
The coefficients of the double poles are read by the contour
integral $\int \sqrt{\phi_2}$ around the degenerating handle, and
coincide as expected.

There are also degeneration limits where punctures on the torus
collide. These are strong coupling limits for the original gauge
groups on the quiver, and weak coupling limits for the appropriate
choice of a S-dual generalized quiver in the ensemble ${\cal
T}_{n,1}[A_1]$. It is easy to see that as the punctured torus
degenerates to a torus with fewer punctures and a punctured sphere,
the SW curve factorizes properly to the SW curves of the appropriate
${\cal T}_{n-m+1,1}[A_1]$ and ${\cal T}_{m+1,0}[A_1]$ theories.
Ultimately,  the set of possible complete degenerations of the $n$
punctured torus is a graph of three punctured spheres with one loop.
It coincides with the set of all possible weakly coupled generalized
conformal quivers of $n$ $SU(2)$ groups with one loop.

By analogy with the case of open and closed linear quivers, we are
ready to associate to the ensemble of generalized quiver gauge
theories with $g$ loops and $n$ $SU(2)$ flavor symmetry groups
${\cal T}_{n,g}$ the canonical SW curve in the cotangent bundle of a
Riemann surface $C_{n,g}$ of genus $g$ and $n$ punctures.
\begin{equation}
x^2 = \phi_2(z)
\end{equation}
We will allow the quadratic differential to have simple poles at the
$n$ punctures of $C_{n,g}$. Mass deformations arise as double poles
for the quadratic differential, of residue $m^2$.

Notice that the linear space of quadratic differential with $n$
simple poles on a Riemann surface of genus $g$ has dimension $3g-3 +
n$, as they are in one-to-one correspondence with the complex moduli
of the Riemann surface itself. Hence the dimension of the Coulomb
branch of the generalized quiver is correctly reproduced: there is a
$u$ parameter for each of the $3g-3+n$ gauge groups. The gauge
coupling moduli space of ${\cal T}_{n,g}$ is the same as ${\cal
M}_{n,g}$. The S-duality group is $\pi_1({\cal M}_{n,g})$. Again
$\pi_1({\cal M}_{n,g})$ acts as the permutation group on the $n$
punctures, and we expect it to act in the same fashion on the
$SU(2)$ factors of the flavor symmetry group, which are in
one-to-one correspondence with the punctures.

The possible degeneration limits of $C_{n,g}$ to a collection of
three-punctured spheres are in perfect correspondence with the set
of all possible generalized quivers with $g$ loops and $n$ flavor
groups. As $C_{n,g}$ degenerates, the quadratic differential develops
double poles at the degeneration points. We can see the SW curve collapse
to the curve for the three-punctured spheres, and ${\cal T}_{n,g}$ take the form of a very weakly coupled
generalized quiver in the shape of the graph of three spheres (see figures in the previous subsection).

Again, it is important not to confuse the base Riemann surface with
the covering SW curve. The moduli of the former are exactly marginal
deformations of the ${\cal N}=2$ SCFT, and only affect the positions
of the poles of $\phi_2$. The period matrix of the SW curve gives
the IR gauge couplings, and it depends on the position of the zeroes
of $\phi_2(z)$.

\subsection{An explicit construction from the $A_1$ $(2,0)$ six
dimensional SCFT} Some features of the SW curves described in the
previous section may appear quite peculiar. The way the punctures on
the base Riemann surface are associated with the factors of the
flavor symmetry group is particularly intriguing. It is also
interesting that a very weakly coupled $SU(2)$ gauge group should
emerge whenever the gauge couplings are adjusted so that a handle of
the surface is pinched off. There is a general construction for
these theories, which explains such peculiarities.

The construction of the SW curves in \cite{Witten:1997sc} starts
from a IIA brane setup whose decoupling limit yields the desired
quiver. A lift to M-theory replaces the brane system with a single
$M5$ brane wrapping a curve\footnote{Notice that often the direct
M-theory lift actually leads to a curve in Taub-NUT space. There is
a simple limiting procedure which brings certain D6 branes away from
the region of interest, and converts Taub-NUT into $\IC \times
\IC^*$ without affecting the complex structure of the curve. See
\cite{GMN} for the detailed limiting procedure} in $\IC \times
\IC^*$ (or $\IC \times \IT^2$ for elliptic models) which coincides
with the SW curve .The KK reduction of the worldvolume $u(1)$
$(2,0)$ six dimensional field theory on the curve leads to the
correct IR ${\cal N}=2$ abelian Lagrangian associated to the SW
curve and appropriate SW differential. This single M5 brane is
really the deformation of a simple set of several M5 branes. In the
case of $SU(2)$ quiver gauge theories, one has two coincident M5
branes wrapping $\IC^*$ (and the four-dimensional space-time), which
intersect at each of the locations $t_a \in \IC^*$  a single
transverse M5 branes wrapping $\IC$ (and spacetime).

From the point of view of the worldvolume theory of the two
coincident M5 branes, the deformation of the brane system to a
single complicated M5 brane corresponds to an expectation value of
the transverse scalars which describe the motion of the two M5
branes in $\IC$. The scalars diverge at the location of the
transverse M5 branes. At low energy, we can take a decoupling limit.
The transverse M5 branes decouple for the four dimensional dynamics,
and only appear as codimension two defects at $t=t_a$ in the $(2,0)$
six-dimensional worldvolume theory of the two coincident M5 branes
wrapping $\IC^*$. The interacting worldvolume theory of two $M5$
branes is quite mysterious, but there are known BPS operators whose
expectation value parameterizes the transverse motion of the two
fivebranes (see for example
\cite{Bhattacharya:2008zy,Aharony:1998an}). The two M5 branes are
indistinguishable, and their positions are determined as the roots
of a polynomial $v^2 = \phi_1(t) v + \phi_2(t)$. $\phi_{1,2}$ are the
expectation values of BPS operators of dimension $2$ and $4$
respectively.

A look at the SW curve ${\cal C}$ for the linear quiver,
\begin{equation}
\prod_{a=0}^{n} (t-t_a) v^2 = M_{n+1}(t) v + U_{n+1}(t)
\end{equation}
indicates that the expectation value of the BPS operators should be
identified with $M/\prod (t-t_a)$ and $U/\prod (t-t_a)$
respectively. The effect of the transverse M5 branes appears to be
that the operators $\phi_1, \phi_2$ are allowed appropriate simple
poles at the intersection points $t_a$. A related type of defect has
been studied in the case of ${\cal N}=4$ SYM \cite{Gukov:2006jk}. It
is useful to manipulate worldvolume theory of the two M5 branes in a
similar way as we did to the SW equation. As we remove the linear term
in $v$ from the SW equation, we can strip off the
center-of-mass degrees of freedom of the M5 brane pair, leaving a
pure $A_1$ $(2,0)$ six dimensional SCFT. Next, we would like to put
the $t=0,\infty$ punctures on the same footing as the $t=t_a$
punctures. This will require a small detour, as we need to learn how
to define the $(2,0)$ theory on the complex sphere. With an eye
towards the general construction, we also want to be able to put the
$(2,0)$ theory on a generic Riemann surface while preserving 8
supercharges (${\cal N}=2$ in four dimensions).

This is readily done by a well known twisting procedure. The
R-symmetry group of the $(2,0)$ SCFT is $SO(5)$. The Poincar\'e
supercharges transform in the $(4\otimes 4)$ representation of
$SO(5,1) \times SO(5)$ with a symplectic Majorana reality
constraint. We twist the holonomy of the theory on a Riemann surface
by a $SO(2)_R$ subgroup of $SO(5)_R$. Under the subgroup
$SO(3,1)\oplus SO(2)_s \oplus SO(3)_R \oplus SO(2)_R$ the
supercharges transform as
\begin{equation}
\left((2,1)_{\frac{1}{2}} \oplus (1,2)_{-\frac{1}{2}}\right) \otimes
\left(2_{\frac{1}{2}} \oplus 2_{-\frac{1}{2}} \right)
\end{equation}
After the twisting, we are left with
\begin{equation}
(2,1;2)_{1} \oplus (2,1;2)_0 \oplus (1,2;2)_0 \oplus (1,2;2)_{-1}
\end{equation}
This includes a $SO(3)_R$ doublet of scalar supercharges, which we
identify with the
 ${\cal N}=2$ four-dimensional supersymmetries. The central charge of the  ${\cal N}=2$ superalgebra is computed as the
anticommutator of the supersymmetries in $(2,1;2)_0$ and turns out
to be the $\bar\partial $ operator on the Riemann surface. It
follows that the chiral operators in four-dimensions are obtained
from holomorphic operators on the Riemann surface. In particular the
operator $\phi_2(z)$ has charge $2$ under $SO(2)_R$, and becomes a
quadratic differential after the twisting.

Protected quantities in the twisted theory should only depend on the
complex structure of the Riemann surface, and not on the specific
choice of metric on it. While the full six dimensional theory, even
twisted, surely depends on the metric, we will make the crucial
assumption that the four dimensional limit of the theory  only
depends on the complex structure of the surface. Notice that the far
infrared information, as the prepotential of the IR abelian gauge
theory, only depends on the complex structure of the Riemann
surface, and even the BPS spectrum of the four dimensional theory is
described by certain string webs \cite{GMN} which are only sensitive
to the complex structure.

When we map the theory from the cylinder $\IC^*$ to the complex
sphere, we have seen that the expectation value of the quadratic
differential
\begin{equation}
\phi_2 = \frac{P_{2n+2}(t)}{t^2 \prod_{a=0}^{n} (t-t_a)^2} dt^2
\end{equation}
has double poles of fixed coefficients both at the original
punctures $t=t_a$ and at the new punctures at $t=0, \infty$.
We take this to be the defining properties of the codimension two defects
in the $A_1$ theory. In their basic form, they allow $\phi_2$ a
simple pole at the location of the defect. They have a mass deformation parameter. Turning on the mass deformation $m$ allows a second order pole for $\phi_2$, with coefficient $m^2$.

Our second crucial assumption is that the four dimensional limit of
the twisted $(2,0)$ theory on the cylinder is equivalent to the four
dimensional limit of the twisted $(2,0)$ theory on the sphere with
extra insertions of defect operators at $t=0,\infty$. S-duality of
the $SU(2)$ gauge theory with four flavors is a strong piece of
evidence towards this assumption, as it requires the four punctures
on the sphere to be indistinguishable. Again, neither the
prepotential of the Abelian theory in the far infrared, nor the BPS
spectrum can distinguish the difference between the punctures.

The direct relation between $SU(2)$ flavor symmetry groups and
punctures suggests that the $SU(2)$  flavor symmetries truly live at
the codimension two defects. If we conformally map the region near
the puncture to a semi-infinite thin tube, we can reduce the $A_1$
$(2,0)$ theory on the tube to a five-dimensional $SU(2)$ Young-Mills
theory on a half line. The $SU(2)$ flavor symmetry at the defect is
probably the four dimensional remnant of this five-dimensional
theory. The mass parameter for this flavor symmetry can be
interpreted as a boundary condition at infinity for the scalar
fields on the half line.

Now we are in position to give a uniform definition of the ${\cal
T}_{n,g}$ theories: the four dimensional limit of the twisted $A_1$
$(2,0)$ six-dimensional field theory on a Riemann surface of genus
$g$, in presence of $n$ defect operators at the punctures, or
equivalently on a non-compact Riemann surface of genus $g$ with $n$
semi-infinite tubular regions. The Coulomb branch of the theory is
parameterized by the choice of quadratic differential $\phi_2(z)$.
The SW curve and differentials take the expected form of a double
cover of the Riemann surface
\begin{equation} x^2 = \phi_2(z) \qquad \qquad \lambda = x dz\end{equation}
and the complex structure moduli space of the Riemann surface ${\cal
M}_{n,g}$ is a space of exactly marginal couplings for the theory.

This definition makes our conclusions regarding the degeneration
limits for the Riemann surface quite obvious.  In particular it
sheds light on why a weakly coupled $SU(2)$ gauge group should
appear in the degeneration limit. Intuitively, a long tube in the
Riemann surface can be replaced by a long segment with a
five-dimensional Young-Mills theory on it, which reduces to a weakly
coupled four dimensional gauge theory.

\section{$SU(3)$ generalized quivers} \label{sec:su3}
\subsection{Argyres-Seiberg duality}

The ${\cal N}=2$ $N_f=6$ $SU(3)$ gauge theory plays a crucial role
in this section. The gauge coupling
\begin{equation}
\tau = \frac{\theta}{\pi} + \frac{8 \pi i}{g^2}
\end{equation}
is exactly marginal, because because the number of flavors is twice
the number of colors. The theory is another simple example of a
${\cal N}=2$ SCFT. The fundamental representation of $SU(3)$ is
complex, hence the flavor symmetry is simply $U(6)$.

\begin{figure}
  \begin{center}
    \includegraphics[width=2.5in]{SL2plotb.pdf} \hfill
     \includegraphics[width=3in]{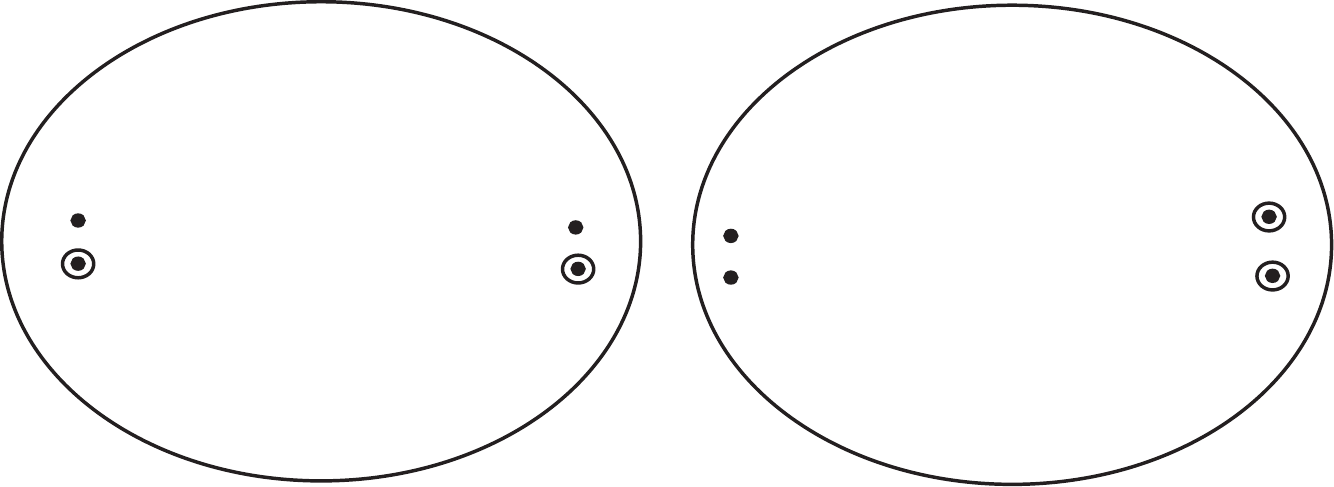}
  \end{center}
  \caption{From left to right: the space of gauge couplings $\tau$ modulo S-duality for $SU(3)$ $N_f=6$. It is the same as ${\cal M}_{(2,2),0}$, the space of spheres with
  two punctures of a type, two of another type. The degeneration limit of the four punctured sphere corresponding to the
  weakly coupled region at $\tau \to i \infty$. The degeneration limit of the four punctured sphere corresponding to the Argyres Seiberg dual weakly coupled region at $\tau \to 1$}
  \label{fig:su3mod}
\end{figure}

The moduli space of the theory has two distinct types of ``weakly
coupled'' cusps (see fig. \ref{fig:su3mod}). The first type of cusps are simply S-dual images of
the obvious weak coupling region $\tau \to i \infty$. At the second
type of cusps, a new weakly coupled $SU(2)$ gauge group emerges.
This phenomenon is the canonical example of Argyres-Seiberg duality
\cite{Argyres:2007cn}. At the second type of cusp, the $SU(2)$ gauge group is coupled
weakly to a single fundamental hypermultiplet and to an $SU(2)$
subgroup of the $E_6$ flavor symmetry group of an interacting
superconformal field theory, the $E_6$ theory.
This $SU(2)$ subgroup is such that it commutes with an $SU(6)$ in
$E_6$. This $SU(6)$ together with the $SO(2)$ flavor symmetry of the
single fundamental hypermultiplet reconstructs the $U(6)$ flavor
symmetry of the full theory. The $E_6$ theory has a one-complex
dimensional moduli space, parameterized by the expectation value
$u_3$ of a dimension $3$ operator. This is identified with the $Tr
\Phi^3$ operator in the $SU(3)$ gauge theory. The dimension two
operators in the $SU(2)$ theory and in the $SU(3)$ theory are
identified with each other.
Notice that the $E_6$ Lie algebra decomposes into the sum of $SU(6)$
and $SU(2)$ Lie algebras, together with a set of generators in the
$20 \times 2$ of $SU(6) \times SU(2)$, i.e. in the three-indices
antisymmetric representation of $SU(6)$ and fundamental of $SU(2)$.

\begin{figure}
  \begin{center}
    \includegraphics[width=2.5in]{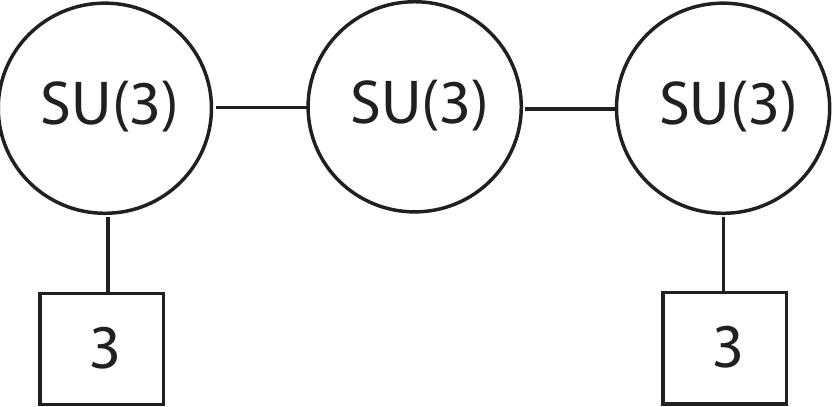} \hfill
     \includegraphics[width=2in]{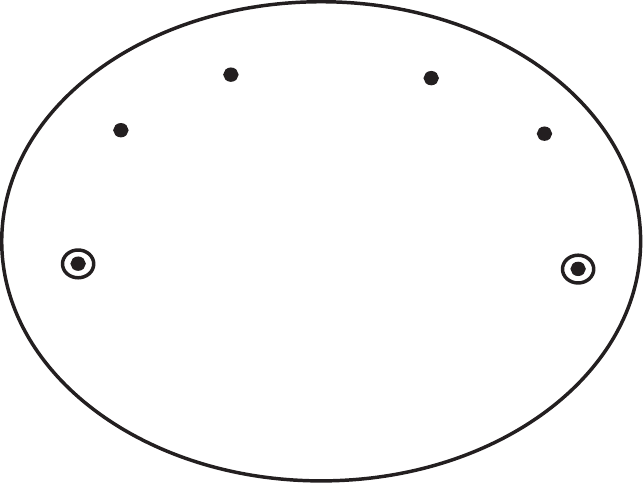}
  \end{center}
  \caption{A simple conformal linear quiver of three $SU(3)$ gauge groups. The flavor symmetry group is $U(1)^4 \times  SU(3)^2 $.
 The gauge coupling parameter space of the theory is parameterized by a sphere with four punctures of the first type, two of the second type.}
  \label{fig:su3line}
\end{figure}

Now, consider a linear quiver of $SU(3)$ gauge groups, with $3$
flavors at each end node (see fig. \ref{fig:su3line}). Such quivers are superconformal, as each
node has $6$ flavors. Each bifundamental hypermultiplet carries a
$U(1)$ flavor symmetry.
We can start with a configuration where all the gauge couplings are
very weak, pick a middle node, and make the gauge coupling at that
node strong. We see a strongly coupled $SU(3)$ gauge theory with six
flavors, and a $SU(3) \times SU(3)$ subgroup of the $U(6)$ flavor
symmetry is very weakly gauged at the neighboring nodes of the
quiver. We expect to be able to apply Argyres-Seiberg duality at the
strongly coupled node. In the dual weakly coupled frame we find an
$E_6$ theory, together with the usual $SU(2)$ weakly coupled gauge
group.

The gauge groups at the neighboring nodes couple to an $SU(3)\times
SU(3)\in SU(6)$ subgroup of the $E_6$ flavor symmetry. The remaining
$U(1)$ in $SU(6)$, together with the $SO(2)$ flavor symmetry of the
single fundamental flavor for $SU(2)$, are a mixture of the $U(1)$
flavor symmetries of the original pair of bifundamentals.

\begin{figure}
  \begin{center}
    \includegraphics[width=1.5in]{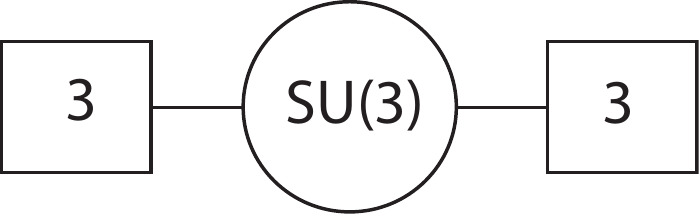} \hfill
    \includegraphics[width=1.6in]{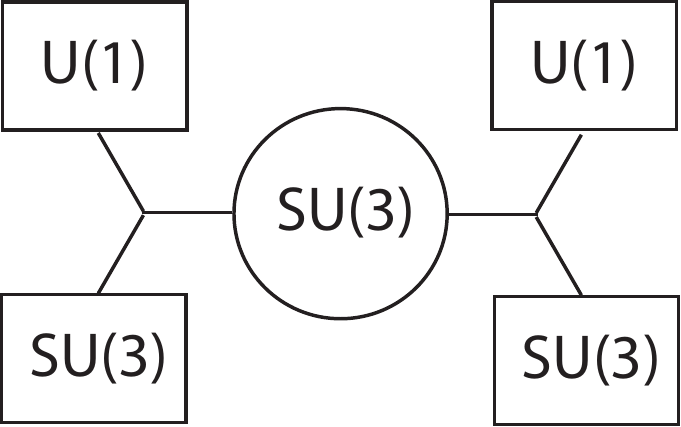}\hfill
    \includegraphics[width=2in]{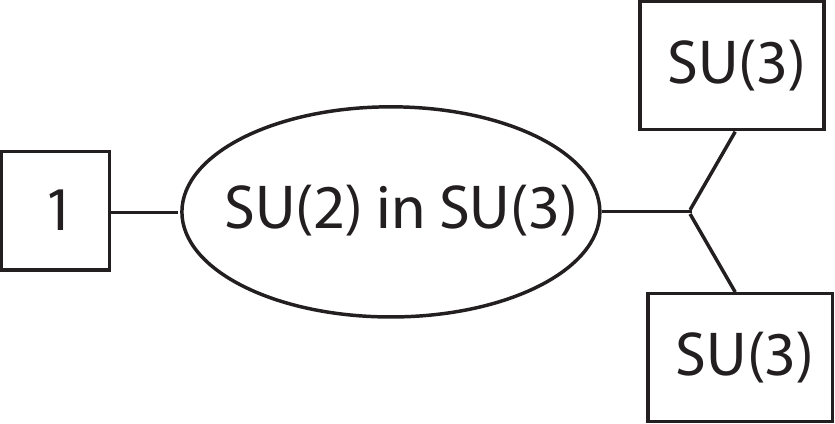}
  \end{center}
  \caption{Different useful ways to depict $SU(3)$ gauge theory with six fundamental flavors. On the left, a quiver diagram where the six flavors have been split into two groups of three flavors;
  each group carries a $U(3)=SU(3) \times U(1)$ flavor symmetry. In the middle, a generalized quiver diagram, depicting separately the two $SU(3)$ and the two $U(1)$
  flavor groups. On the right: Argyres-Seiberg S-dual frame: an $SU(2)$ gauge theory coupled to the $E_6$ theory and a single fundamental.
  We focus on a $SU(3) \times SU(3) \times SU(3)$ subgroup of $E_6$, and represent the $E_6$ theory as a three pronged junction.}
  \label{fig:su3nf6}
\end{figure}

What happens if the gauge coupling of the new $SU(2)$ node is
completely turned off? The single fundamental of $SU(2)$ will
decouple. Furthermore, the flavor symmetry of the $E_6$ theory will
grow to the commutant of the gauged $SU(3) \times SU(3)$  inside $E_6$. This
includes the $SU(2)$ we just ungauged and the $U(1)$ in $SU(6)$, but
it it is larger: the $20 \times 2$ of $SU(6) \times SU(2)$
decomposes under  $SU(3) \times SU(3)\times U(1) \times SU(2)$
according to
\begin{equation}
20 = (1\otimes 1)_3 \oplus (\bar 3\otimes 3)_1 \oplus (3 \otimes \bar 3)_{-1} \oplus (1 \otimes 1)_{-3}
\end{equation}
The $SU(3)\times SU(3)$ singlets reassemble a $SU(3)$ out of $U(1)
\times SU(2)$. Indeed $E_6$ contains a $SU(3)_a\times SU(3)_b \times
SU(3)_c$ subgroup. The $(\bar 3\otimes 3)_1 \otimes 2$ pieces
together with the remaining generators in $SU(6)$ can be recast in
the pleasant form $(3_a \otimes 3_b \otimes 3_c) + c.c.$ (Here we
replaced $\bar 3 \to 3_a$ for the first gauge group). The three
$SU(3)$ subgroups of $E_6$ are identical. As the quiver is
conformal, we learn that if any of the three $SU(3)$ subgroups of
the $E_6$ is gauged, it will provide half of the amount of matter
(same as $3$ fundamentals) required for conformal symmetry. Hence
the $E_6$ theory can play the same role in a generalized quiver of
$SU(3)$ gauge groups as the blocks of four hypermultiplets plays in
generalized quivers of $SU(2)$ groups. This motivates the depiction
of fig. \ref{fig:su3nf6}

\begin{figure}
  \begin{center}
    \includegraphics[width=1.2in]{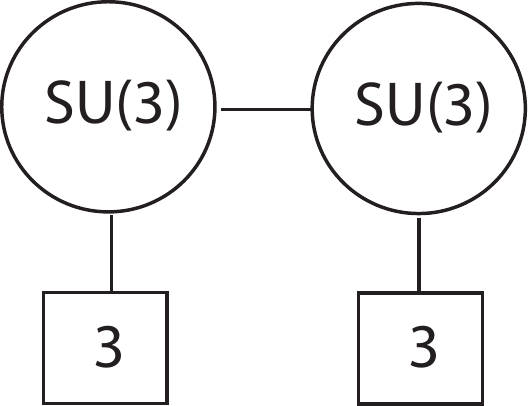} \hfill
    \includegraphics[width=2.3in]{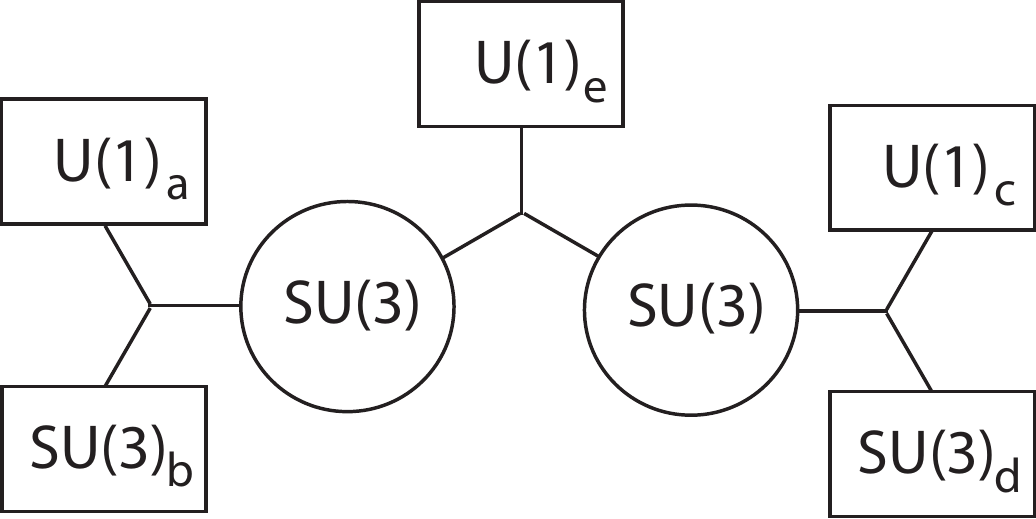} \hfill
     \includegraphics[width=1.6in]{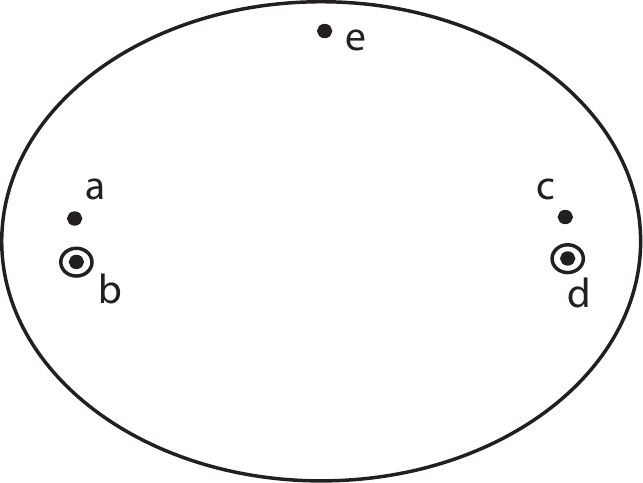}
      \end{center}
  \caption{Left: A simple conformal linear quiver of two $SU(3)$ gauge groups. The flavor symmetry group is $U(1)^3 \times  SU(3)^2 $.
 Middle: same quiver, depicted as one of the generalized quivers with $f_1=3,f_2=2,g=0$. Right: the cusp in the gauge coupling parameter space
 where this quiver description is weakly coupled.}
  \label{fig:su23a}
\end{figure}

\begin{figure}
  \begin{center}
      \includegraphics[width=3in]{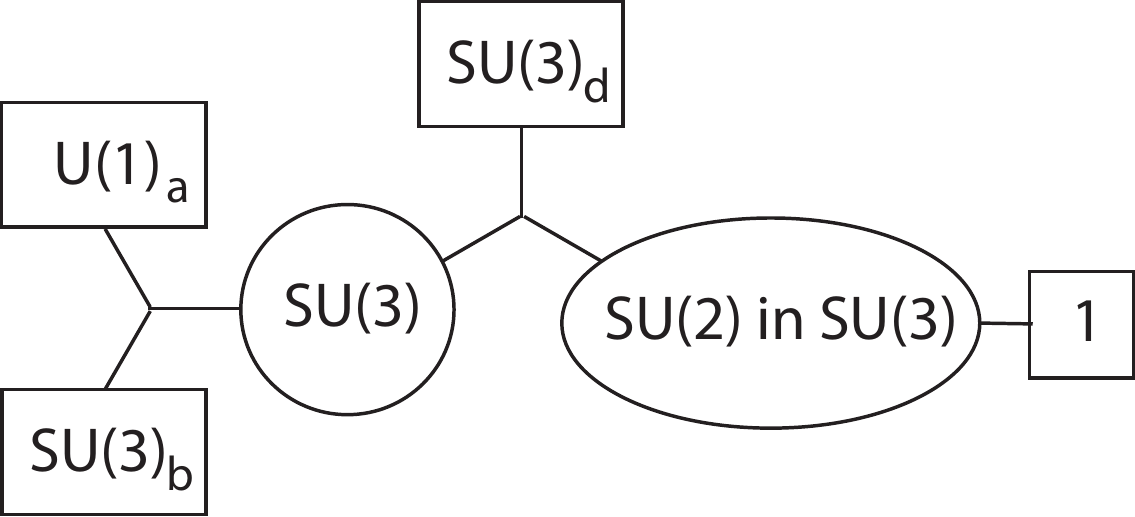} \hfill
     \includegraphics[width=2in]{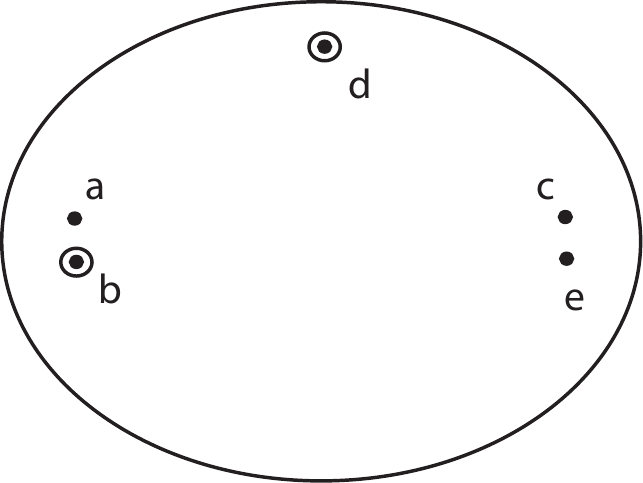}
           \end{center}
  \caption{Left: Another generalized quiver with $f_1=3,f_2=2,g=0$. The $U(1)_c\times U(1)_e$ factors are mixed in the flavor symmetry of the $SU(2)$ fundamental,
  and the commutant of $SU(2)$ in $SU(3)$. Right: the cusp in the gauge coupling parameter space
 where this generalized quiver description is weakly coupled.}
  \label{fig:su23b}
\end{figure}

\begin{figure}
  \begin{center}
      \includegraphics[width=1.6in]{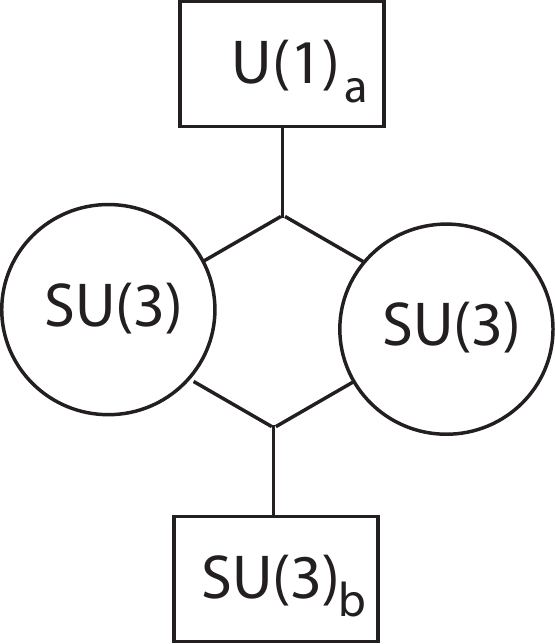} \hfill
     \includegraphics[width=2.7in]{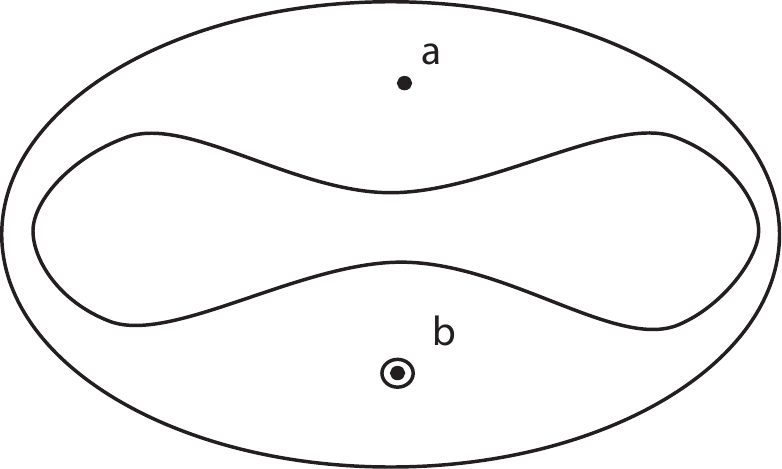}
           \end{center}
  \caption{Left: A generalized quiver with $f_1=1,f_2=1,g=1$. Right: the cusp in the gauge coupling parameter space
 where this generalized quiver description is weakly coupled.}
  \label{fig:su110a}
\end{figure}

\begin{figure}
  \begin{center}
      \includegraphics[width=1.8in]{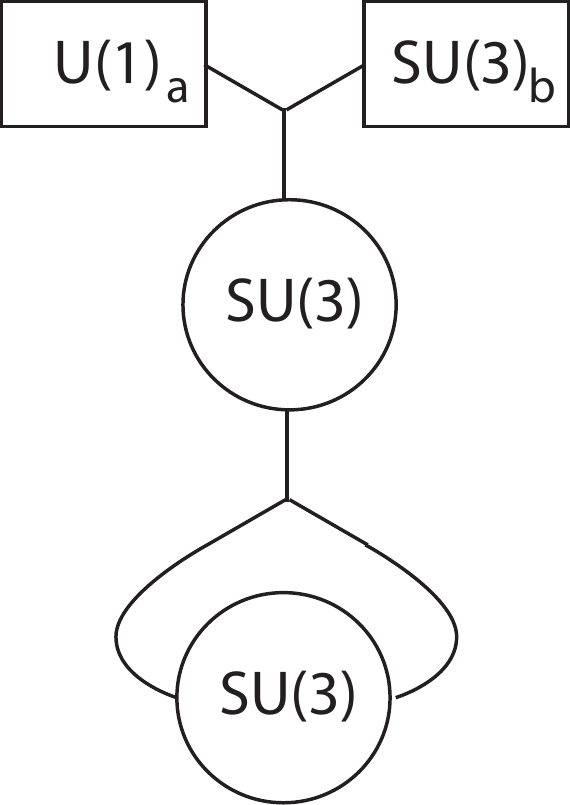} \hfill
     \includegraphics[width=2.6in]{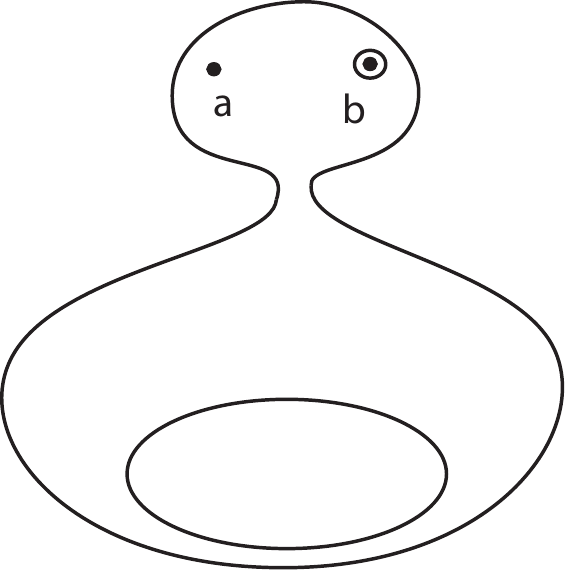}
  \end{center}
  \caption{Left: The second and last generalized quiver with $f_1=1,f_2=1,g=1$. Right: the cusp in the gauge coupling parameter space
 where this generalized quiver description is weakly coupled.}
  \label{fig:su110b}
\end{figure}

We cannot directly produce the set of S-dual generalized quivers by
simple dualities at each node, because we do not know the strong
coupling limits of the nodes where $SU(3)$ is coupled to the $E_6$
theory. Instead, we will produce directly a large set of generalized
quivers with similar properties which are candidates to be various
S-dual weakly coupled descriptions of the same theory, and then
verify through the SW curve that they exhaust the set of possible
``cusps'' of the gauge coupling parameter space. One can start with
a product of $m_2$ bifundamental blocks and $m_3$ $E_6$ theories
with a choice of an $SU(3)_a\times SU(3)_b \times SU(3)_c$ subgroup,
and introduce $n_3$ $SU(3)$ gauge groups, each coupled diagonally to
any two of the $SU(3)$ flavor symmetries of the matter system. All
the $n_3$ gauge couplings will be exactly marginal. Moreover, one
can introduce $n_2$ $SU(2)$ gauge groups, each coupled to a single
fundamental flavor and to an $SU(2)$ subgroup of an $SU(3)$ flavor
symmetry group. Again, all the $n_2$ gauge couplings will be exactly
marginal. The resulting generalized quiver gauge theory possesses
$f_3=3m_3 + 2 m_2 - 2 n_3 - n_2$ $SU(3)$ residual flavor symmetry
groups, $f_1=m_2 + 2 n_2$ $U(1)$ residual flavor symmetry groups.
The number of loops of the graph is $g=n_3 + 1 - m_3 - m_2$. Notice
that the three numbers $f_1,f_3,g$ determine the dimension of the
gauge coupling parameter space and of the Coulomb branch of the
theory. The number of gauge couplings is $n_2 + n_3 = 3g -3 + f_1 +
f_3$.  The Coulomb branch of the theory is parametrized by the
expectation value of  $n_3 + n_2=3g-3+f_1 + f_3$ dimension $2$
operators and $n_3 + m_3= 5g-5 + f_1 + 2 f_3$ dimension $3$
operators. See fig. \ref{fig:su23a},\ref{fig:su23b},\ref{fig:su23b}
for the three generalized quivers with $f_1=3,f_3=2,g=0$. See fig.
\ref{fig:su110a},\ref{fig:su110b} for the two generalized quivers
with $f_1=1,f_3=1,g=1$.

\begin{figure}
  \begin{center}
      \includegraphics[width=3in]{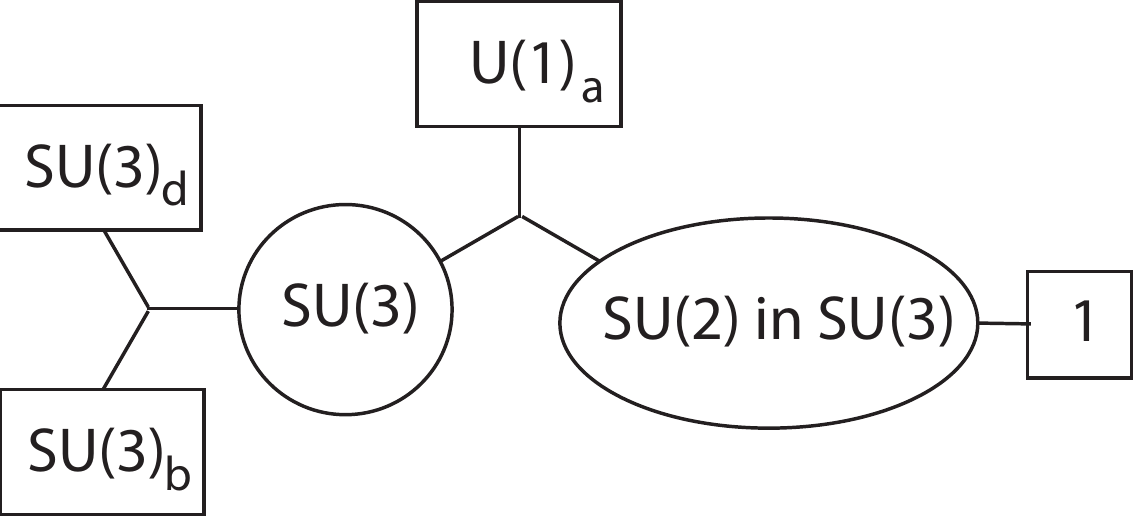} \hfill
     \includegraphics[width=2in]{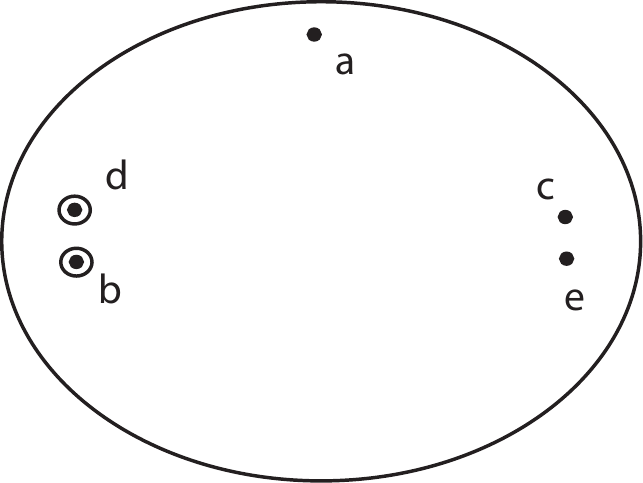}
  \end{center}
  \caption{Left: The third and last generalized quiver with $f_1=3,f_2=2,g=0$. The $U(1)_c\times U(1)_e$ factors are mixed in the flavor symmetry of the $SU(2)$ fundamental,
  and the commutant of $SU(2)$ in $SU(3)$. Right: the cusp in the gauge coupling parameter space
 where this generalized quiver description is weakly coupled.}
  \label{fig:su23c}
\end{figure}

The rest of this section will be devoted to show that the set of all
generalized quivers with $g$ loops, $f_1$ and $f_3$ flavor symmetry
groups coincides with the set of all weakly coupled duality frames
for a single ${\cal N}=2$ SCFT ${\cal T}_{(f_1,f_3),g}[A_2]$ with
flavor symmetry group (at least) $SU(3)^{f_3} \times U(1)^{f_1}$. We
will see that the gauge coupling parameter space of this theory
coincides with the moduli space ${\cal M}_{(f_1,f_3),g}$ of Riemann
surfaces of genus $g$ with two types of puncture.

\begin{figure}
  \begin{center}
    \includegraphics[width=5.5in]{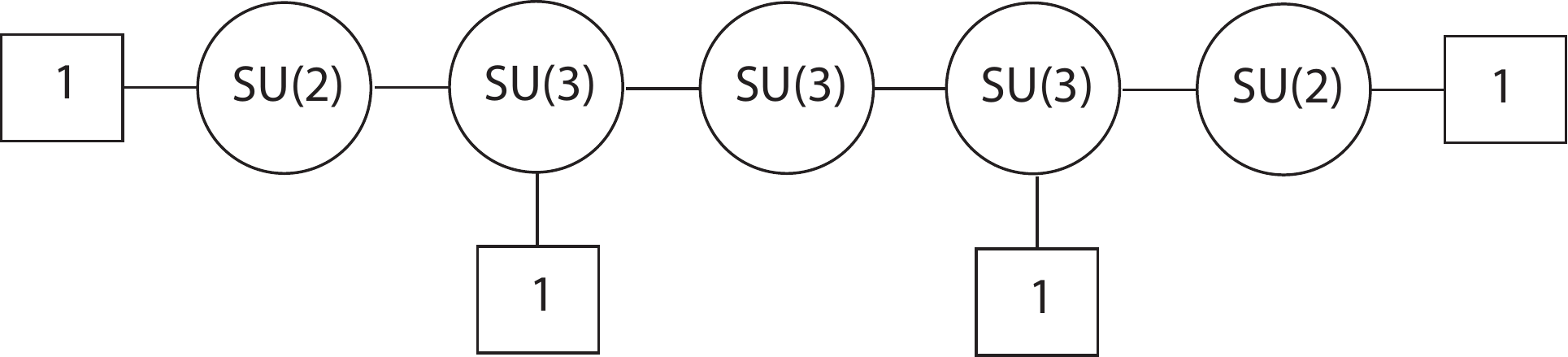}
     \includegraphics[width=5.5in]{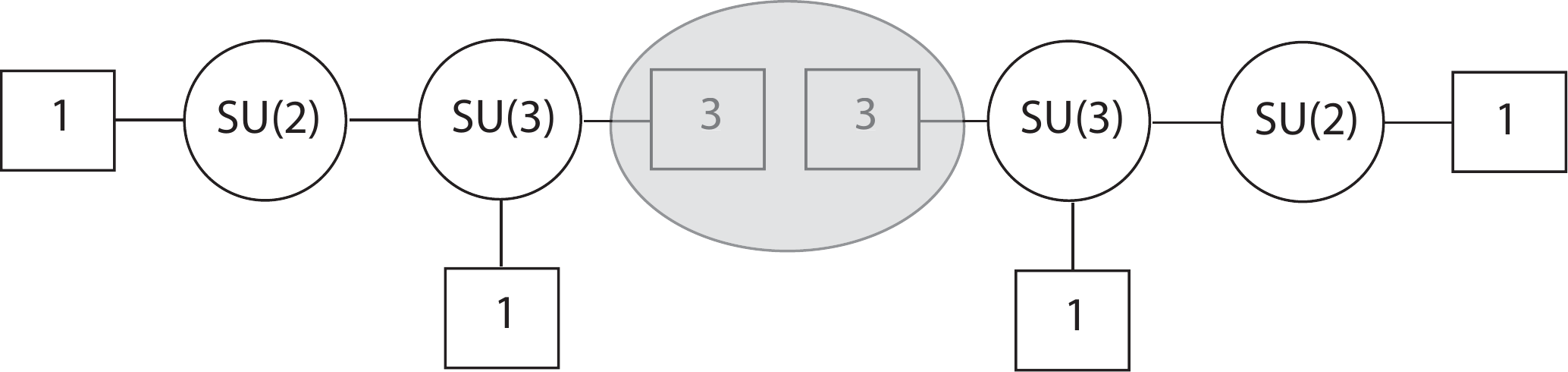}
  \end{center}
  \caption{A simple conformal linear quiver of flavor symmetry group $U(1)^8$. and one simple decoupling limit. }
  \label{fig:su3bline}
\end{figure}

Let's sketch some steps of the derivation: we consider a linear
quiver which has no $SU(3)$ flavor groups, and defines ${\cal
T}_{(n+3,0),0}[A_2]$: a conformal linear quiver theory with gauge
groups (in order) $SU(2) \times SU(3)^{n-2} \times SU(2)$(see fig.
\ref{fig:su3bline}). The gauge coupling moduli space of this theory
is the moduli space of a sphere with $n+3$ identical punctures. It
has an S-duality group which permutes all punctures and $U(1)$
flavor symmetry groups freely.

\begin{figure}
  \begin{center}
   \includegraphics[width=2in]{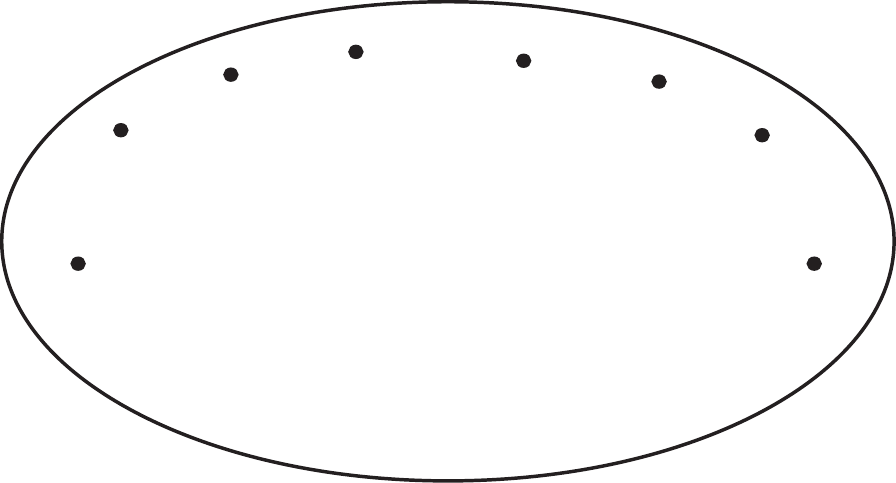} \hfill
     \includegraphics[width=3.5in]{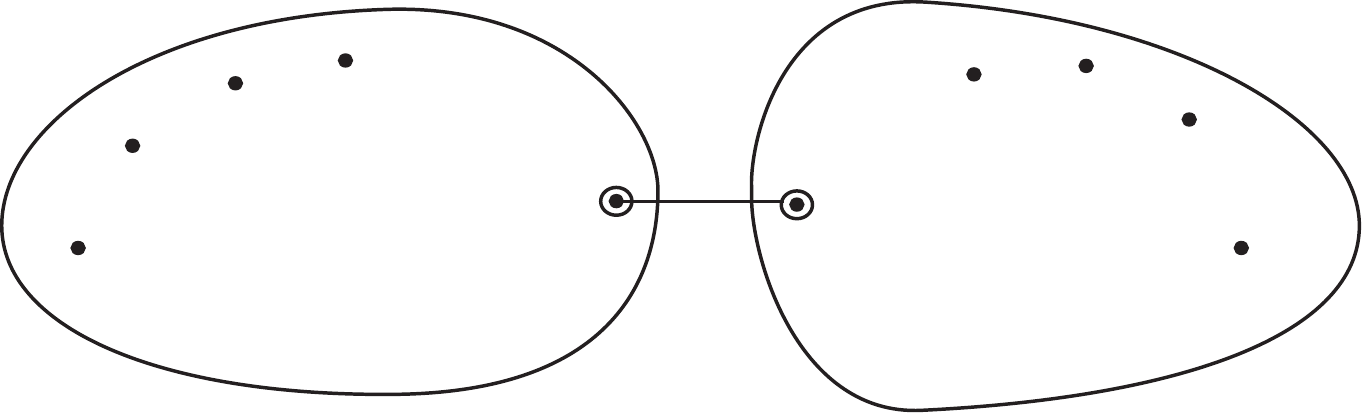}
  \end{center}
  \caption{
  The gauge coupling parameter space for the fig. \protect \ref{fig:su3bline} is parameterized by a sphere with eight punctures of the first type, none of the second type. }
  \label{fig:su3blinedeg}
\end{figure}
We will see that the region in parameter space where the sphere
degenerates, and $m+1$ punctures on the sphere collide, corresponds
in some S-dual frame to the weak coupling limit of the $m$-th gauge
group of the quiver. If $m>1$ the decoupling gauge group is a
$SU(3)$. The quiver decomposes into two subquivers (see fig.
\ref{fig:su3blinedeg}) labeled as ${\cal T}_{(m+1,1),0}[A_2]$ and
${\cal T}_{(n+2-m,1),0}[A_2]$. As a rule, punctures of the second
type appear at degeneration points. The collision of two punctures
of the first type is a bit special, in that only a $SU(2)$ gauge
group becomes very weakly coupled. The nearby $SU(3)$ node is still
coupled to three fundamental flavors with a $SU(3)$ flavor symmetry
group. The remaining quiver is labeled as ${\cal
T}_{(n+1,1),0}[A_2]$: the two punctures of the first type has
coalesced to a single  puncture of the second type with a $SU(3)$
flavor symmetry group. The decoupling $SU(2)$ gauge fields were
coupled to a single fundamental hypermultiplet and to a $SU(2)$
subgroup of this $SU(3)$ flavor group.

\begin{figure}
  \begin{center}
    \includegraphics[width=4in]{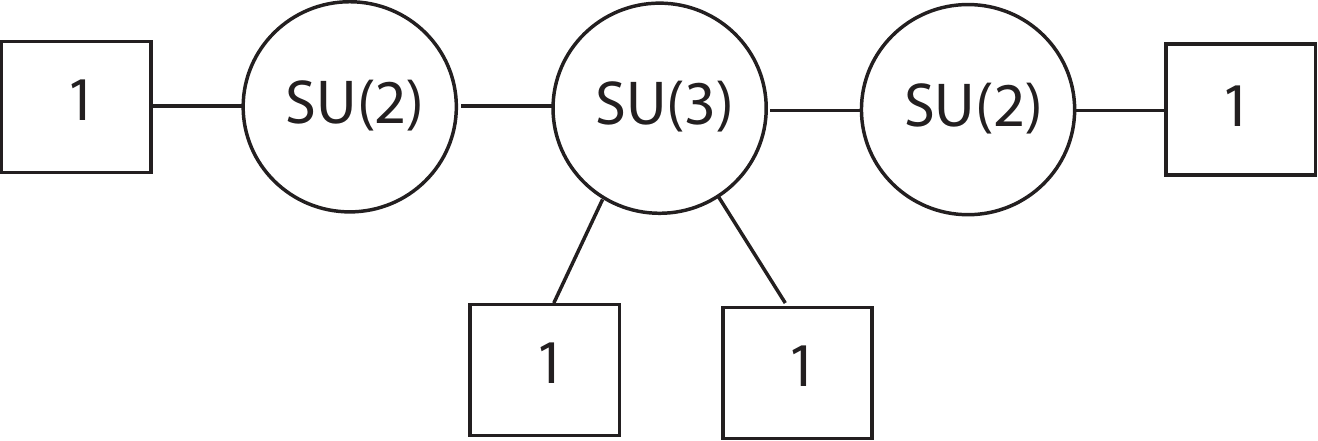} \\  \vspace{.3in}
     \includegraphics[width=5.5in]{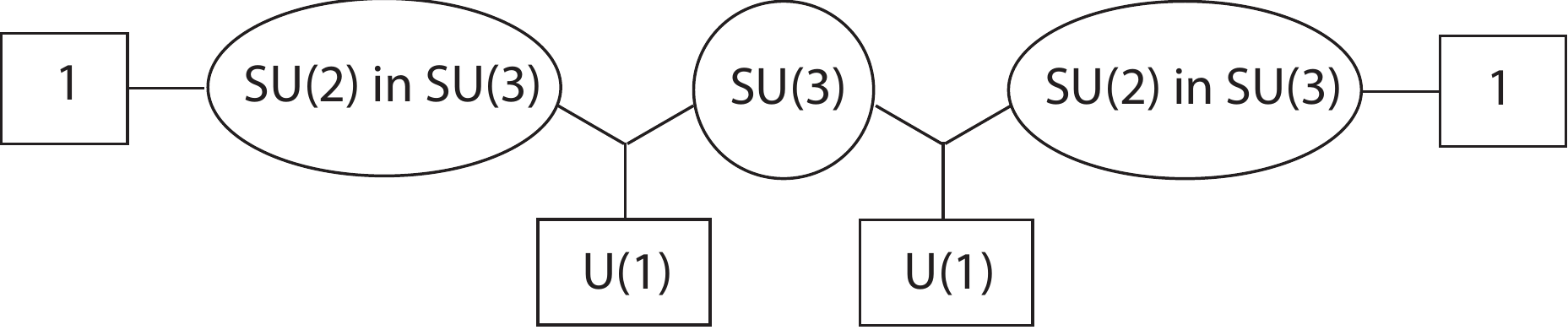}\vspace{.3in}
     \includegraphics[width=4.5in]{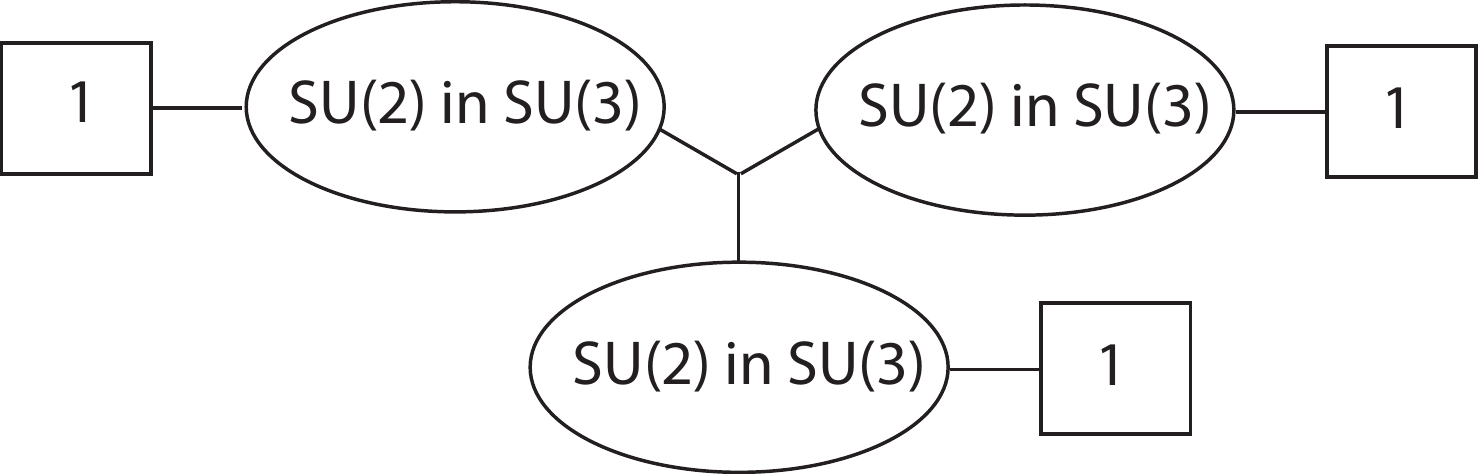}
  \end{center}
  \caption{Top: A simple conformal linear quiver realization of ${\cal
T}_{(6,0),0}[A_2]$. Colliding punctures pairwise decouples three $SU(2)$ gauge groups.
Two are visible in the original quiver, the last must emerge in the very strong coupling region of the middle $SU(3)$ node. Middle: the same quiver, depicted as a generalized quiver.
Bottom: the result of the pairwise collision of punctures must take this symmetric form. The three-pronged theory ${\cal
T}_{(0,3),0}[A_2]$ is identified with the $E_6$ theory}
  \label{fig:ASproof}
\end{figure}

We can start from the quiver for ${\cal T}_{(6,0),0}[A_2]$, and
collide the six punctures pairwise. Because of S-duality all
punctures are equivalent and three $SU(2)$ gauge groups must
decouple in total. Two are simply the two manifest $SU(2)$ gauge
groups of the quiver (see fig. \ref{fig:ASproof}). The third is the
hidden $SU(2)$ in the strong coupling region of the $SU(3)$ node.
Decoupling the three $SU(2)$ gauge groups must leave behind a theory
with three $SU(3)$ flavor symmetry groups. As any pair of $SU(3)$ is
actually part of an $SU(6)$, one can derive the full $E_6$ flavor
symmetry group of the theory. Hence S-duality of ${\cal
T}_{(6,0),0}[A_2]$ implies Argyres-Seiberg duality. The $E_6$ theory
is represented by a sphere with three punctures of the second type,
and can be denoted as ${\cal T}_{(0,3),0}[A_2]$.

\begin{figure}
  \begin{center}
    \includegraphics[width=2.6in]{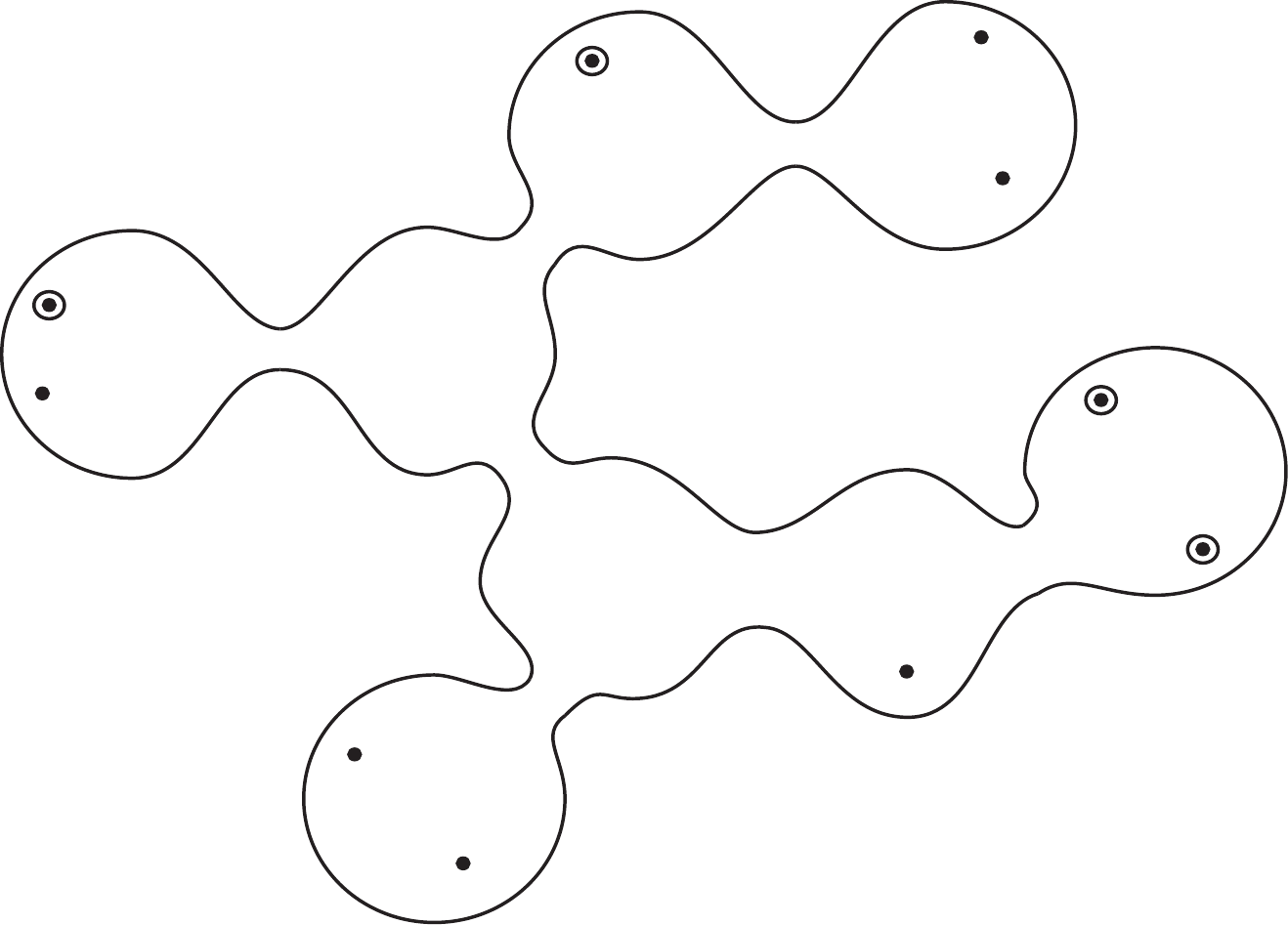} \hfill
     \includegraphics[width=3.1in]{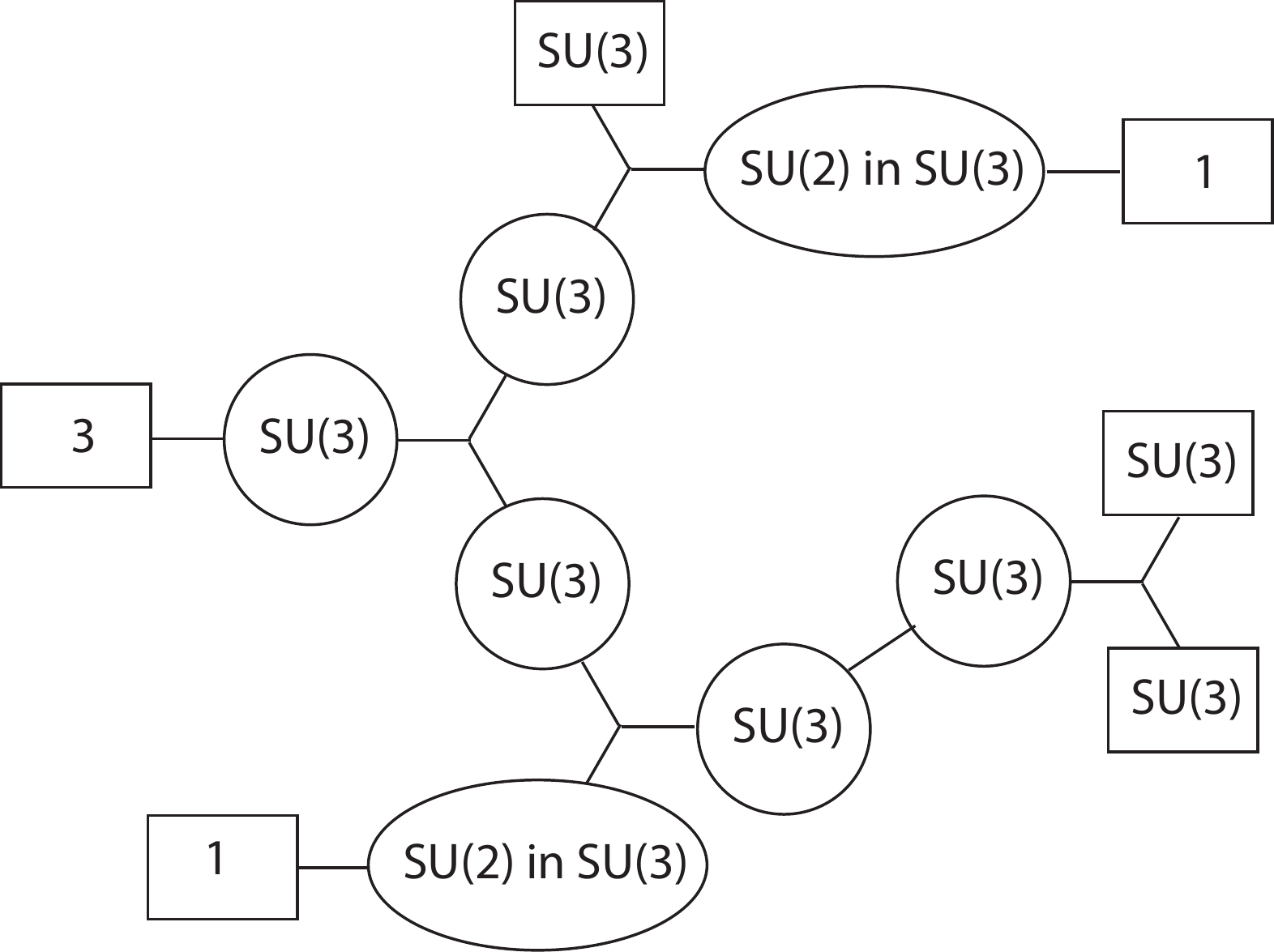}
  \end{center}
  \caption{From left to right: A generic degeneration limit of ${\cal T}_{(6,4),0}[A_2]$ and the corresponding weakly coupled generalized quiver description. For clarity we suppressed
  the $U(1)$ flavor groups, and represented fundamentals and bifundamentals simply as a line.}
  \label{fig:su3gen}
\end{figure}

We can define ${\cal T}_{(f_1,f_3),0}[A_2]$ from a very strongly coupled limit of ${\cal T}_{(f_1+2 f_3,0),0}[A_2]$, by colliding $f_3$ pairs of punctures of the first type. $f_3$ $SU(2)$ gauge groups decouple, and leave behind $f_3$ $SU(3)$ flavor symmetry groups.
Combining the information about all possible degeneration limits of ${\cal T}_{(n+3,0),0}[A_2]$,
we see that at a generic cusp, where the $f_2+f_3$ punctured sphere degenerates to a binary tree of three-punctured spheres, ${\cal T}_{(f_1,f_3),0}[A_2]$ also decomposes into elementary building blocks
coupled by $SU(3)$ or $SU(2)$ gauge groups. Furthermore, we see the three possible building blocks: ${\cal T}_{(0,3),0}[A_2]$ represents the $E_6$ theory, ${\cal
T}_{(1,2),0}[A_2]$ bifundamentals for $SU(3) \times SU(3)$. Spheres
with two punctures of the first type are a bit of an exception, as
they are ``connected'' to the rest of the theory by an $SU(2)$ gauge
group only, and they simply represent the lone fundamental for $SU(2)$
(See an example in fig. \ref{fig:su3gen}). Every cusp in the gauge coupling parameter space of
${\cal T}_{(f_1,f_3),0}[A_2]$ has a weakly coupled generalized quiver description.

We can also gauge pairs of $SU(3)$ flavor symmetry groups to produce generalized quivers with loops ${\cal T}_{(f_1,f_3),g}[A_2]$.
In that case, we have to conjecture a Seiberg-Witten curve, and there is a very natural conjecture based on a Riemann surface $C_{(f_1,f_3),g}$.
At the boundary of ${\cal M}_{(f_1,f_3),g}$ either a handle pinches off, and ${\cal T}_{(f_1,f_3),g}[A_2]$ degenerates to ${\cal T}_{(f_1,f_3+2),g-1}[A_2]$, or the Riemann surface splits in two, and correspondingly ${\cal T}_{(f_1,f_3),g}[A_2]$ degenerates to ${\cal T}_{(f'_1,f'_3+1),g'}[A_2]$ and ${\cal T}_{(f_1-f'_1,f_3-f'_3+1),g-g'}[A_2]$. At cusps of ${\cal M}_{(f_1,f_3),g}$ we find  all possible degeneration limits of a Riemann
surface with $f_1$ punctures of the first type, $f_3$ of the second
type into of a graph of three-punctured spheres, joined at nodal
singularities, corresponding to all generalized quivers labeled by $f_1,f_3,g$.
In particular all pictures of generalized $SU(2)$ quivers ${\cal T}_{n,g}[A_1]$ in the section
\ref{sec:su2} can be adapted to represent generalized $SU(3)$
quivers ${\cal T}_{(0,n),g}[A_2]$, simply replacing the blocks of
four hypermultiplets with an $E_6$ theory, and the $SU(2)$ gauge
groups with $SU(3)$ gauge groups.

\begin{figure}
  \begin{center}
    \includegraphics[width=3.5in]{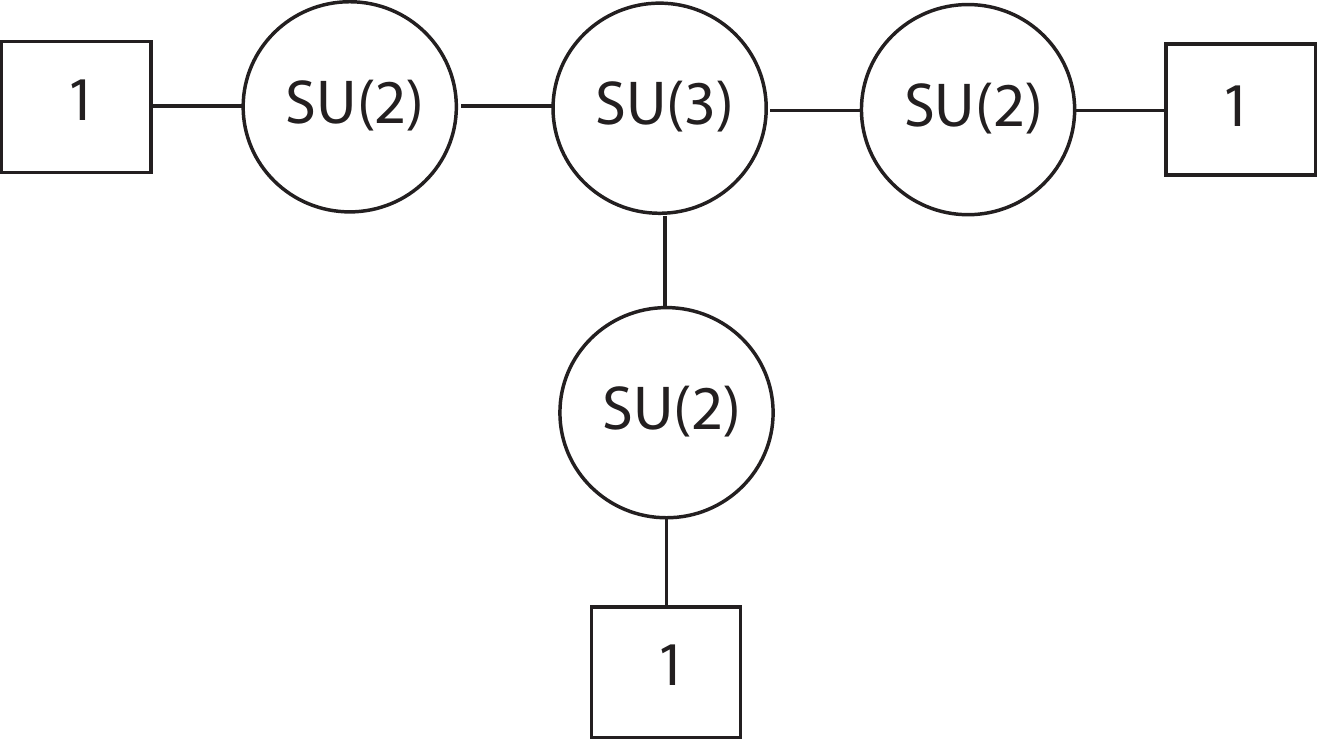}
  \end{center}
  \caption{The exceptional quiver. }
  \label{fig:exception}
\end{figure}

The reader should not think that our definition of generalized
quivers and computations of SW curves, S-dualities etc. covers all
conformal quivers of $SU(2)$ and $SU(3)$ gauge groups. There is a
single exception, depicted in fig. \ref{fig:exception}, where we
gauge the enhanced $U(2)$ flavor symmetry of ${\cal
T}_{(6,0),0}[A_2]$ to produce a result which falls outside the class
${\cal T}_{(f_1,f_3),g}[A_2]$. Notice that the theory has only $6$
$U(1)$ flavor symmetry groups, but a four dimensional gauge coupling
parameter space. Any member of the class ${\cal T}_{(6,0),g}[A_2]$
would have had $3g+3$ dimensional gauge coupling parameter space.
Furthermore, Argyres-Seiberg duality at the middle node produces an
$E_6$ theory where four identical $SU(2)$ subgroups have been
gauged.

Quivers of this shape, a D-type Dynkin diagram, are still amenable
of a brane analysis using orbifold fiveplanes (see section
\ref{sec:last} for more details)

\subsection{SW curves}
Again we will use the final result of \cite{Witten:1997sc},
specialized to a product of $n$ $SU(3)$ gauge groups with three
hypermultiplets at each end. We will follow closely the analysis for
the quiver of $SU(2)$ gauge groups.
The curve is given as a polynomial equation in two variables
$(v,t)$, defined in $\IC \times (\IC^*-\{t_0, \cdots, t_n\})$. The
polynomial is of third order in $v$ and order $n+1$ in $t$. If we
collect the same powers of $t$ it can be written as
\begin{equation}
v^3 t^{n+1} + c_1 (v^3 - u^{(2)}_1 v - u^{(3)}_1 ) t^n + \cdots + c_{n} (v^3 - u^{(2)}_n v - u^{(3)}_n ) t + c_{n+1}v^3=0
\end{equation}
The $u^{(d)}_i$ coefficients parameterize the Coulomb branch of the
theory. At weak coupling, $u_i^{(d)}$ should be identified with the
expectation value $Tr \Phi_i^d$ for the $i$-th $SU(3)$ gauge group
in the quiver. The $c_i$ parameterize the space of gauge couplings
of the theory. We can collect the same powers of $v$ as
\begin{equation}
\prod_{a=0}^{n} (t-t_a) v^3 = U^{(2)}_{n-1}(t)t v + U^{(3)}_{n-1}(t) t
\end{equation}
then at weak coupling
\begin{equation}\label{eq:couplingb} \tau_a = \frac{1}{ i \pi} \log \frac{t_{a-1}}{t_{a}}\end{equation}
The overall rescaling of $t_a$ is immaterial. We can set, for
example, $t_0=1$.

The Seiberg Witten differential has the familiar expression (See
section \ref{sec:su2} for more details)
\begin{equation}
\lambda = v \frac{dt}{t}
\end{equation}
Let us specialize to the case $n=1$. The curve equation is
\begin{equation}
(t-1)(t-t_1) v^3 = u^{(2)}t v + u^{(3)} t
 \end{equation}
This form of the equation needs some improvements. Following the
example of the $SU(2)$ analysis,  we write $v= t x$, so that
$\lambda = x dt$, and the curve becomes
\begin{equation}
x^3 = \frac{u^{(2)}}{(t-1)(t-t_1) t} x + \frac{u^{(3)}}{(t-1)(t-t_1) t^2}
\end{equation}

This is the first example of a canonical form for the SW curve of a
generalized $SU(3)$ quiver.
\begin{equation}
x^3 = \phi_2(z) x + \phi_3(z)
\end{equation}
$(x,z)$ are local coordinates in the cotangent bundle of some
punctured Riemann surface. The SW differential is the canonical one
form $x dz$, $\phi_2(z) dz^2$ is a quadratic differential and
$\phi_3(z) dz^3$ a cubic differential on the punctured Riemann
surface with appropriate poles at the puncture. In the case of the
$SU(3)$ $N_f=6$ the quadratic differential $\phi_2(t)dt^2$ has
simple poles at $t=0,1,t_1, \infty$. The cubic differential
$\phi_3(t) dt^3$ has simple poles at $t=1,t_1$, and double poles at
$t=0, \infty$. We see that the punctures at $t=0, \infty$ have
different properties than the punctures at $t=1, t_1$. We will call
punctures where the cubic differential has a simple pole ``punctures
of the first type'', and punctures where the cubic differential has
a double pole ``punctures of the second type''.

The original weak coupling region for the $SU(3)$ gauge theory is
$t_1 \gg 1$. Traditional S-duality acts by the fractional linear
transformation $t \to 1/t, x \to - t^2 x$. The S-duality invariant
moduli space is the moduli space of a sphere with two punctures of
one type, and two of another type, indicated here as ${\cal
M}_{(2,2),0}$.

The Argyres-Seiberg dual frame is weakly coupled in the degeneration
limit where $t_1$ is close to $1$. If we set $t_1=1$ and turn off
the $u^{(2)}$ parameter, we decouple the new $SU(2)$ degrees of
freedom and the SW curve becomes the curve for the $E_6$ theory:
\begin{equation}
x^3 = \frac{u^{(3)}}{(t-1)^2 t^2}
\end{equation}
There is a perfect symmetry between the punctures at $t=0, 1,
\infty$. The cubic differential has a double pole at all punctures.
We will associate the three punctures with the three $SU(3)$ flavor
symmetry subgroups of $E_6$ which played a key role in the previous
subsection. Let us consider the more general linear quiver of
$SU(3)$ gauge groups.

\begin{equation}
x^3 = \frac{U^{(2)}_{n-1}(t)}{\prod_{a=0}^{n} (t-t_a)t} x + \frac{U^{(3)}_{n-1}(t)}{\prod_{a=0}^{n} (t-t_a)t^2}
\end{equation}
The quadratic differential $\phi_2(t)dt^2$ has simple poles at all
punctures $t=0,\cdots,t_a,\cdots, \infty$. The cubic differential
$\phi_3(t) dt^3$ has simple poles at the $n+1$ punctures  $t=t_a$,
and double poles at $t=0, \infty$. The gauge coupling moduli space
is the moduli space of a sphere with $n+1$ punctures of one type,
$2$ of another type ${\cal M}_{(n+1,2),0}$. Notice that the quiver
gauge theory has $n+1$ $U(1)$ flavor symmetry groups, and $2$
$SU(3)$ flavor symmetry groups.

Can we modify this quiver in such a way that all punctures are
equivalent?  In the case of $SU(3)$ $N_f=6$,  ungauging the $SU(2)$
group in the Argyres-Seiberg dual frame coincided with the
degeneration limit where the two punctures of first type at
$t=1,t_1$ merge into a single puncture of second type, identical to
the punctures at $t=0, \infty$. We would like now to do the
opposite: replace each of the punctures of second type at
$t=0,\infty$ with a pair of punctures of the first type. It is
surprisingly simple to archive that result. We will add two new
$SU(2)$ nodes to the quiver. The new superconformal linear quiver,
${\cal T}_{(n+3,0),0}[A_2]$, has gauge groups $SU(2) \times
SU(3)^{n-2} \times SU(2)$. (The order of the factors indicates the
position along the quiver). Superconformal invariance requires a
single fundamental hypermultiplet at the $SU(2)$ nodes, and a single
fundamental hypermultiplet at the first and last $SU(3)$ nodes.

The SW curve for such a quiver can be computed in the general
framework of \cite{Witten:1997sc}. It can be written as
\begin{align}
v^3 t^{n+1} + &c_1 v (v^2 - u^{(2)}_1) t^n + c_{2} (v^3 - u^{(2)}_2 v - u^{(3)}_2 ) t^{n-1}  \cdots +\\ &c_{n-1} (v^3 - u^{(2)}_{n-1} v - u^{(3)}_{n-1} ) t^2 + c_{n}v (v^2 - u^{(2)}_n) t + c_{n+1}v^3=0
\end{align}
with the usual SW differential. We can collect the same powers of $v
= t x$ to get
\begin{equation}
x^3 = \frac{U^{(2)}_{n-1}(t)}{\prod_{a=0}^{n} (t-t_a) t }x + \frac{U^{(3)}_{n-3}(t)}{\prod_{a=0}^{n} (t-t_a) t }
\end{equation}
As desired, both $\phi_2(t) dt^2$ and $\phi_3(t) dt^3$ have now
simple poles at all $n+3$ punctures $0,\cdots, t_a,\cdots \infty$.
This indicates that the theory has a large S-duality group, similar
to the one for $SU(2)$ linear quivers, which can permute the
punctures freely. The correct gauge moduli space of this theory
should be ${\cal M}_{n+3,0}$, the moduli space of a sphere with
$n+3$ identical punctures.

We can use a fractional linear transformation to put all punctures
on the same footing.
\begin{equation}
x^3 = \frac{U^{(2)}_{n-1}(z)}{\Delta_{n+3}(z)}x + \frac{U^{(3)}_{n-3}(z)}{\Delta_{n+3}(z)}
\end{equation}
Notice that the theory has a flavor symmetry $U(1)^{n+3}$.  It is
natural to assume that each $U(1)$ group is associated to a puncture
of the first type. We will verify that momentarily by looking at
mass deformations. First though lets write down the SW curve
naturally associated to a theory ${\cal T}_{(f_1,f_3),0}[A_2]$, with
$f_1$ punctures of the first type, and $f_2$ of the second type.
\begin{equation}
x^3 = \frac{U^{(2)}_{f_1 + f_3 -4}(z)}{\Delta_{f_1}(z) \Delta_{f_3}(z) } x + \frac{U^{(3)}_{f_1 + 2f_3 -6}(z)}{\Delta_{f_1}(z) \Delta_{f_3}^2(z) }
\end{equation}

Now we should introduce mass deformations. The mass-deformed SW
curve for ${\cal T}_{(n+3,0),0}$ is given in \cite{Witten:1997sc} as
\begin{align} \label{eq:su3qm}
&(v-m_1)(v-m_2)^2 t^{n+1} + c_1 (v-m_2) (v^2 - m_3 v- u^{(2)}_1) t^n + \notag \\ +&c_{2} (v^3 - m_4 v^2- u^{(2)}_2 v - u^{(3)}_2 ) t^{n-1}  \cdots +
c_{n-1} (v^3 - m_{n+1} v^2- u^{(2)}_{n-1} v - u^{(3)}_{n-1} ) t^2 + \notag \\+&c_{n}(v-m_{n+3}) (v^2 - m_{n+2} v- u^{(2)}_n) t + c_{n+1}(v-m_{n+3})^2 (v-m_{n+4})=0
\end{align}
 Because of the freedom $v \to v+ v_0$ only $n+3$ mass parameters are physical.
We can collect the same powers of $v$
\begin{equation}
\prod_{a=0}^{n} (t-t_a) v^3 = M_{n+1}(t) v^2 + V_{n+1}^{(2)}(t) v + V_{n+1}^{(3)}(t)
\end{equation}
The coefficients of the polynomials $M, V^{(d)}$ will not be quite
generic, because of the specific pattern of $(v-m_i)$ factors in
\ref{eq:su3qm}. This pattern affects the asymptotic behavior of the
roots $v_{\alpha}(t)$ at $t=0, \infty$. As $t\to \infty$
\begin{equation}v_{\alpha}(t) \sim (m_1 + O(1/t), m_2 + O(1/t), m_2 +
O(1/t)) \end{equation} and similarly as $t\to 0$,
\begin{equation}v_{\alpha}(t) \sim (m_{n+4} + O(t), m_{n+3} + O(t), m_{n+3} +
O(t)) \end{equation}

The SW differential has poles at all punctures $t=0,
\cdots,t_a,\cdots,\infty$. At $t=t_a$ a single root $v_{\alpha}(t)$
out of three has a pole, leading directly to a pole in the SW
differential. The residue is identified with the mass parameter for
the $U(1)$ flavor symmetry of the $a$-th bifundamental
hypermultiplet. The identification of the mass parameters of the
fundamental hypermultiplets requires a little bit of care. For
example the residue at $t=t_1$ is the mass parameter for an overall
rotation of both the $2 \times 3$ bifundamentals between the first
and second node, and the fundamental at the second node. $m_1$ and
$m_2$ are mass parameters for $U(1)$ rotations of the fundamentals
at the first and second node respectively.

Things become a bit cleaner if we remove the linear term in $v$ by
shifting $v$ by $\frac{M}{\Delta} dt$ and set $v= t x$. As in the
case of $SU(2)$ quivers we keep $\lambda = x dt$, and accept the
corresponding shift in the definition of the $U(1)$ flavor
symmetries as welcome. After the shift of $v$ the SW differential
has poles on all three branches of $x_{\alpha}(t)$ at all ${n+3}$
punctures. The shift of $v$ can only modify the the residues of the
poles at $t=t_a$ from some $(3 \tilde m_a,0,0)$ to $(2 \tilde
m_a,-\tilde m_a,-\tilde m_a)$. Similarly the residues of the poles
at $t=0,\infty$ will go from something of the form $(m_1,m_2,m_2)$
to something of the form $(2 \tilde m, - \tilde m, -\tilde m)$.

We see once more that the behavior of the SW curve and differential
at all $n+3$ punctures is equivalent. We can do the standard
manipulations to bring it to the symmetric form
\begin{equation}
x^3 = \frac{P^{(2)}_{2n+2}(z)}{\Delta_{n+3}(z)^2} x + \frac{P^{(3)}_{3n+3}(z)}{\Delta_{n+3}(z)^3}
\end{equation}
The condition on the residue of $\lambda = x dt$ at the zeroes of
$\Delta_{n+3}$ can be written as a factorization of the
discriminant:
\begin{equation} \label{eq:disc}
4 P^{(2)}_{2n+2}(z)^3 - 27 P^{(3)}_{3n+3}(z)^2 = \Delta^2_{n+3}(z) Q_{4n}(z)
\end{equation}
which guarantees that there are two roots whose difference does not
diverge at the punctures. A simple check is in order: $(2n+3) +
(3n+4)$ coefficients of $P^{(2,3)}$ with $2n+6$ relations from the
factorization of the discriminant gives $3n+1$ independent numbers.
$n+3$ mass parameters, $n$ $u^{(2)}_a$ and $n-2$ $u^{(3)}_a$ moduli
also give a total of $3n+1$.

To understand the mass deformation of a theory with punctures of the
second type, we can observe the mass deformations of the linear
quiver of $SU(3)$ groups which define ${\cal T}_{(f_1,2),0}$. The
polynomials of $v$ in
\begin{equation}
\prod_a(t-t_a) v^3 = M_{n+1}(t) v^2 + V_{n+1}^{(2)}(t) v + V_{n+1}^{(3)}(t)
\end{equation}
are now unrestricted. In particular $v$ assumes three generic values
at $t=0$ and at $t=\infty$. After removing the term linear in $v$,
the residues of the SW differential at $t=0, \infty$ become of the
generic form $(m_1+m_2, -m_1, -m_2)$. This is identified with the
Cartan of the $SU(3)$ flavor symmetry group of the three fundamental
flavors at either end of the quiver.

We are in position to write the mass deformed SW curve for the
${\cal T}_{(f_1,f_3),0}$ theory
\begin{equation}
x^3 = \frac{P^{(2)}_{2 f_1 + 2 f_2 -4}(z)}{\Delta_{f_1}(z)^2\Delta_{f_3}(z)^2} x + \frac{P^{(3)}_{3f_1 + 3f_2 -6}(z)}{\Delta_{f_1}(z)^3\Delta_{f_3}(z)^3}
\end{equation}
Only the residues at the $f_1$ punctures are constrained
\begin{equation}
4 P^{(2)}_{2 f_1 + 2 f_2 -4}(t)^3 - 27 P^{(3)}_{3f_1 + 3f_2 -6}(t)^2 = \Delta^2_{f_1}(t) Q_{4f_1 + 6f_2 -12}(t)
\end{equation}
We would like to caution the reader that there is a marked
difference between a puncture of the first type where the residues
of $\lambda$ are of the form $(2 m_1, -m_1, -m_1)$ and the $m_2\to
m_1$ limit of a puncture of the second type where the residues are
generic $(m_1+m_2, -m_1, -m_2)$. Careful inspection shows that in
the latter case the difference between the two roots of $\lambda$
will still not be finite, but diverge as $(t-t_a)^{-1/2}$. $\lambda$
is not even single valued around $t_a$ in that case. Notice that the
limit $m_2\to m_1$ gives a single zero in the discriminant, instead
of the double zero required in (\ref{eq:disc}).

Now we have all tools needed to study the degeneration limits in the
gauge coupling moduli space, to recover the generalized quiver
construction for ${\cal T}_{(f_1,f_3),0}[A_2]$. Let us start again
from the basic theory ${\cal T}_{(n,0),0}[A_2]$ with punctures of
the first type only, and mass parameters turned off. If we consider
the boundary of the gauge coupling parameter space where any number $m$
of punctures come together, we can always go in a duality frame
where the $m$ punctures are at $t=0,t_0,\cdots t_{m-2}$. As in the
$SU(2)$ examples, when the punctures are brought together to zero,
$2 \pi i \tau_{m-1} \sim \log t_{m-2}/t_{m-1}$ diverges and the
$m-1$-th node of the quiver becomes infinitely weakly coupled.

If $m=2$, the decoupling gauge group is $SU(2)$, coupled to a single
fundamental hypermultiplet and to the remaining part of the quiver,
which now has an $SU(3)$ end node with three fundamental flavors.
The SW curve for the latter quiver is easily written down with the
usual methods, and brought to the canonical form ${\cal
T}_{(n+1,1),0}$ with $n-2$ punctures of the first type, and one of
the second type associated with the $SU(3)$ flavor symmetry of the
three fundamental flavors. As $t_0 \to 0$ two simple poles of the
quadratic and cubic differentials converge to a double pole. This is
the expected mass deformed puncture of the second type: the lack of
an order three pole for the cubic differential indicates that one of
the mass eigenvalues for the $SU(3)$ flavor symmetry is zero, and
the SW differential will have residues $(m_1,-m_1,0)$. $m_1^2$ is
the coefficient of the double pole of $\phi_2$, and is naturally
identified with the $u$ parameter of the decoupling $SU(2)$ gauge
group.

If $n-1>m>2$ then the decoupling gauge group is an $SU(3)$ and the
quiver breaks down to two subquivers, ${\cal T}_{(n+3-m,1),0}[A_2]$
and ${\cal T}_{(m,1),0}[A_2]$. The $n+3$ punctured sphere is indeed
degenerating to the union of two spheres joined at a nodal
singularity. We see that at the degeneration points, punctures of
the second type appear at the node. This is very natural: the
decoupling $SU(3)$ gauge group leaves behind two $SU(3)$ flavor
symmetry groups. It is simple to verify that if we start from the
general mass deformed expression
\begin{equation}
x^3 = \frac{P^{(2)}_{2 f_1 + 2 f_2 -4}(z)}{\Delta_{f_1}(z)^2\Delta_{f_3}(z)^2} x + \frac{P^{(3)}_{3f_1 + 3f_2 -6}(z)}{\Delta_{f_1}(z)^3\Delta_{f_3}(z)^3}
\end{equation}
and we collapse any number of punctures, while keeping the residues
of $\lambda$ fixed, the quadratic and cubic differentials will have
generic poles of degree $2$ and $3$ respectively at the resulting
new puncture.

Notice that if $m=3$, the decoupling $SU(3)$ gauge group was the
first of last $SU(3)$ node of the quiver. The decoupling limit
leaves behind three fragments: ${\cal T}_{(n,1),0}[A_2]$, a
fundamental of $SU(3)$ and a $SU(2)$ gauge theory coupled to $3+1$
fundamentals. We expected ${\cal T}_{(3,1),0}[A_2]$. Notice that the
overall degree of the poles for $\phi_3$ is only $3 \times 1 + 1
\times 2=5$, and is too small for a non-zero cubic differential on
the sphere (which has an overall degree at least $6$). The
Seiberg-Witten curve for ${\cal T}_{(3,1),0}[A_2]$ is degenerate
\begin{equation}
x( x^2 - \frac{u_2}{t(t-1)(t-t_1)})=0
\end{equation}
and truly reduces to the SW curve for $SU(2)$ $N_f=4$. It is
interesting though to keep the mass deformations for the type $2$
puncture (say at $t = \infty$), which came from the $u_2,u_3$
parameters of the decoupling $SU(3)$. The $\phi_3$ is not zero
anymore
\begin{equation}
x^3 = \frac{u_2+ \tilde m_2 t}{t(t-1)(t-t_1)}x + \frac{\tilde m_3}{t(t-1)(t-t_1)}
\end{equation}
This is an unusual form a mass-deformed $SU(2)$ $N_f=4$ which makes
manifest the $SU(3)$ subgroup of flavor symmetry. It takes some work
to bring it to a standard form. The simplest way is to go back to $
\IC \times \IC^*$ by the $x \to v/t$
\begin{equation}
(t-1)(t-t_1) v^3 = (u_2+ \tilde m_2 t)t v + \tilde m_3 t^2
\end{equation}
Collect the same powers of $t$
\begin{equation}
(v^3 - \tilde m_2 v - \tilde m_3) t^2 - ((t_1+1)v^2 +u_2) v t + t_1 v^3
\end{equation}
If we factor \begin{equation}(v^3 - \tilde m_2 v - \tilde m_3) =
(v-m_1)(v-m_2)(v+m_1+m_2)\end{equation} then we can redefine
$(v+m_1+m_2) t = v \tilde t$ and get the familiar form
\begin{equation}
(v-m_1)(v-m_2) \tilde t^2 - (t_1+1)(v^2 + \tilde u_2) \tilde t + t_1 v (v+m_1+m_2)
\end{equation}
The coordinate transformation $(v+m_1+m_2) t = v \tilde t$ does not
preserve the fibration structure of the Seiberg-Witten curve. ${\cal
T}_{(3,1),0}[A_2]$ is a construction of $SU(2)$ $N_f=4$ from the
$A_2$ (2,0) SCFT, which makes manifest a $U(1)^3) \times SU(3)$
subgroup of the flavor group.

It is clear that the theory ${\cal T}_{(f_1,f_3),0}[A_2]$ can be
engineered from the ${\cal T}_{(f_1+2 f_3,0),0}[A_2]$ quiver by
colliding $f_3$ pairs of punctures of the first type. In terms of
the original quiver, this operation will typically push some $SU(3)$
gauge coupling to the very strongly coupled region, so that a total
of $f_3$ $SU(2)$ gauge groups decouple.
The gauge coupling parameter space ${\cal M}_{(f_1,f_3),0}$ of ${\cal
T}_{(f_1,f_3),0}[A_2]$ has various cusps, where the $f_1+f_3$
punctured sphere degenerates to a tree of three-punctures spheres
joined by nodal singularities. Our construction shows that the
Seiberg Witten curve dutifully factorizes in the limit, and confirms
that each cusp corresponds to a weakly coupled  generalized $SU(3)$
quiver built out of $E_6$ theories, bifundamental field and the
occasional $SU(2)$ end-nodes, as discussed in the previous
subsection.

Next we would like to conjecture a SW curve for generalized $SU(3)$
conformal quiver gauge theories with loops, the ${\cal
T}_{(f_1,f_3),g}$. For inspiration we can use again the result of
\cite{Witten:1997sc} for the solution for a close linear loop of
unitary groups, which has $U(1)$ flavor symmetries only, and should
be one of the weak coupling descriptions of ${\cal T}_{(f_1,0),1}$.
For zero masses, the SW equation for the closed loop of $n$ $SU(2)$
gauge groups is given simply as curve in $\IC \times \IT^2$, which
we parameterize with coordinates $v$ and $z$. The curve is defined
by a degree $2$ polynomial in $v$
\begin{equation}
v^3 = f_2(z) v + f_3(z)
\end{equation}
The functions $f_{2,3}(z)$ are a meromorphic function on $\IT^2$,
with simple poles at the $n$ punctures $z=z_a$. The SW differential
is again $v dz$. The space of couplings of the theory is the moduli
space of a torus with $n$ marked points. In the weak coupling limit,
the gauge couplings are simply given as $\tau_a \sim z_{a+1}-z_a$.
The sum of gauge couplings is the modular parameter of the torus.
The space of possible $f_{2,3}(z)$ is $2n$ dimensional: constant
shifts are allowed, but the sum of the residues of $f_{2,3}$ is
zero, because the torus is compact.

The solution with mass deformations given in \cite{Witten:1997sc}
involves the usual slight complication: morally, one would just set
\begin{equation}
v^2 = f_1(z) v^2 + f_2(z) v + f_3(z)
\end{equation}
and identify the residues of $f_1$, which coincide with the residues
of $\lambda$, with the mass parameters of the theory, but the sum of
the residues of $f_1$ is zero if $f_1$ is a standard meromorphic
function, so one has to value $v$ in some affine bundle. The shift
of $v \to x+1/2 f_1(t)$ which  eliminates the linear term also
eliminates the need of affine bundles, bringing us back to the form
\begin{equation}
x^3 = \phi_2(z) x + \phi_3(z)
\end{equation}
Again we keep $\lambda=x dz$, as the shift produces a useful shift
in the definition of the flavor charges. $\phi_2(z)$ is now allowed
double poles at $z=z_a$, and $\phi_3(z)$ triple poles. As for the
genus zero case, before the shift only one branch of $v(z)$ had a
pole at $z=z_a$, hence after the shift the residues of the poles of
$\lambda$ on the three branches $x(z)$ must take the form $(2 \tilde
m_a,-\tilde m_a,-\tilde m_a)$. Alternatively, the discriminant
\begin{equation}
4 \phi_2(z)^3 - 27 \phi_3(t)^2
\end{equation}
is allowed poles of order at most $4$.

By analogy with the case of open and closed linear quivers, we would
to associate the ensemble of generalized quiver gauge theories with
$g$ loops,$f_1$ $U(1)$ flavor symmetry groups and $f_3$ $SU(3)$
flavor symmetry groups ${\cal T}_{(f_1,f_3),g}$ with the canonical
``rank $3$'' SW curve on a Riemann surface $C_{(f_1,f_3),g}$ of genus $g$ and
$(f_1,f_3)$ punctures.
\begin{equation}
x^3= \phi_2(z)x + \phi_3(z)
\end{equation}
We will allow the quadratic differential to have simple poles at all
punctures, and the cubic differential to have simple poles at the
punctures of first type, double at the punctures of second type.

Mass deformations will appear as double poles for the quadratic
differential, triple for the cubic differential. The two types of
punctures are distinguished by the fact that at punctures of the
first type the residues of $\lambda$ on the three branches should be
again of the form $(2 \tilde m_a,-\tilde m_a,-\tilde m_a)$, i.e. the
discriminant
\begin{equation}
4 \phi_2(z)^3 - 27 \phi_3(t)^2
\end{equation}
is allowed poles of order at most $4$.

Notice that the linear space of quadratic differential with
appropriate poles on a Riemann surface of genus $g$ has $3g-3 + f_1
+ f_3$, as they are in one-to-one correspondence with the complex
moduli of the Riemann surface itself. The space of cubic
differentials with appropriate poles is of dimension $5g-5 + f_1 + 2
f_3$. The dimension of the Coulomb branch of the generalized quiver
is correctly reproduced, as each puncture of the first type
corresponds to an $U(1)$ factor in the flavor symmetry group, each
puncture of the second type to a $SU(3)$ factor.

The possible degeneration limits of $C_{(f_1,f_3),g}$ to a collection of
three-punctured spheres are again in perfect correspondence with the
set of all possible generalized quivers with $g$ loops and
appropriate flavor groups.

\subsection{An explicit construction from the $A_2$ $(2,0)$ six
dimensional SCFT}
 In the case of $SU(3)$ quiver gauge theories, the brane construction involves three coincident M5 branes  wrapping $\IC^*$ or $\IT^2$ and the four-dimensional space-time.

We can immediately give a uniform definition of the ${\cal
T}_{(f_1,f_3),g}$ theories: the four dimensional limit of the
twisted $A_2$ $(2,0)$ six-dimensional field theory on a Riemann
surface of genus $g$, in presence of $f_1+f_3$ defect operators of
appropriate type at the punctures, or equivalently on a non-compact
Riemann surface of genus $g$ with $n$ infinite tubular regions and
appropriate boundary conditions at the end of the tubes.

The $A_2$ theory has two protected operators, of R-charge $2$ and
$3$ respectively, which after the twist become a quadratic and a
cubic differential. We identify their expectation values with
$\phi_{2,3}$. We will not try to define the defect operators in
detail. Through our examples, we have specified the  admissible
poles for $\phi_{2,3}$ at the defects of the two types, and the
flavor symmetries associated with the defects. We refer the reader
to section 3 of \cite{GMN} for a more detailed discussion of the
relation between these defects and similar codimension $2$ defects
in 5d Young-Mills theory.
\section{$SU(N)$ generalized quivers}\label{sec:sun}
\subsection{Intermission}
Before we continue, it is useful to ask a simple question: what are
the possible superconformal quivers with unitary gauge groups? If we
denote with $(n_i)$ the vector of ranks of the nodes, and $(m_i)$
the number of fundamental flavors at the nodes, then the condition
$N_f = 2 N_c$ at each node can be written as $C n = m$, where
$C_{ij}=-1$ if the nodes $i$ and $j$ are contiguous, and $C_{ii}=2$.

\begin{figure}
  \begin{center}
    \includegraphics[width=3.8in]{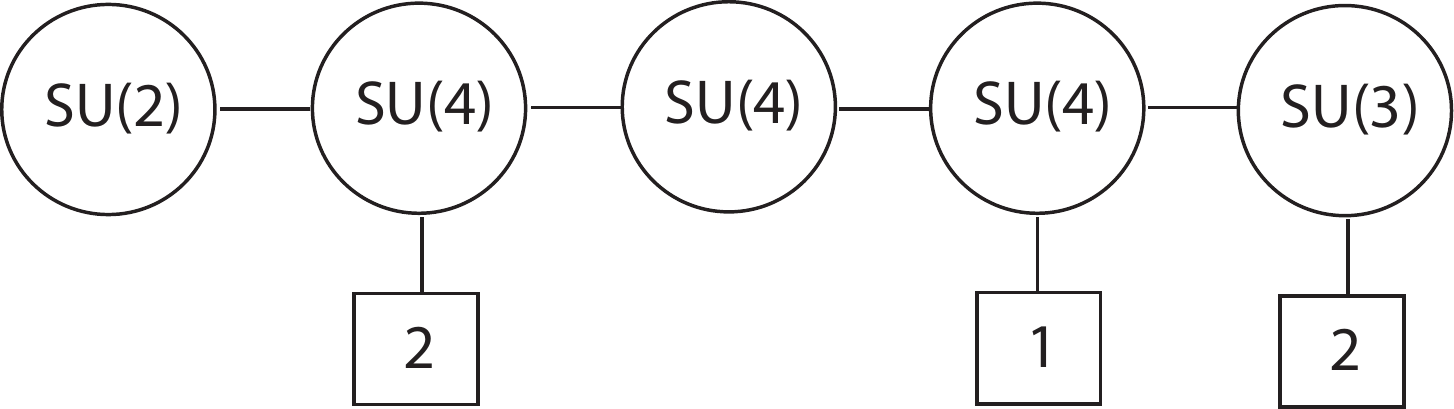}
  \end{center}
  \caption{A conformal quiver of unitary gauge groups in the shape of an $A_5$ Dynkin diagram}
  \label{fig:typeA}
\end{figure}

\begin{figure}
  \begin{center}
    \includegraphics[width=3.8in]{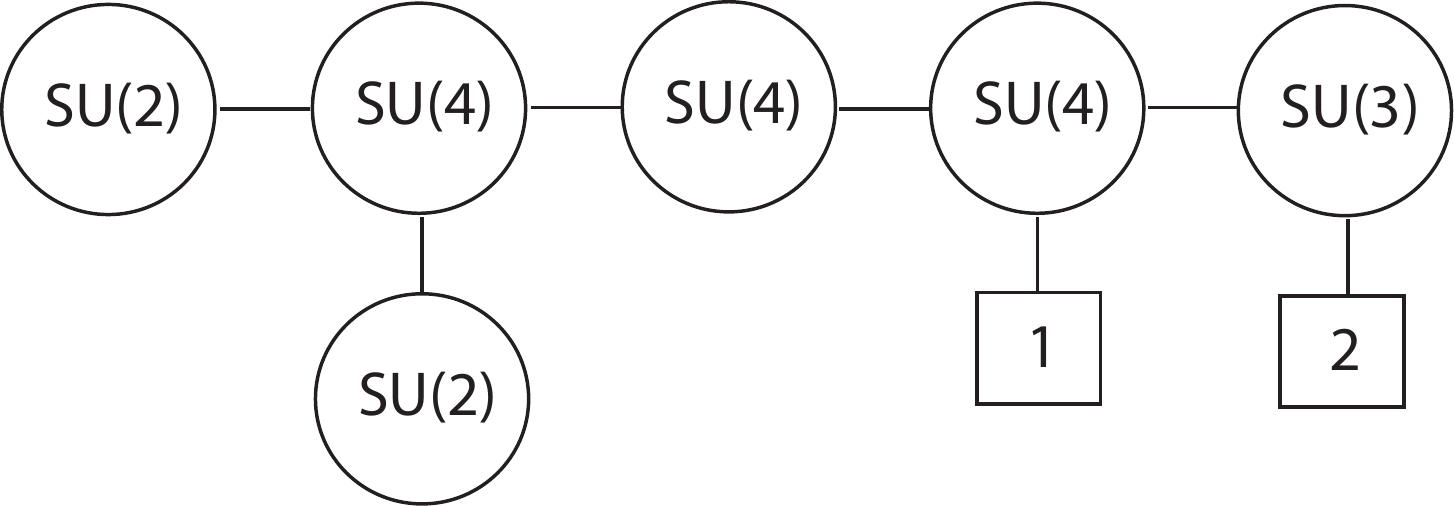}
  \end{center}
  \caption{A conformal quiver of unitary gauge groups in the shape of an $D_6$ Dynkin diagram}
  \label{fig:typeD}
\end{figure}

\begin{figure}
  \begin{center}
    \includegraphics[width=5.5in]{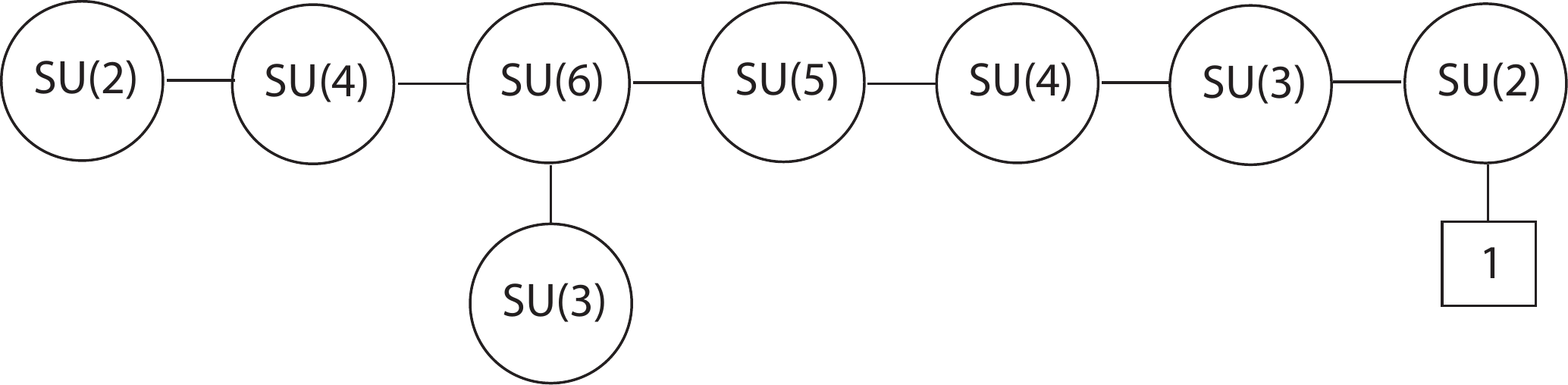}
  \end{center}
  \caption{A conformal quiver of unitary gauge groups in the shape of an $E_8$ Dynkin diagram}
  \label{fig:typeE}
\end{figure}

The graphs which admit a solution with positive non-zero $n$ and
non-negative $m$ have been identified in \cite{Gaiotto:2008ak} as
ADE graphs (see fig.
\ref{fig:typeA},\ref{fig:typeD},\ref{fig:typeE}). The $A$ type of
quivers coincide with the linear quiver which are solved in
\cite{Witten:1997sc}. The $D$ types of quivers can be probably
solved with similar methods, including orbifold fiveplanes planes in
the brane construction. (See \cite{Gaiotto:2008ak} for a similar
three-dimensional construction). The $E$ type quivers may lie
outside our reach. It would be interesting to compute the
Seiberg-Witten curve and differential for those.

\begin{figure}
  \begin{center}
    \includegraphics[width=4.7in]{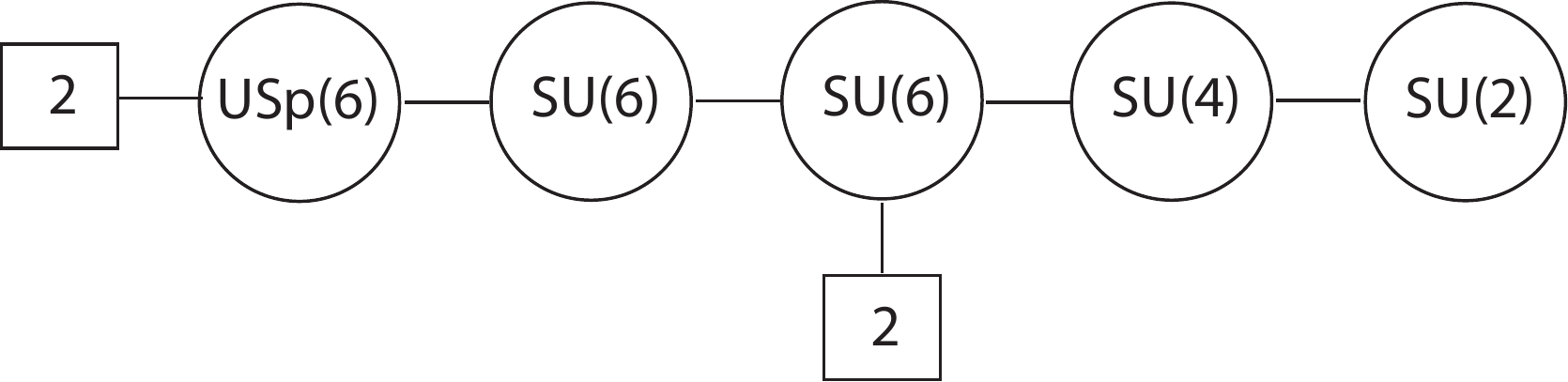}
  \end{center}
  \caption{A conformal linear quiver of unitary gauge groups terminated by an $USp$ gauge group. The two flavors at the $USp$ node have a $SO(4) = SU(2) \times SU(2)$ flavor symmetry}
  \label{fig:typeSp}
\end{figure}

We will meet another special class of superconformal quiver gauge
theories. A linear quiver of unitary groups can be terminated by a
symplectic group $USp(2n)$ at one or either ends (see fig.
\ref{fig:typeSp}) \cite{Argyres:2002xc}. The condition for conformal
invariance of the $USp(2n)$ node is that it should be coupled to a
total of $2n+2$ fundamental flavors (we use a notation where $m$
flavors of $USp(2n)$ have a $SO(2m)$ flavor symmetry). It is also
possible to end a quiver of unitary groups by coupling to matter in
an antisymmetric representation. This case could probably also be
treated through the orientifold construction in
\cite{Argyres:2002xc}.
\subsection{Canonical form}
In the $SU(N)$ case we do not have the guidance of S-duality or
Argyres-Seiberg duality. We will need to immediately plunge in the
analysis of examples of SW curves. We expect to find a structure
akin to the $SU(2)$ and $SU(3)$ generalized quivers. For that
reason, we will set up a convenient framework.

The canonical form for SW curves of generalized $SU(N)$ quivers will
always be of the now familiar form: a curve in the cotangent bundle
of a Riemann surface of genus $g$ and $n$ punctures, with canonical
one form $x dz$ and
\begin{equation}
F(x,z) = x^N - \sum_{i=2}^{N} x^{N-i} \phi_i(z)
\end{equation}
The $\phi_i(z) dz^i$ are degree $i$ differentials with appropriate
poles at the punctures.

There is a rich zoo of possible choices at the punctures. Without
mass deformations, there are two extreme possibilities: simple poles
for all $\phi_i$, or poles of order up to  $i-1$ for $\phi_i$. More
general conditions are possible, labeled by the order $p_i < i$ of
admissible poles for $\phi_i$. As long as $p_i < i$, $x \sim
(z-z_a)^{-c/q}$ at punctures, with $c<q$ both positive integers.  If
we write $(z-z_a) \sim w^q$ then $\lambda$ has locally the form
$w^{q-1-c} dw$ and has no poles at the puncture. In practice, we
will soon find that the punctures we need have $c=q-1$, hence
$\lambda$ is not forced to have zeroes either. Then the roots of $x$
at $z_a$ will come in groups of $q_s$ $x \sim
(z-z_a)^{\frac{1-q_s}{q_s}}$, and one possible labeling of the type
of punctures will be the set of $q_s>0$ such that $\sum_s q_s =N$.

With mass deformations turned on, we take $\phi_i$ to have order
$i$, so that $\lambda$ has simple poles at the punctures. The $N$
residues of $\lambda$ at the puncture may not be all distinct, and
we can require various groups of $h_t$ identical residues. Thus we
can consider a puncture labeled by a partition of $N=\sum_t h_t$.
Our first task will be to find out which partitions $h_t$ of $N$
label mass deformations of which pole structures $(p_i)$ and which
partitions $q_i$, and which flavor symmetry group is associated with
such puncture. Luckily, we will be able to find sufficient examples
among superconformal linear quivers to derive the complete
dictionary.

\begin{figure}
  \begin{center}
    \includegraphics[width=2in]{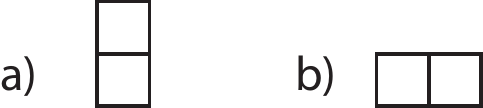} \hfill
     \includegraphics[width=2.5in]{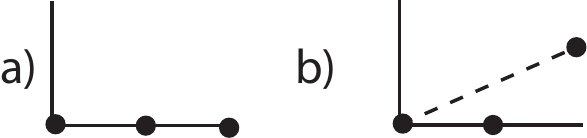}
  \end{center}
  \caption{Young tableaux labeling of $SU(2)$ punctures. Corresponding Newton polygon.\newline
  a) Absence of a puncture. $p_1=1-1=0$, $p_2=2-2=0$. Two equal mass parameters adding to zero. \newline
  b) Basic puncture: $p_1=1-1=0$, $p_2=2-1=1$. Mass parameters $(m,-m)$. $SU(2)$ flavor symmetry. \newline}
  \label{fig:Young2}
\end{figure}

\begin{figure}
  \begin{center}
    \includegraphics[width=2.6in]{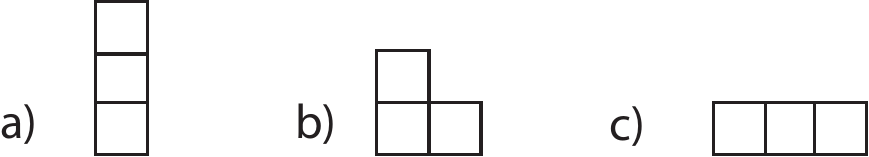} \hfill
     \includegraphics[width=3in]{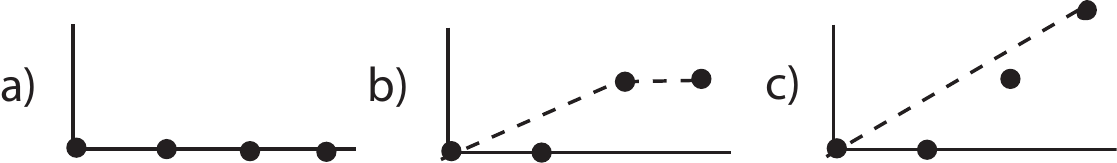}
  \end{center}
  \caption{Young tableaux labeling of $SU(3)$ punctures. Corresponding Newton polygon.\newline
  a) Absence of a puncture. $p_i=i-i=0$. Three equal mass parameters adding to zero. \newline
  b) Type 1 puncture: $p_1=1-1=0$, $p_2=2-1=1$, $p_3=3-2=1$. Mass parameters $(m,m,-2m)$. $S(U(1)\times U(1))=U(1)$ flavor symmetry. \newline
  c) Type 2 puncture: $p_i = i-1$. Generic mass parameters $(m_1,m_2,-m_1-m_2)$. $SU(3)$ flavor symmetry}
  \label{fig:Young3}
\end{figure}

\begin{figure}
  \begin{center}
    \includegraphics[width=5.5in]{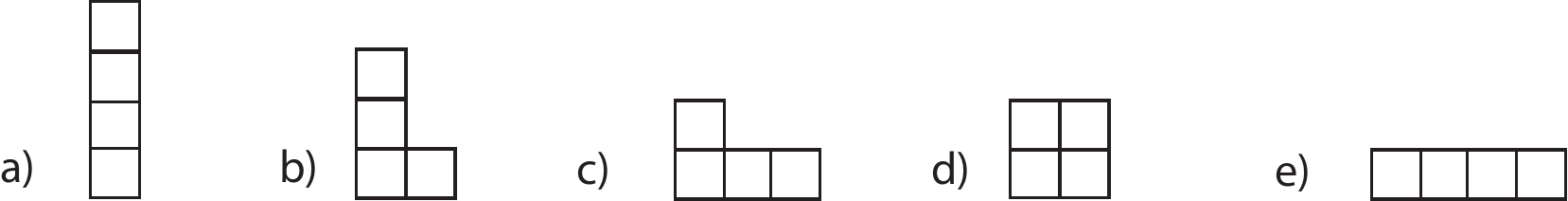} \\ \vspace{.2in}
    \includegraphics[width=5.5in]{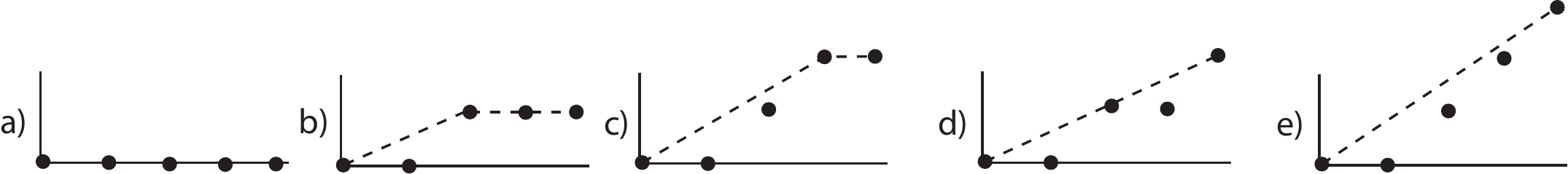}
  \end{center}
  \caption{Yang tableaux labeling of $SU(4)$ punctures. Corresponding Newton polygon.\newline
  a) Absence of a puncture. $p_i=i-i=0$. Four equal mass parameters adding to zero. \newline
  b) Basic puncture: $p_i=1$. Mass parameters $(m,m,m,-3m)$. $S(U(1)^2)=U(1)$ flavor symmetry. \newline
  c) $p_1=1-1=0$, $p_2=2-1=1$, $p_3=3-1=2$, $p_4=4-2=2$. Mass parameters $(m_1,m_1,m_2,-2m_1-m_2)$. $S(U(1)\times U(2))$ flavor symmetry. \newline
  d) $p_1=1-1=0$, $p_2=2-1=1$, $p_3=3-2=1$, $p_4=4-2=2$. Mass parameters $(m_1,m_1,-m_1,-m_1)$. $SU(2)$ flavor symmetry. \newline
  e) Full puncture: $p_i = i-1$. Generic mass parameters $(m_1,m_2,m_3,-m_1-m_2-m_3)$. $SU(4)$ flavor symmetry}
  \label{fig:Young4}
\end{figure}

The simple answer is that if $q_s$ are the lengths of rows of a
Young Tableau of $N$ boxes, then $h_t$ are the height of the
columns. The flavor symmetry is $S(\prod U(N_h))$, where $N_h$ is
the number of columns of height $h$. If we write the Young Tableaux
with columns of height decreasing left to right, and rows of length
increasing top to bottom, the numbers $i-p_i$ are monotonically
non-decreasing, and correspond to the height of the $i-th$ box of
the diagram (see fig. \ref{fig:Young2},\ref{fig:Young3} and
\ref{fig:Young4}). The relation between the $p_i$ and the $q_s$ is
easily understood through the Newton polygon.

The Newton polygon predicts the number of roots of $x$ with given
behavior at $z_a$ in terms of the slopes of sides of the convex
envelope of the $p_i$. The basic idea is to solve for $x$
approximately by only considering terms in the polynomial for which
$p_i$ lie on a given side of the convex envelope. The approximate
solution for $x$ behaves like $(z-z_a)^{-1}$ to the power of the
slope of that side. Every other term in the polynomial is a
subleading correction to this approximation for $z\to z_a$. It can
be easily shown that this method identifies all the roots of the
polynomial.

From $i=1$ up, the $p_i$ grow with slope $1$ except at the beginning
of each row of the Young Tableau, where they remain constant for one
step. Hence each row of length $q_s$ of the Young Tableau
corresponds to a side of the convex envelope, of slope
$\frac{q_s-1}{q_s}$.

As the case of $SU(2)$, $SU(3)$ quiver gauge theories, we can
 give a uniform definition of certain ${\cal T}_{(f_a),g}[A_{N-1}]$ theories:
the four dimensional limit of the twisted $A_{N-1}$ $(2,0)$
six-dimensional field theory on a Riemann surface of genus $g$, in
presence of $(f_a)$ defect operators of appropriate type at the
punctures, or equivalently on a non-compact Riemann surface of genus
$g$ with $n$ infinite tubular regions and appropriate boundary
conditions at the end of the tubes. The $A_{N-1}$ theory has $N-1$
protected operators, of R-charge $i=2\cdots N$, which after the
twist become differentials of degree $i$. We identify their
expectation values with $\phi_{i}$. We will not try to define the
defect operators in detail, but we are satisfiesd to have given the
form of admissible singularities for $\phi_i$ at the defects of
various types, and the flavor symmetries associated with the
defects. We refer the reader to section 3 of \cite{GMN} for a more
detailed discussion of the relation between these defects and
similar codimension $2$ defects in 5d Young-Mills theory.

\begin{figure}
  \begin{center}
    \includegraphics[width=3.6in]{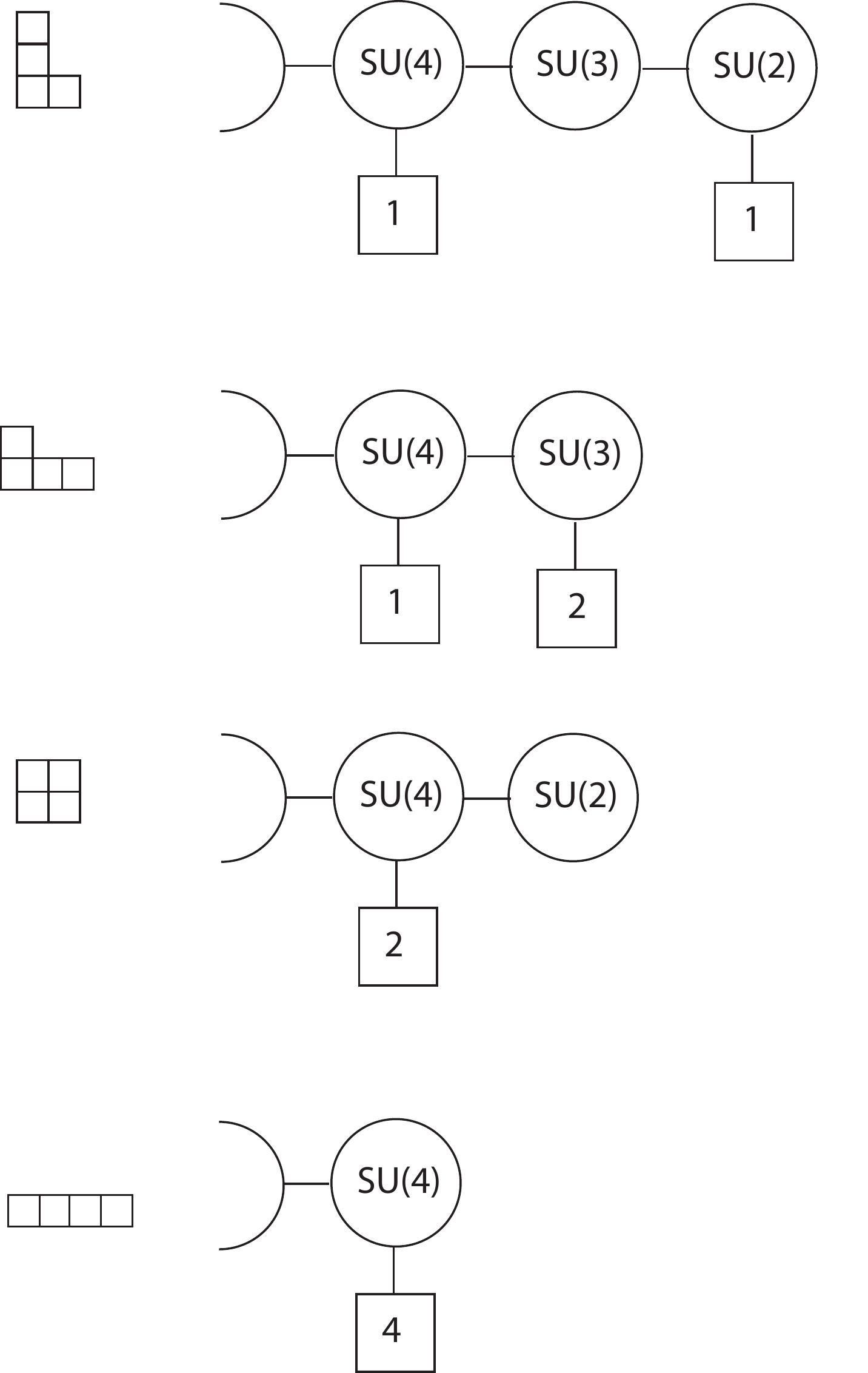}
  \end{center}
  \caption{Different ways to end a conformal linear quiver of $SU(4)$ groups, and corresponding punctures.}
  \label{fig:quiverend}
\end{figure}

Because of the $N_f=2 N_c$ condition, general conformal linear
quivers have a convex profile of gauge group ranks $2n_i -n_{i+1} -
n_{i-1}=m_i$. From one end to the other, the ranks $n_i$ grow up to
a  maximum value $N$, then decrease again. The Seiberg-Witten curve
for such a quiver is associated to a sphere with several basic
punctures $p_i=1$ and two generic punctures, whose data encode the
specific profile of the ranks $n_i$ up to $N$ at either end of the
quiver. More precisely, the differences $n_{i+1}-n_i$ at the
beginning of the quiver coincide with the lengths of the rows of the
Young Tableaux for the first generic puncture. The differences
$n_{i-1}-n_i$ at the end of the quiver coincide with the lengths of
the rows of the Young Tableaux for the second generic puncture (see
fig. \ref{fig:quiverend} for all $N=4$ cases).

\begin{figure}
  \begin{center}
    \includegraphics[width=5.3in]{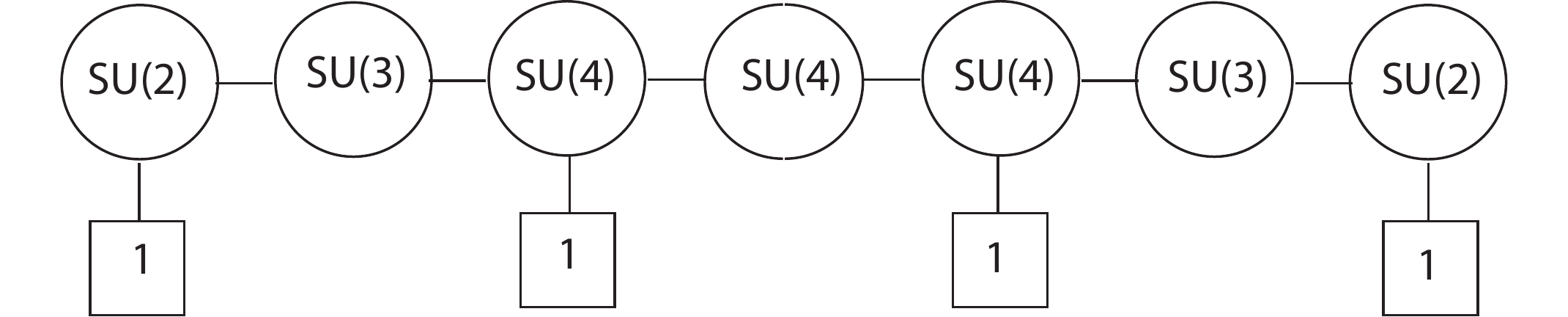}
  \end{center}
  \caption{Linear conformal quiver associated to a sphere with $10$ basic $p_i=1$ punctures for $SU(4)$}
  \label{fig:squiver}
\end{figure}

We can build a quiver whose Seiberg-Witten curve is associated to a
sphere with $n$ basic $p_i=1$ punctures only (see fig.
\ref{fig:squiver}). Its gauge coupling parameter space is ${\cal
M}_{n,0}$, as all punctures are equivalent.
This quiver  has the largest possible S-duality group, because all
punctures can be permuted among themselves. We can use this
S-duality to learn a lot of useful information.  For example the
degeneration limit where two punctures $p_i=1$ collide corresponds
to the weak coupling limit of an $SU(2)$ gauge group, coupled to a
single fundamental hypermultiplet and to the $SU(2)$ flavor symmetry
of a puncture associated to the partition $N= (N-2)+1+1$.

A simple consequence is that $SU(N)$ with $N_f=2N$ should enjoy a
form of Argyres-Seiberg duality. Indeed the theory is associated to
a sphere with two basic $p_i=1$ punctures and two full $p_i=i-1$
punctures (as for $N=2,3$ only a $U(1)^2 \times SU(N)^2$ subgroup of
the flavor group is manifest). The usual weak coupling limit
corresponds to the degeneration of the sphere which brings two
different types of punctures together. The very strongly coupled
region corresponds to the limit where the two basic punctures
collide. In that region, the theory must have a realization as a
$SU(2)$ dual gauge theory coupled to a single fundamental
hypermultiplet and to an interacting SCFT with $SU(2) \times SU(2N)$
flavor symmetry (no further enhancement is expected for $N>3$).

More generally, the collision of $m<N-1$ of the $p_i=1$ punctures
corresponds to the weak coupling limit of an $SU(m)$ gauge group,
and leaved behind a puncture associated with the partition $N=
(N-m)+1+1+\cdots 1$. We will call such partitions ``L-shaped'' Young
Tableaux. The collision of $N-1$ or more $p_i=1$ punctures
corresponds to the weak coupling limit of an $SU(N)$ gauge group,
and leaves behind a full puncture $p_i=i-1$. We can identify the
behavior of this simple quiver theory at all strongly coupled cusps
of moduli space, and see a dual weakly coupled description involving
a generalized quiver, with special unitary gauge groups and several
distinct building blocks, which are irreducible theories associated
to three-punctures spheres with L-shaped Young Tableaux labeling the
punctures. We can use decoupling limits to produce a general theory
${\cal T}_{(f_a),0}$ involving only L-shaped Young Tableaux at
punctures, and we can gauge pairs of $SU(N)$ flavor symmetries at
full punctures $p_i=i-1$ to engineer ${\cal T}_{(f_a),g}$, again
with L-shaped Young Tableaux only.

\begin{figure}
  \begin{center}
    \includegraphics[width=2.2in]{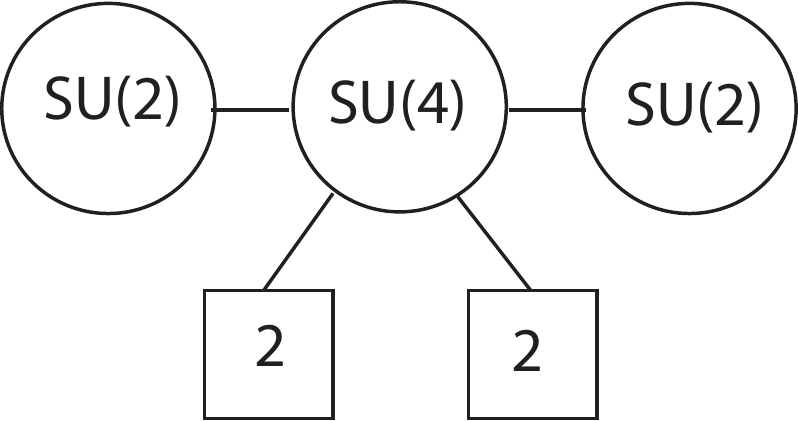} \hfill  \includegraphics[width=2.5in]{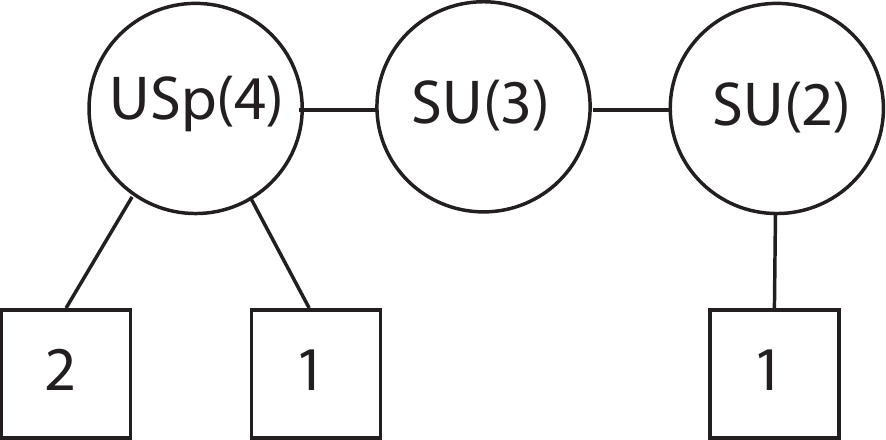}
  \end{center}
  \caption{Left: a quiver associated to a sphere with $4$ basic $p_i=1$ punctures and two punctures labeled by Young Tableaux with two columns of height $2$
  The theory has a $U(1)^4 \times SU(2)^2$ natural flavor symmetry, actually enhanced to $U(1)^3 \times SU(4)$. Right: In the very strongly coupled limit where the two special punctures
  are closed together, an S-dual weakly coupled quiver emerges, with a $USp$ gauge group.
  Notice that the $SU(4)$ flavor symmetry is realized as the $SO(6)$ flavor symmetry of the three fundamentals of $USp(4)$}
  \label{fig:su4sp1}
\end{figure}

\begin{figure}
  \begin{center}
    \includegraphics[width=2.7in]{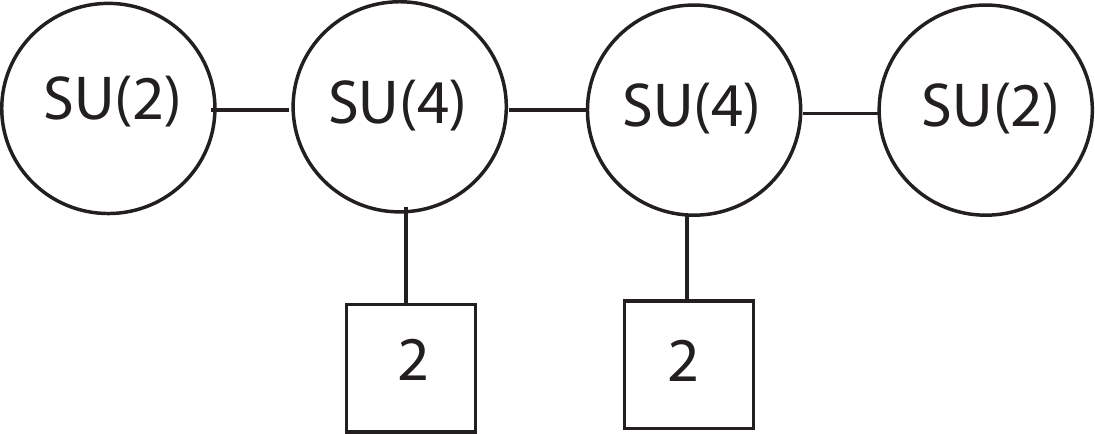} \hfill  \includegraphics[width=2.7in]{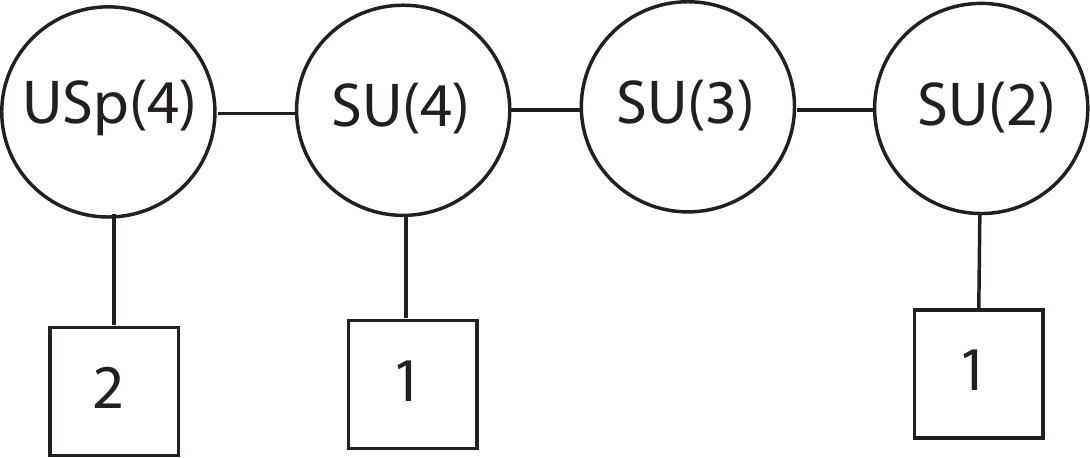}
  \end{center}
  \caption{Another pair of S-dual quivers associated to a sphere with $5$ basic $p_i=1$ punctures and two punctures labeled by Young Tableaux with two columns of height $2$}
  \label{fig:su4sp2}
\end{figure}

\begin{figure}
  \begin{center}
    \includegraphics[width=5in]{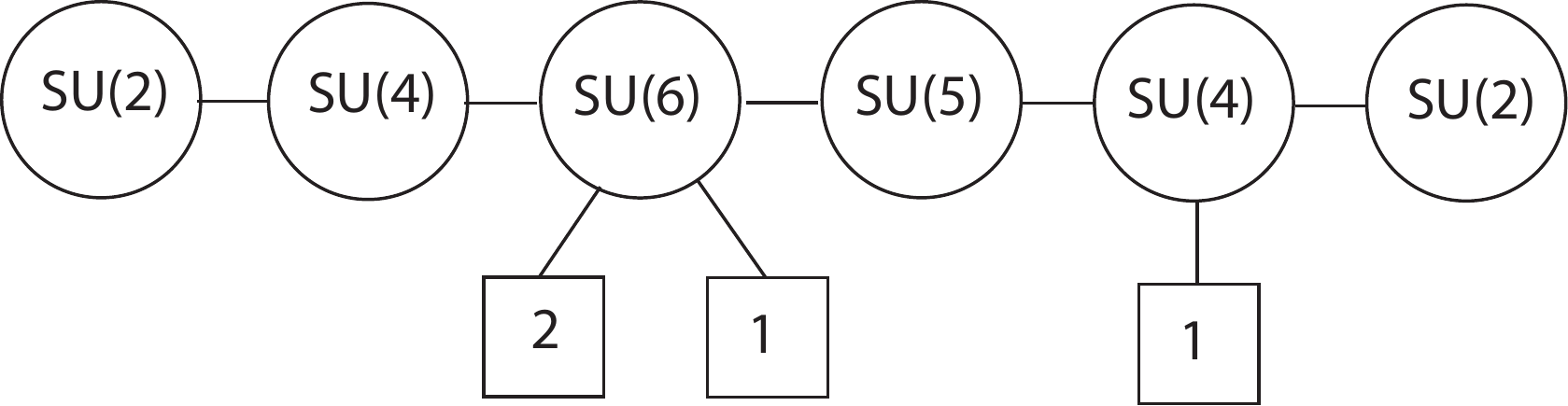} \\ \vspace{.2in}  \includegraphics[width=5in]{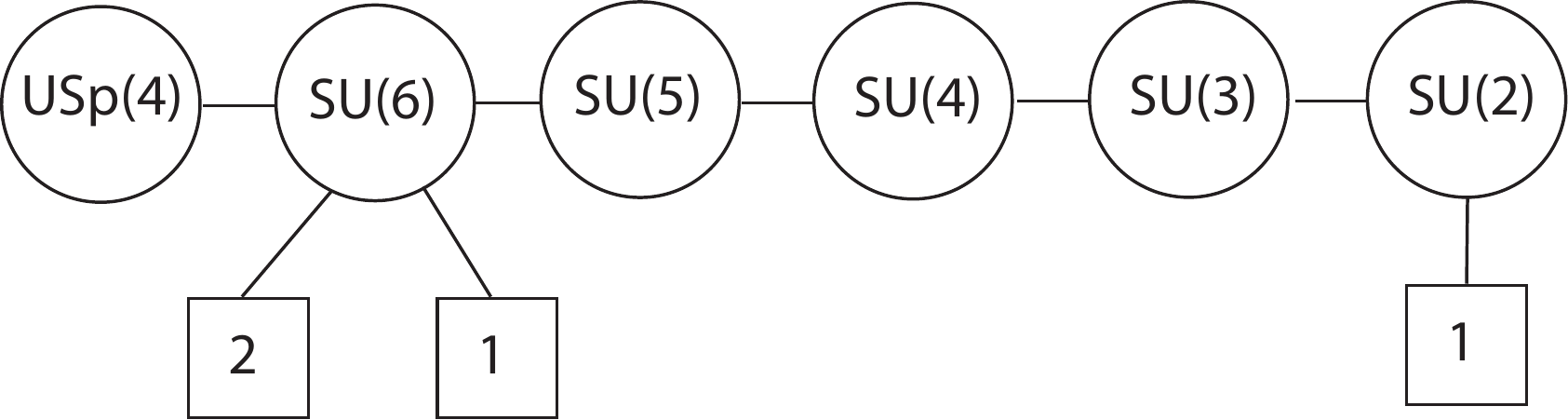}
  \end{center}
  \caption{A third example of S-dual quivers, associated to a sphere with $7$ basic $p_i=1$ punctures,
  one puncture labeled by a Yang Tableau with two columns of height $3$, one puncture labeled by a Yang Tableau with a column of height $4$, one of height $2$ }
  \label{fig:su4sp3}
\end{figure}

What about theories with more general Young Tableaux at the
punctures? We have linear quivers representing theories with up to
two general punctures. S-duality is not strong enough to predict the
behavior at all cusps in parameter space, and there is no way to
study the collision of the two generic punctures directly. Still, we
can look carefully at the degeneration limit of the Seiberg-Witten
curve, and hazard some educated guesses. We find a surprise. The
dual weakly coupled gauge groups arising at the very strongly
coupled region of the parameter space where the two generic
punctures collide is typically $SU(N)$ or some other $SU(I)$ with
$I<N$, but there is a special possibility: the collision of two
punctures associated to Young Tableaux with two columns only. Then
the decoupling gauge group appears to be a symplectic group,
possibly coupled to some extra fundamentals. Among the examples, we
see linear quivers of unitary groups, terminated by the symplectic
group. See figures \ref{fig:su4sp1}, \ref{fig:su4sp2},
\ref{fig:su4sp3} for some nice examples. These quivers admit a brane
realization involving O6 planes \cite{Argyres:2002xc}. The M-theory
lift should still involve the $A_{N-1}$ theory on a sphere. The
geometry of the O6 planes will lead to some appropriate defects. Our
result indicates that the ``O6 defect'' leading to symplectic groups
must coincide with a pair of standard defects with two columns.

The educated guess on the decoupling gauge group gives us a good
control on the cusps in parameter space for general linear quiver
gauge theories. Through decoupling limits, we can enlarge further
the set of irreducible SCFTs corresponding to three punctured
spheres which admit a direct construction, and can be used as
elementary building blocks for generalized quivers. Unfortunately,
not all conceivable theories associated to several punctures of
generic type can be constructed from decoupling limits of linear
quiver gauge theories. Clearly we can start with a general linear
quiver, and produce theories associated with two general punctures,
and one or more L-shaped Young Tableaux. On the other hand we expect
that an interacting SCFTs should exist, associated to spheres with
three punctures labeled by any three Young Tableaux with $N$ boxes.
A six-dimensional construction involving the compactification of the
$A_{N-1}$ $(2,0)$ theory on the three-punctured sphere gives us a
definition, but we would prefer a purely four dimensional
construction. There are indications that an alternative construction
may lie in a rich network of RG flows between such theories,
initiated by expectation values in appropriate Higgs branches. We
will not pursue this idea further in this paper.

\begin{figure}
  \begin{center}
    \includegraphics[width=5.5in]{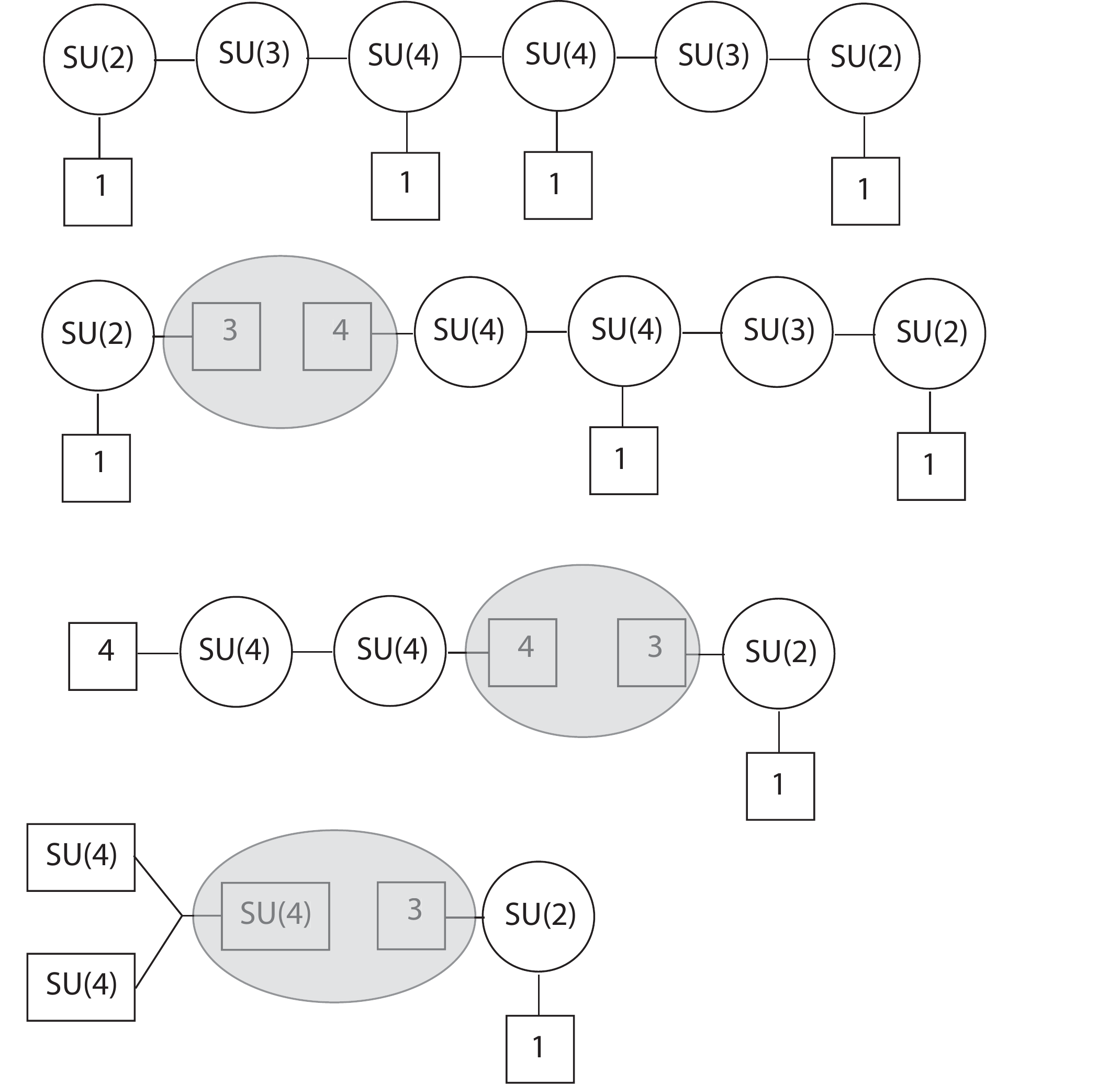}
  \end{center}
  \caption{The construction of $T[A_3]$ from the linear quiver associated to the sphere with $9$ basic punctures. \newline
  Collision of three punctures decouples an $SU(3)$ gauge group, coupled to a $SU(2)$ $N_f=4$ theory and to the rest of the quiver. \newline
  Collision of three more punctures decouples a second $SU(3)$ gauge group, coupled to a $SU(2)$ $N_f=4$ theory and to the rest of the quiver. \newline
  Collision of the last three punctures happen in a very strongly coupled cusp of the original quiver, and
  must decouple a third $SU(3)$ gauge group, coupled to a $SU(2)$ $N_f=4$ theory and to a $SU(3)$ subgroup of a $SU(4)$ flavor group of $T[A_3]$.}
  \label{fig:tn}
\end{figure}

The ``unconstructible'' theories are a finite minority for each
given $N$. Most theories admit a realization as generalized
quivers of simpler building blocks. There is an elementary building block that
really stands out of the crowd.  It is a $SU(N)$ generalization of the theory four hypers for $SU(2)$, or
the $E_6$ theory for $SU(3)$: an interacting SCFT with three $SU(N)$ flavor
symmetry groups which, when gauged, provide half the amount of
matter needed by a conformal $SU(N)$ gauge theory. This theory is
associated to a sphere with three full $p_i=i-1$
punctures. We will simply denote it as $T[A_{N-1}]$. We can build it
from the linear quiver with $3(N-1)$ basic $p_i=1$ punctures, in the
limit where three groups of $N-1$ punctures collapse. (see fig.
\ref{fig:tn} for an $SU(4)$ example)

Starting from $2g-2$ $T[A_{N-1}]$ theories and gauging $3g-3$ pairs
of $SU(N)$ flavor symmetry groups we can define a ${\cal
T}_{g}[A_{N-1}]$ theory which we associate to a Riemann surface of
genus $g$. We identify this theory as the theory of $N$ M5 branes
(or better, the twisted $A_{N-1}$ $(2,0)$ theory) wrapping a Riemann
surface of genus $g$.

\subsection{The linear quivers}
In the formalism of \cite{Witten:1997sc}, $SU(N)$ with $N_f = 2N$
has a SW curve of the form
\begin{equation}
v^N t^2+ c_1 (v^N - u^{(2)} v^{N-2} - u^{(3)} v^{N-3}  \cdots - u^{(N)}) t + c_2 v^N
\end{equation}
and SW differential $\lambda = v \frac{dt}{t}$. The usual
manipulations lead us to
\begin{equation}
x^N = \sum_{d=2}^N \frac{u^{(i)}}{(t-1)(t-t_1)t^{i-1}} x^{N-i}
\end{equation}
We see that $\phi_i$ have simple poles at the $t=1,t_1$ punctures,
and poles of order $i-1$ at the $t=0,\infty$ punctures. Remember
that the brane construction only makes a $U(N) \times U(N)$ subgroup
of the full flavor symmetry group manifest. Following the example of
the $SU(2)$, $SU(3)$ case, we would like to associate the punctures
at $t=1,t_1$ with the two $U(1)$ factors, and the punctures at $t=0,
\infty$ with the $SU(N)$ factors. We can verify the identification
through the mass deformed curve. The mass-deformed curve is given as
a general polynomial of degree $N$ in $v$ and $2$ in $t$.
\begin{equation}
(t-1)(t-t_1) v^N = M_{2}(t) v^{N-1} + \sum_{d=2}^N P_2^{(i)}(t) v^{N-i}
\end{equation}
The equation can be brought to the canonical form by usual shift of
$v$ to remove the coefficient of $v^{N-1}$, and map to coordinates
$(x,z)$.

At the punctures $t=1,t_1$ in the original curve, $\lambda$ has a
pole on a single branch of $v(t)$. The residue is identified with
the mass parameter of a $U(1)$ factor. After the shift of $v$, the
residues of $\lambda$ at $t=1,t_1$ on the $N$ branches of $x(z) dz$
take the form $((N-1) m, -m, \cdots,-m)$. We conclude that a
puncture with all $p_i=1$ has a mass deformation associated to the
partition of $N$ $N=(N-1)+1$ and $U(1)$ flavor symmetry. On the
other hand the limiting values of $v$ at $t=0, \infty$ were generic.
After the shift the residues of $\lambda$ at $t$ zero or infinity
take a generic form $(\sum m_a, -m_1,-m_2 \cdots,-m_{N-1})$ and are
identified with the mass parameters of an $SU(N)$ factor of the
flavor symmetry group. We learn that a puncture with $p_i = i-1$ is
associated to the partition of $N=1+1 \cdots 1$, and a $SU(N)$
flavor symmetry.

We do not gain much by considering a simple linear quiver of $n$
$SU(N)$ gauge groups with $N$ fundamentals at each end. The SW curve
of the theory is associated to a sphere with $n+1$ punctures of the
$p_i=1$ type, and $2$ of the $p_i=i-1$ type. Both the mass deformed
SW curve and the flavor symmetry group $SU(N)^2 \times U(1)^{n+1}$
confirm the conclusions of the $SU(N)$ $N_f = 2N$ analysis. We gain
more insight from a linear conformal quiver associated to a sphere
with $n+3$ identical, simple punctures. The quiver has gauge groups,
in order, $SU(2)\times SU(3) \times SU(4) \cdots SU(N)^{n-2N+4}
\cdots SU(4) \times SU(3) \times SU(2)$. Conformal symmetry requires
a fundamental hypermultiplet at the $SU(2)$ nodes and at the first
and last $SU(N)$ node. This quiver has the desired $U(1)^{n+3}$
flavor symmetry. The SW curve for such a quiver is described in
\cite{Witten:1997sc}. It can be written as
\begin{align}
&v^N t^{n+1} + c_1 v^{N-2} (v^2 - u^{(2)}_1) t^n + c_2 v^{N-3} (v^3 - u^{(2)}_2 v - u^{(3)}_2 ) t^{n-1} +\cdots \notag \\ &+c_{N-1} (v^N - u^{(2)}_{N-1} v^{N-2} - \cdots - u^{(N)}_{N-1}) + \cdots \notag \\
& + c_{n-N+2} (v^N - u^{(2)}_{n-N+2} v^{N-2}- \cdots - u^{(N)}_{n-N+2})+ \cdots \notag \\ &+c_{n-1} v^{N-3} (v^3 - u^{(2)}_{n-1} v - u^{(3)}_{n-1} ) t^2 + c_{n}v^{N-2} (v^2 - u^{(2)}_n) t + c_{n+1}v^N=0
\end{align}
with the usual SW differential. We can collect the same powers of $v
= t x$ to get
\begin{equation}
\prod_{a=0}^n (t-t_a) t x^N = \sum_{i=2}^N U^{(i)}_{n+3-2i}(t)x^{N-i} \end{equation}

All the differentials of various degrees have simple poles at all
$n+3$ punctures $0,\cdots, t_a,\cdots \infty$. This indicates that
the theory has a large S-duality group, similar to the one for
$SU(2)$ linear quivers, which can permute the punctures freely. The
correct gauge coupling parameter space of this theory should be ${\cal
M}_{n+3,0}$. We can use a fractional linear transformation to put
all punctures on the same footing.
\begin{equation}
\Delta_{n+3}(t) x^N = U^{(2)}_{n-1}(t)x^{N-2} + U^{(3)}_{n-3}(t) x^{N-3} + \cdots U^{(N)}_{n+3-2N}
\end{equation}

The mass deformation of the curve takes the form
\begin{align}
&(v-m_1)(v-m_2)^{N-1} t^{n+1} + c_1 (v-m_2)^{N-2} (v^2 +\cdots) t^n + \notag \\ &+c_2 (v-m_2)^{N-3} (v^3 +\cdots) t^{n-1} +\cdots +c_{N-1} (v^N +\cdots) + \cdots\notag \\
&  c_{n-N+2} (v^N +\cdots)+ \cdots c_{n-1} (v-m_3)^{N-3} (v^3 +\cdots) t^2 + \notag \\ &+ c_{n}(v-m_3)^{N-2} (v^2 +\cdots) t + c_{n+1}(v-m_3)^{N-1} (v-m_4)=0
\end{align}
The form of the curve is quite forbidding, but what matters is the
pattern of $(v-m_a)$ factors at the beginning and at the end of the
polynomial. We have seen a similar pattern in the $SU(3)$ case. That
pattern has a very simple meaning. It is necessary and sufficient to
insure that the asymptotic behavior of the $N$ branches of $v(t)$ at
$t=0$ is $(m_4 + O(t), m_3 + O(t), m_3 + O(t), m_3 + O(t) \cdots,
m_3 + O(t))$ and similarly at $t =\infty$. This means that the SW
differential $\lambda$ has simple poles at all $n+3$ punctures
$t=0,\cdots,t_a,\cdots \infty$, and that after the usual shift of
$v$, the residues of $\lambda$ take the expected form $((N-1) \tilde
m_a, -\tilde m_a, \cdots,-\tilde m_a)$. All punctures have $p_i=1$
and $U(1)$ flavor symmetry.

We can also produce linear quivers with $n+2$ punctures of this
simple type, and one completely general. The idea is to tweak one of
the ``tails'' of the basic quiver, to get a more general gauge group
$SU(n_1)\times SU(n_2) \times SU(n_3) \cdots \times SU(n_k) \times
SU(N)^{n-k-N+2} \cdots SU(4) \times SU(3) \times SU(2)$. The quiver
requires $F_i = 2 n_i - n_{i-1} - n_{i+1}$ flavors at the $SU(n_i)$
node for conformal invariance, and $N-n_k$ at the first $SU(N)$
node. Here we assume $n_i<N$ and $2n_i \geq n_{i-1} + n_{i+1}$. It
is useful to associate the rows of a Young Tableaux of $N$ boxes to
the nonincreasing sequence of positive numbers $q_i = n_{i+1}-n_i$.
Notice that $\sum q_i = N$.

The curve from \cite{Witten:1997sc} takes the intuitive form
\begin{align}
&v^N t^{n+1} + c_1 v^{N-n_1} (v^{n_1} - u^{(2)}_1 v^{n_1-2} + \cdots +u^{(n_1)}_1) t^n + \notag \\
&+c_2 v^{N-n_2} (v^{n_2} - u^{(2)}_1 v^{n_2-2} + \cdots +u^{(n_2)}_1)t^{n-1} +\cdots c_{k+1} (v^N - u^{(2)}_{k+1} v^{N-2} - \cdots - u^{(N)}_{k+1}) + \cdots \notag \\
& + c_{n-N+2} (v^N - u^{(2)}_{n-N+2} v^{N-2}- \cdots - u^{(N)}_{n-N+2})+ \cdots \notag \\ &+c_{n-1} v^{N-3} (v^3 - u^{(2)}_{n-1} v - u^{(3)}_{n-1} ) t^2 + c_{n}v^{N-2} (v^2 - u^{(2)}_n) t + c_{n+1}v^N=0
\end{align}
We can collect the same power of $v$. The structure of the $v$
factors implies that the coefficient of $v^{N-i}$ has degree
$d_i=n+1-s$ in $t$ if and only if $i\leq n_s$ and $i>n_{s-1}$. The
coefficient contains a factor of $t^{i-1}$
\begin{equation}
\prod_{a=0}^n (t-t_a) t^N x^N = \sum_{i=2}^N
U^{(i)}_{d_i-i+1}(t)t^{N-i+i-1} x^{N-i} \end{equation} We can erase
a common $t^{N-1}$ factor. We see that $\phi_i$ has a simple pole at
all $t=0,\cdots t_a$, but a pole of order $p_i = d_i-i+1-n-2+2i
=d_i+i-n-1=i-s$ at $t=\infty$ if and only if $i\leq n_s$ and
$i>n_{s-1}$. In other words, there are $L_s=n_{s+1}-n_s$ $p_i$ with
value $i-1-s$. The string of numbers $i-p_i$ is just the height of
the $i-th$ box in the Young Tableaux.

Now we can introduce mass deformations, following
\cite{Witten:1997sc}.
\begin{align}
&\prod_{s}(v-m_s)^{l_s} t^{n+1} + c_1 \prod_{s}(v-m_s)^{l_s-1} (v^{n_1} + \cdots) t^n + \notag \\
&+c_2 \prod_{s |_s>1}(v-m_s)^{l_s-2}(v^{n_2} + \cdots)t^{n-1} +c_3 \prod_{s |_s>2}(v-m_s)^{l_s-3}(v^{n_3} + \cdots)t^{n-2}+\cdots \notag \\ &+c_{k+1} (v^N + \cdots) + \cdots
+ c_{n-N+2} (v^N +\cdots)+ \cdots \notag \\&+c_{n-1} (v-\tilde m_1)^{N-3} (v^3 +\cdots) t^2 + c_{n}(v-\tilde m_1)^{N-2} (v^2 +\cdots) t + c_{n+1}(v-\tilde m_1)^{N-1} (v-\tilde m_2)=0
\end{align}
The powers $l_s$ are such that
\begin{equation} \sum_{s | l_s > r-1}(l_s-r) = N-n_r
\end{equation}
It is easy to verify that the $l_s$ are the lengths of columns of
the Young Tableaux whose rows are $q_r$. We recognize that the mass
deformed puncture at infinity corresponds to the partition $N=\sum
l_s$. As a final check, we see from the linear quiver that a whole
$S(\prod U(F_i))$ factor of the flavor symmetry group should be
associated to this puncture, as the other punctures only account for
the mass parameters of $n+2$ $U(1)$ gauge groups.

We found that we need to consider in general punctures labeled by a
Young Tableaux of $N$ boxes, from which we can read the admissible
poles of $\phi_i$ at the puncture, the structure of the
mass-deformed puncture and the flavor symmetry factors associated to
that puncture. Notice that each $SU(F_i)$ factor of the flavor
group, if weakly gauged, would have $n_i$ flavors. $n_i$ is strictly
bigger than $F_i$, with a single exception. Only the ``largest''
type of puncture with $p_i=i-1$ has an $SU(N)$ flavor symmetry which
would contribute to an $SU(N)$ beta function the same way as $N$
flavors.
\subsection{Degeneration limits}
It is easy to see that linear quivers with two generic tails are
associated to a sphere with two generic punctures and several basic
ones. Can we produce theories associated to a sphere with several
generic punctures? We can surely write down a tentative
Seiberg-Witten curve for the theory, starting from degree $i$
differentials with appropriate $p_i$ poles at each puncture.
Following the example of the $SU(2)$ and $SU(3)$ cases, we can use
the brane construction for the linear quiver examples to argue that
the $A_{N-1}$ $(2,0)$ theory possesses codimension two defect
operators associated with all possible types of punctures. Then we
can consider the twisted compactification of the $A_{N-1}$ theory on
the Riemann surface, in the presence of appropriate defect operators
at all punctures.

We would like to accomplish two goals from a purely four dimensional
point of view: determine the nature of the ``very strongly coupled
regions'' of moduli space of familiar theories, possibly in terms of
generalized quivers, and characterize the ensemble of generalized
quivers which corresponds to S-dual frames of the same theory. What
happens when punctures collide? The linear quivers we have available
can only tell us what happens when a generic puncture collide with a
$p_i=1$ basic puncture. Up to S-duality, we can always pick the
basic puncture at $t_n$ as the one colliding with the puncture at
infinity. This makes the gauge coupling at the last $SU(n_1)$ node
of the quiver very weak.

After decoupling, the differences $n_i-n_{i-1}$ are unchanged,
except at the last node of the quiver. The new Young Tableaux has
rows of the same length as the one for the original quiver, except
that the two longest rows, of length $n_1$ and $n_2-n_1$, have been
combined into a single row of length $n_2$. Correspondingly, the
degree $p_i$ of the poles of the differentials $\phi_i$ increases by
one, except for $i\leq n_1$, where it stays $i-1$. The result is
unsurprising: if a pole of order $p_i$ and a pole of order $1$ are
brought together naively, the resulting pole will have degree
$p_i+1$. The first $n_1$ differentials had $p_i=i-1$, and after the
collision they have poles of degree $p_i=i$. The coefficients of the
poles are naturally associated to the $u_i$ parameters of the
decoupling $SU(n_1)$ gauge group, which become mass parameters after
decoupling. This covers essentially all degeneration limits of a
linear quiver with zero or one generic punctures. It will be quite
harder to determine the result of a collision between generic
punctures, which is really needed to understand the very strongly
coupled region of general conformal linear quivers where the two
generic punctures collide.

Lets start with the simplest theory of rank $N$, with $n+3$ simple
punctures on the sphere. The gauge coupling parameter space is the
moduli space of $n+3$ points on the sphere, and has boundaries where
$k+2$ punctures come together, or better, the sphere develops a
nodal singularity, separating two spheres with, say, $k+3$ punctures
and $n-k+2$ ones, joined at a puncture. Lets assume that $k+3 \leq
n-k+2$. Again, without loss of generality, we can assume that the
punctures are the $t_a$ for $a\leq k$. In that case, the gauge
coupling of the $k+1$-th node of the linear quiver is becoming very
weak. We can distinguish several possibilities. If $k+2>N$, the
gauge group is just an $SU(N)$. The quiver gauge theory decomposes
into two superconformal linear quivers with the general structure
$SU(2)-SU(3)-\cdots-SU(N)-SU(N)-\cdots SU(N)$. These are associated
respectively to a sphere with $k+2$ basic punctures and one full
one, and a sphere with $n-k+1$ basic punctures and a full one.

Hence if more than $N$ basic punctures come together, the simple
picture where a sphere degenerates into two spheres, as a long tube
pinches off and is replaced by two new full $p_i=i-1$ punctures,
works well. This is a very natural result. Both fragments are still
associated with rank $N$ theories, and have a full puncture with a
$SU(N)$ flavor symmetry. The anomaly for the flavor symmetry is such
that, when gauged, contributes to half of the amount of matter
needed for a conformal $SU(N)$ gauge theory. This simple picture
needs a refinement for collisions involving $N$ or fewer basic
punctures.

If we bring together $N$ punctures, the gauge group which decouples
is the first $SU(N)$ gauge group of the quiver. The theory
decomposes into three pieces: a long linear quiver with gauge groups
$SU(2)-SU(3)-\cdots-SU(N)-SU(N)-\cdots SU(N)$, a ``triangular''
quiver with gauge groups $SU(2)-SU(3)-\cdots-SU(N-1)$ and a single
fundamental hypermultiplet for the $SU(N)$ gauge group. The long
quiver, as usual, is a rank $N$ theory associated to a sphere with
$n-N+5$ simple punctures and a full one. From the punctured sphere
picture, we would have expected to see a second theory appear,
associated to a sphere with $N$ basic punctures and a full one. This
tentative theory has a bit of a problem: a nonzero degree $i$
differential on the sphere must have poles of a total degree $2i$.
$N$ simple punctures and a full one give poles of total degree
$2N-1$ for $\phi_N$, hence in general the last coefficient of the
defining polynomial $x^N - \sum x^{N-i} \phi_i=0$ is zero and the SW
curve simplifies to the curve for a rank $N-1$ theory. The resulting
rank $N-1$ curve, with $N$ basic punctures and a single full
puncture indeed coincides with the curve for the triangular quiver
with gauge groups $SU(2)-SU(3)-\cdots-SU(N-1)$. It is a bit harder
to ``see'' the fundamental hypermultiplet. If we turn on mass
parameters for the $SU(N)$ flavor symmetry group, $\phi_N$ can be
non-zero, as the total degree of the poles for $\phi_N$ raises to
$2N$. The SW curve does not factorize anymore, and appears very
different from the mass-deformed rank $N-1$ SW curve for the
triangular quiver. Inspection of specific, simple examples for low
$N$ show that the two SW curves actually coincide, but the map
between the two mixes the coordinates $x,t$ on the base and the
fibre of $T^* \IC\IP^1$. It is interesting that the rank $N$
realization of the theory makes the $SU(N)$ flavor symmetry group of
the quiver manifest, while it would appear as an accidental
enhancement in the rank $N-1$ realization, which has a full puncture
with naive flavor symmetry $SU(N-1)$ and $N$ simple punctures.

If we bring together $N-1$ punctures, a $SU(N-1)$ gauge group
becomes weakly coupled. The main part of the quiver still terminates
on a $SU(N)$ group with $N$ flavors and a full $SU(N)$ flavor
symmetry, of which a $SU(N-1)$ subgroup is weakly coupled. The
remaining gauge groups form a triangular quiver of gauge nodes
$SU(2)-SU(3)-\cdots-SU(N-2)$. Now if mass parameters are absent,
both $\phi_N$ and $\phi_{N-1}$ must be zero, because $N-1$ simple
punctures and a full one still cannot provide poles of sufficiently
high overall rank. The SW curve simplifies to a rank $N-2$ one. Even
if one allows for mass parameters for the $SU(N-1)$ flavor symmetry,
$\phi_N$ still has to be zero, and we find again the rank $N-1$
realization of the rank $N-2$ triangular quiver.

Finally, if we bring together $k+2<N-1$ punctures, a $SU(k+2)$ gauge
group becomes weakly coupled. The main part of the quiver terminates
on a $SU(k+3)$ group with $k+2$ fundamental flavors. If we bring
together $k+2$ basic punctures we expect to find poles of degree up
to a maximum of $k+2$ for the $\phi_i$. Indeed the big fragment of
the linear quiver is associated to a sphere with $n-k+1$ basic
punctures and one with $p_i=min(i-1,k)$, and an L-shaped Young
Tableaux. The remaining gauge groups form a triangular quiver of
gauge nodes $SU(2)-SU(3)-\cdots-SU(k+1)$. Indeed if mass parameters
are absent, the SW curve simplifies to a rank $k+1$ one.

Now we are in position to explore any cusp in the gauge coupling
prameter space, where the sphere factorizes into a generic tree of
three-punctured spheres. The nodes between spheres are associated to
weakly coupled $SU(k+2)$ gauge groups with $k$ up to $N-2$. Each
node splits the sphere into a sphere with $k+3$ punctures and one
with $n-k+2$ punctures for some $k+3 \leq n-k+2$, and is assigned a
$SU(k+2)$ gauge group if $k+2<N$, $SU(N)$ otherwise. If $k+2=N$ the
$SU(N)$ gauge group is coupled to an extra fundamental
hypermultiplet.

Correspondingly, each three-punctured sphere will have punctures
with $p_i=min(i-1,k_a)$, $a=1,2,3$. The corresponding theory has
rank $min((k_1+k_2+k_3)/2,N)$ (or $min((k_1+k_2+k_3-1)/2,N)$,
whatever is integer), as the differentials of higher degree are
necessarily zero. If one of the punctures is basic, say $k_3=1$,
comparison with the linear quiver identifies the theory as a set of
free hypermultiplets for $SU(k_1) \times SU(k_2)$. Otherwise, the
theory is an interacting SCFT, with a non-trivial Coulomb branch and
flavor symmetry $U(k_1) \times U(k_2) \times U(k_3)$.

Notice that the anomaly for the $SU(k_3)$ flavor currents of such a
theory is typically the same as the anomaly of $k_3+1$ fundamentals
of $SU(k_3)$. Indeed the theory can be produced as the degeneration
limit of a linear quiver with one puncture $p_i=min(i-1,k_a)$ and
$k_1+k_2$ basic punctures. If $k_1+k_2>k_3+1$, $SU(k_3)$ is the
flavor symmetry of $k_3$ fundamentals of $SU(k_3+1)$. The anomaly is
unaffected by the degeneration limit. Hence two ``typical'' theories
cannot be coupled together by a conformal invariant $SU(k)$ gauge
group. We saw two exceptions to this obstruction. If $k_1+k_2=k_3$,
the theory has rank $k_3$, and can be produced as the degeneration
of a quiver with a full puncture of $SU(k_3$ flavor symmetry and
$k_1+k_2=k_3$ basic punctures. This is a triangular quiver of gauge
groups $SU(2) \times SU(3) \times \cdots \times SU(k_3-1)$, with
$k_3$ fundamentals at the $SU(k_3-1)$ node. The currents for the
$SU(k_3)$ flavor symmetry of the triangular quiver have the anomaly
of $k_3-1$ fundamentals of $SU(k_3)$ only. That must also be the
anomaly of the $SU(k_3)$ flavor currents for the three-sphere theory
with $k_1+k_2=k_3$. This allows a conformal $SU(k_3)$ coupling to a
theory with the anomaly of $k_3+1$ fundamentals of $SU(k_3)$. A
second exception are full $SU(N)$ punctures, which have the anomaly
of $N$ fundamentals of $SU(N)$, and can always be coupled pairwise.

The tree will have a ``core'' of three-spheres connected by $SU(N)$
gauge groups, which gauge the flavor symmetries of pairs of full
punctures, each contributing half of the required beta function for
conformality. Outer branches of the tree are connected by $SU(k)$,
$k<N$ gauge groups, and the beta function receives a contribution
proportional to $k-1$ from the puncture on the side of the tree
core, and $k+1$ from the puncture on the outer side. They must be
made by three spheres with $k_3=k_1+k_2$ only. Clearly, starting
from the linear quiver with basic punctures and colliding them, one
can generate a rank $N$ theory with an arbitrary punctures
$p_i=min(i-1,k_a)$, corresponding to L-shaped Young Tableaux. The
degeneration limits of such a theory involves the same building
blocks as above. Further gauging pairs of $SU(N)$ flavor symmetries
leads to theories which are naturally associated to Riemann surfaces
of higher genus, with punctures labeled by L-shaped Young Tableaux.

Out of this crowd of building blocks, a special role is clearly
played by the theory with three full punctures of flavor symmetry
$SU(N)$ on the sphere. It is relatively easy to ``build'' this
crucial theory. Consider the linear quiver theory with $3N-3$
punctures of the $p_i=1$ type, gauge group $SU(2) \times SU(3)
\cdots SU(N-1) \times SU(N)^{N-2} \times SU(N-1) \times SU(N-2)
\cdots SU(2)$. If we bring $N-1$ punctures together, we know that we
can work in a duality frame where a $SU(N-1)$ gauge group is
becoming weakly coupled, splitting the quiver into a triangular
$SU(2)\times SU(3) \cdots \times SU(N-2)$ and the main part of the
quiver with $2N-2$ gauge groups $SU(N)^{N-2} \times SU(N-1) \times
SU(N-2) \cdots SU(2)$. This remaining quiver has one puncture with
$SU(N)$ flavor symmetry and $2N-2$  with $p_i=1$. Next we can bring
together other $N-1$ basic punctures. Again, there is a duality
frame in which the other $SU(N-1)$ gauge group becomes weakly
coupled, another triangular quiver decouples, and leaves behind a
chain of $N-2$ $SU(N)$ groups. This quiver has two punctures with
$SU(N)$ flavor symmetry, and $N-1$ $p_i=1$ punctures. Finally, we
can bring these $N-1$ punctures together. By symmetry, we know that
another hidden $SU(N-1)$ must become weakly coupled in this
degeneration limit. This group must be coupled to a triangular
quiver $SU(2)\times SU(3) \cdots \times SU(N-2)$, and to a $SU(N-1)$
subgroup of one of the three $SU(N)$ flavor groups of the
interacting SCFT we are after. We will denote this special $SCFT$ as
${\cal T}_N$. If we place a puncture at $t=0$, one at $t=\infty$,
$T_N$ has a SW curve
\begin{equation}
x^N = \sum_{i=3}^N \frac{P_{i-3}(t)}{t^{i-1} (t-1)^{i-1}}
\end{equation}

Now we can follow the example of $SU(2)$ and $SU(3)$ theories, and
define simple theories associated to Riemann surfaces of genus $g$
and $n$ punctures with $SU(N)$ flavor symmetry, as a generalized
quiver of $g$ loops whose matter consists of an appropriate number
of the ${\cal T}_N$ theories. We can naturally identify such
generalized quiver gauge theories with the twisted compactification
of the $A_{N-1}$ $(2,0)$ theory on the Riemann surface, possibly in the
presence of appropriate defect operators or semi-infinite tubes. All
possible degeneration limits in the moduli space ${\cal M}_{n,g}$
will reduce to an appropriate generalized quiver of $T_N$ theories,
with $g$ loops and $n$ $SU(N)$ flavor groups. Of course we would
like to build theories associated with Riemann surfaces and
punctures of a general type, and give a generalized quiver
description of all possible degeneration limits of the gauge
coupling moduli space.

What is the result of a degeneration limit which brings two generic
punctures of degrees $p_i$, $p'_i$ together? The first question to
ask is which gauge group may become weakly coupled in the limit. The
three punctured sphere will have the original two punctures, and a
third one. Remember that a degree $i$ differential must have poles
of a total degree $2i$ on the sphere. Whenever $p_i+p'_i<i$, even if
the third puncture is a full puncture, with mass parameters turned
on, the $i$-th differential $\phi_i$ will have to be zero. Out of
the naive mass parameters for the full puncture at infinity,
 the dimension $i$ combination can be turned on only if $p_i+p'_i\geq
 i$. If $p_i+p'_i>i$ and we turn off the mass parameters,
$\phi_i = \frac{P_{p_i + p'_i - i -1}}{t^{p_i} (t-1)^{p'_i}}$ and
the Coulomb branch has $p_i + p'_i - i$ operators of dimension $i$.

From the point of view of the remaining Riemann surface, the
differentials $\phi_i$ have naive degree $p_i + p'_i$. Turning off
the Coulomb branch parameters of the theory associated to the
three-punctured sphere naturally reduces that degree to
$\mathrm{min}(p_i+p'_i, i)$. Again, for each $p_i+p'_i\geq i$ the
coefficient of the degree $i$ pole is the degree $i$ combination of
the mass parameters for the  flavor symmetry group. These mass
parameters should be identified with the Coulomb branch fields for
the decoupling gauge group.

Everything would be clear if $p_i+p'_i\geq i$ were true for all $i
\leq I$. Then we would just conclude that an $SU(I)$ gauge group is
becoming weakly coupled. That is almost always true. We know that
$i-p_i$ are the heights of boxed in the Young Tableaux. We can write
$p_i+p'_i-i = i-(i-p_i)-(i-p'_i)$. Observe that this quantity can
drop as $i$ increases by one only if the heights $(i-p_i)$,
$(i-p'_i)$ simultaneously grow. If it drops twice in a row, it means
we are at the top of the tableaux, where rows have length one. Then
the heights of boxes will keep growing, and $p_i+p'_i-i$ have to
keep decreasing. If $p_i+p'_i-i$ drops by one and then remains
constant, or viceversa, it means one of the heights grew twice in a
row, and then it will keep growing. Hence if $p_i+p'_i-i$ drops by
one and then remains constant, or viceversa, it can only remain
constant or decrease as $i$ increases. Hence the only way
$p_i+p'_i-i$ can go from $0$ to $-1$ without remaining negative
afterwards is through a sequence $-1,0,-1,0,-1,\cdots$. In all other
cases, $p_i+p'_i-i$ will be nonnegative until $i=I$ and then remain
negative. The exception corresponds to the sequences $p_i =
(0,1,1,2,2,3,3,\cdots)$ and $p'_i = (0,1,1,2,2,3,3,\cdots)$ which
add to $p_i + p'_i = (0,{\bf 2},2,{\bf 4},4,{\bf 6},6,\cdots)$. We
marked the values of $p_i + p'_i=i$. We see combinations of mass
parameters for even dimensions only.

Now we want to demonstrate that the decoupling gauge group in this
case is a $USp(I)$ group. Let us take $N=2M$ and a collision between
punctures associated with two Young Tableaux with two columns of
height $M$. If we turn off the mass parameters, the differentials on
the three punctured sphere have overall degree of the poles
$1+1+1<4,1+1+2<6, 2+2+3<8, 2+2+4<10,\cdots$, and are all zero! The
theory has no Coulomb branch, and we are led to think it may be a
free theory of hypermultiplets. If we turn on mass parameters at the
full puncture, the differentials have degrees $1+1+2=4,1+1+3<6,
2+2+4=8, 2+2+5<10,\cdots$, and are nonzero, with coefficients given
by Casimirs of a flavor group of even degree only. Call that group
$G_M$. The theory should have an overall $SU(2) \times SU(2) \times
G_M\sim SO(4) \times G_M$ symmetry. It would be very natural to have
$G_M = USp(2M)$. If $G_M$ was an orthogonal group, the
hypermultiplets would have an extra $SU(2)$ flavor symmetry. The
main constraint we need to satisfy is that $G_M$ should be
conformal. It is coupled to flavor currents from the remaining part
of the theory, which had the anomaly of $2M$ fundamentals of
$SU(2M)$. The hypothesis $G_M=USp(2M)$ passes the test! $USp(2M)$ is
conformal when coupled to $2M+2$ fundamentals (which have $SO(4M+4)$
flavor symmetry).

As a simple example we can take the quiver with gauge groups
$SU(2)\times SU(4) \times SU(6) \cdots \times SU(2M)^{n-2M+4} \times
\cdots SU(4) \times SU(2)$, which has $n=1$ basic punctures and two
punctures of the type just analyzed. The very strongly coupled
region where the two special punctures collide will give a new
quiver gauge theory with gauge groups $SU(2) \times SU(3) \times
SU(4) \cdots SU(2M)^{n-2M+3} \times USp(2M)$, as depicted in figure
\ref{fig:su4sp2} in the previous subsection. Such quiver admits a
brane realization including a $O6$ plane, which could be used to
test our claim. See section \ref{sec:last} for more references.

Next we should consider a setup with a Young Tableaux with two
columns, each of length $M$, and a second Young Tableaux with two
columns of different length $M-m$ and $M+m$. The decoupling gauge
group should be a $USp(2M-2m)$. The new puncture created by the
collision has a Young Tableaux with $2M-2m$ columns of height one,
and $2$ of height $m$. Consequently, it has a $U(1) \times
SU(2)\times SU(2M-2m)$ flavor symmetry. The $SU(2)$ flavor group
matches the $SU(2)$ flavor symmetry of the original puncture with
two columns of equal height. From a linear quiver realization of
this puncture, we see that the $SU(2M-2m)$ flavor symmetry has an
anomaly corresponding to $2M-2m+2$ fundamentals. We are led to gauge
a $USp(2M-2m)$ subgroup of $SU(2M-2m)$, as depicted in figure
\ref{fig:su4sp2} in the previous subsection.

For more general pairs of Young Tableaux with two columns of
different length and general $N$, the story is quite similar. The
poles from the two punctures, after we remove the parameters for the
$USp(I)$ gauge group add to the $p_i$ for a Young Tableaux with
several columns of height one, and two of different height.
Typically this Tableau has either a $SU(I)$ flavor symmetry and
$I+2$ anomaly, and we gauge the $USp(I)$ subgroup of $SU(I)$.

As anticipated in the previous subsection, we can interpret this
calculation as a prediction that conformal linear quivers of unitary
groups terminated at one or both ends by symplectic groups are part
of the $A_{N-1}$ class of theories, and are associated to spheres with
two or four punctures with two columns, and several basic punctures.
It would be interesting to verify this statement through the brane
construction of \cite{Argyres:2002xc} involving the M-theory lift of O6
planes.

As a simple example, we can produce the curve for $USp(2M)$ with
$2M+2$ flavors. The curve will make manifest a $SO(4) \times U(2M)$
subgroup of the theory. Take a sphere with two punctures associated
to Young Tableaux with two columns of height $m$, a basic puncture
and a full puncture. We will place the full puncture at infinity,
the basic puncture at $t_1$, the others at $0,1$
\begin{equation}
x^{2M} = \sum_i x^{2M-2i} \frac{u_{2i}}{t^i (t-1)^i (t-t_1)}
\end{equation}
To bring this curve to a more familiar form, we can map $4t = s + 2
+ 1/s)$, $y=x t (t-1)=1/16 (s-1/s)^2 x$
\begin{equation}
(s+1/s+2-4 t_1) y^{2M} = \sum_i 4 u_{2i} y^{2M-2i}
\end{equation}

In the $M=2$ case we can have some fun with another Argyres-Seiberg
duality. $USp(4)$ with $6$ flavor should have a strong coupling
limit dual as a $SU(2)$ gauge theory coupled to the $E_7$ SCFT.
Indeed, if we collide the basic puncture with one of the two special
punctures, we get $p_i = 1+1= \bf{2},1+1=2,2+1=3$, and we see the
$u$ parameter for the decoupling $SU(2)$ gauge group, which gauges a
subgroup of the $SU(4)$ subgroup of a full puncture. The $SU(2)$
commutes with a $SU(2) \times U(1) \in SU(4)$. We deduce that the
$A_3$ theory with two full punctures and a puncture associated to a
Young Tableaux with two columns of height $2$ is the $E_7$ SCFT! The
manifest flavor symmetry subgroups groups are a nice $SU(4)\times
SU(4) \times SU(2)\in E_7$.


All in all, we found that the degeneration limits of a general
linear superconformal quiver, possibly with symplectic groups at the
end, are quite under control. They will in general require building
blocks which correspond to three punctured spheres with two generic
punctures and a full puncture. Such theories can be constructed from
the linear quivers with two generic punctures and $N-1$ basic ones,
bringing together the $N-1$ basic punctures into a full one.

On the other hand, not all possible theories associated with Riemann
surfaces and generic punctures can be built that way. A theory
associated to a three-punctured sphere with three generic punctures
admits no construction of the type outlined until now. There are
indications that such a theory could be produced from $T_N$ by
giving expectation values to fields in the Higgs branch, and flowing
to the IR. We will not follow that lead in this paper. Similarly, we
could consider a theory associated to a four-punctured sphere, with
punctures chosen in such a way that every degeneration limit will
produce a theory with three non-full punctures. For a fixed rank
$N$, these unconstructible theories are a finite set. Most choices
of a Riemann surface and punctures will lead to theories which in
every degeneration limit decompose into three punctured sphere
theories with at least a full puncture, which are constructible.

\section{Further developements}
\label{sec:last}

It would be very natural to extend this work to other class of
theories which admit a brane construction, and can be associated to
the six-dimensional $(2,0)$ theory. The work of \cite{Witten:1997sc}
has been extended in various ways. O6 planes have been used to add
symplectic gauge groups at the end(s) of linear quivers of unitary
gauge groups, but also to add matter in the antisymmetric
representation \cite{Argyres:2002xc}. The M-theory lift still involves
M5 branes with a $A_{N-1}$ $(2,0)$ theory on the worldvolume. We found
in our analysis that the addition of symplectic groups at the end(s)
of a quiver indeed leads to standard theories of the $A_{N-1}$ type,
with appropriate punctures. It would be interesting to check that
the same is true for adding antisymmetry matter representations.

Furthermore, linear quivers of (alternating) orthogonal and
symplectic groups have been engineered by adding a O4 plane \cite{Brandhuber:1997cc}\cite{Landsteiner:1997vd}. There are really two cases,
one involving even orthogonal groups, the other involving odd
orthogonal groups. The former case appears to involve the D type
$(2,0)$ theory. It would be straightforward to classify the possible
defects for the D type theory, along analogous lines to our work.
The latter case appears to involve a SW curve constrained by $t \to
-t$. A map $t' =t^2$ may lead to a SW curve of the $A_{N-1}$ type, with
defects at $t'=0, \infty$ which make the differentials of odd degree
antiperiodic. Again, more work would be needed to fully understand
the situation.

In principle one could also consider E type of $(2,0)$ theory. At
the degeneration limits of the base Riemann surface weakly coupled
exceptional gauge symmetries should appear, conformally coupled to
non-trivial SCFTs with exceptional flavor symmetries. It would be
interesting to study the problem, and identify admissible punctures,
and corresponding flavor symmetries.

Finally, one can try adding orbifold fiveplanes to the standard
brane setup. This should lead to quivers of unitary groups in the
shape of D-type Dynkin diagrams. (see \cite{Gaiotto:2008ak} for
related, three-dimensional setup). It would be interesting to
realize those constructions in detail, and identify which sort of
defects in the $A_{N-1}$ theory they lead to.

We have no idea of how to attack the case of quivers of unitary groups in the shape of E type Dynkin diagrams.

\subsection*{Acknowledgments} We would like to thank J. Maldacena, G. Moore, A. Neitzke,
Y. Tachikawa, N. Seiberg and E. Witten for discussions. D.G. is
supported in part by the DOE grant DE-FG02- 90ER40542 and in part by
the Roger Dashen membership in the Institute for Advanced Study.

\bibliography{all}{}
\end{document}